\documentclass[11pt]{article}

\title{Tight Bounds for Sparsifying Random CSPs\footnote{An extended abstract of this work will appear in the proceedings of SODA 2027. This full version contains complete proofs of all claims made in the paper.}}
\author{Joshua Brakensiek\thanks{University of California, Berkeley. Supported in part by a Simons Investigator award of Venkatesan Guruswami, and NSF awards CCF-2211972 and DMS-2503280. Contact: \href{mailto:josh.brakensiek@berkeley.edu}{josh.brakensiek@berkeley.edu}} \and Venkatesan Guruswami\thanks{Simons Institute for the Theory of Computing and the University of California, Berkeley. Supported in part by a Simons Investigator award and NSF award CCF-2211972. Contact: \href{mailto:venkatg@berkeley.edu}{venkatg@berkeley.edu}} \and Aaron Putterman\thanks{Harvard University. Supported in part by the Simons Investigator awards of Madhu Sudan and Salil Vadhan, NSF award CCF 2152413 and AFOSR award FA9550-25-1-0112. Contact: \href{mailto:aputterman@g.harvard.edu}{aputterman@g.harvard.edu}}}

\date{\today}

\usepackage{amsmath}
\usepackage{amssymb}
\usepackage{amsthm}
\usepackage{mathabx}
\usepackage{graphicx} %
\usepackage[margin=1.1in]{geometry}
\usepackage{hyperref}
\usepackage{thmtools}
\usepackage[nameinlink,capitalise]{cleveref}
\usepackage{xcolor}
\usepackage{multicol}
\usepackage{url}
\usepackage[utf8]{inputenc}
\usepackage{enumitem}

\newcommand{\zo}{\{0,1\}}
\newcommand{\Ham}{\operatorname{Ham}}

\newcommand{\supp}{\operatorname{supp}}
\newcommand{\Supp}{\operatorname{Supp}}

\newcommand{\VCSP}{\operatorname{VCSP}}

\newcommand{\sym}{\operatorname{sym}}
\newcommand{\set}{\operatorname{set}}
\newcommand{\ar}{\operatorname{ar}}
\newcommand{\sT}{\mathsf{T}}
\newcommand{\st}{\mathsf{t}}

\newcommand{\sat}{\mathrm{sat}}

\newcommand{\poly}{\operatorname{poly}}

\newcommand{\cS}{\mathcal S}

\newcommand{\AND}{\mathbf{AND}}
\newcommand{\vD}{\vec{D}}
\newcommand{\vE}{\vec{E}}

\newcommand{\eps}{\varepsilon}

\newcommand{\E}{\mathbb E}
\newcommand{\one}{{\bf 1}}
\newcommand{\wt}{\operatorname{wt}}

\newcommand{\hist}{\operatorname{hist}}

\newcommand{\Hist}{\operatorname{Hist}}
\newcommand{\tot}{\leftrightarrow}

\newcommand{\bR}{\mathbf{R}}
\newcommand{\bS}{\mathbf{S}}

\newcommand{\R}{\mathbb{R}}
\newcommand{\Z}{\mathbb{Z}}

\newcommand{\Sym}{\operatorname{Sym}}

\newtheorem{theorem}{Theorem}
\numberwithin{theorem}{section}
\newtheorem{lemma}[theorem]{Lemma}
\newtheorem{claim}[theorem]{Claim}
\newtheorem{proposition}[theorem]{Proposition}
\newtheorem{corollary}[theorem]{Corollary}

\theoremstyle{definition}
\newtheorem{definition}[theorem]{Definition}
\newtheorem{remark}[theorem]{Remark}

\newtheorem{fact}[theorem]{Fact}

\allowdisplaybreaks

\begin{document}
\pagenumbering{gobble}

\maketitle

\begin{abstract}
The problem of CSP sparsification asks: for a given CSP instance, what is the sparsest possible reweighting such that for every possible assignment to the instance, the number of satisfied constraints is preserved up to a factor of $1 \pm \eps$? Many papers over the years have studied the worst-case sparsifiability for a variety of CSP predicates including the problems of cut (Bencz{\'u}r--Karger), hypergraph cut (Kogan--Krauthgamer) and linear equations (Khanna--Putterman--Sudan). Recently, Brakensiek and Guruswami showed that the size of a worst-case sparsifier for a predicate is precisely governed by the combinatorial \emph{non-redundancy} of the predicate. However, determining the non-redundancy of many simple, concrete predicates is wide open due to the existence of highly-intricate worst-case instances. 

In order to understand the ``typical'' bottlenecks for CSP sparsification, we initiate the study of the sparsification of \emph{random} CSPs. In particular,  we consider two natural random models: the $r$-partite model and the uniform model. In the $r$-partite model, CSPs are formed by partitioning the variables into $r$ parts, with constraints selected by randomly picking one variable out of each part. In the uniform model, $r$ distinct variables are chosen at random from the pool of variables to form each constraint.

In the $r$-partite model, we exhibit a sharp threshold phenomenon. For every predicate $P$, there is an integer $k$ such that a random instance on $n$ vertices and $m$ edges cannot (essentially) be sparsified if $m \le n^k$ and can be sparsified to size $\approx n^k$ if $m \ge n^k$. Here, $k$ corresponds to the largest copy of the AND which can be found within $P$. Furthermore, these sparsifiers are simple, as they can be constructed by i.i.d. sampling of the edges.

In the uniform model, the situation is more complex. For every predicate $P$, there is an integer $k$ such that a random instance on $n$ vertices and $m$ edges cannot (essentially) be sparsified if $m \le n^k$ and can be sparsified to size $\approx n^k$ if $m \ge n^{k+1}$. However, for some predicates $P$, if $m \in [n^k, n^{k+1}]$, there may or may not be a nontrivial sparsifier. In fact, we show that there are predicates where the sparsifiability of random instances is \emph{non-monotone}, i.e., as we add more random constraints, the instances become \emph{more} sparsifiable. We give a precise (efficiently computable) procedure for determining which situation a specific predicate $P$ falls into.

The methods involve a variety of combinatorial and probabilistic techniques. In both models, we prove combinatorial decomposition theorems which identify the dominant structure leading to the optimal sparsifier size. Along the way, we also prove suitable generalizations of the cut-counting bound used in the study of cut sparsifiers.
\end{abstract}

\pagebreak

\setcounter{tocdepth}{2} 
\tableofcontents

\pagebreak

\pagenumbering{arabic}

\section{Introduction}

\subsection{Background}

Sparsification is a powerful algorithmic technique used to reduce the size of a combinatorial object while still providing strong theoretical guarantees on the similarity of the resulting sparsified object to the original object. Since its introduction in the seminal work of Bencz\'ur and Karger \cite{BK96} in the context of \emph{cut-sparsification} of graphs, sparsification has developed into a rich, multifaceted field, with works studying various generalizations of cut sparsification like spectral sparsification \cite{BSS09, ST11}, hypergraph sparsification \cite{KKTY21a, KKTY21b, Lee23, JLS22}, sparsifying sums of positive semi-definite matrices \cite{basu2025sparsifying, hsieh2025sparsifying},
and code sparsification \cite{KPS24, khanna2025efficient, brakensiek2025redundancy}, as well as algorithms for designing sparsifiers in restricted, sublinear models of computation like streaming and fully dynamic algorithms \cite{AGM12, AGM12b, KLMMS14, ADKKP16, khanna2024near}.

Recently, a prominent focus in the study of sparsification has shifted to the framework of \emph{CSP sparsification}, as first introduced in the work of Kogan and Krauthgamer \cite{KK15}. In this setting, one is provided with a constraint satisfaction problem (CSP), specified by a universe of variables $x_1, \dots x_n$, $x_i \in D$, a relation $R \subseteq D^r$ (sometimes referred to as a \emph{predicate}), and the constraints themselves, which are the result of applying the relation $R$ to ordered subsets of $r$ variables, along with a non-negative weight. Formally, the $i$th constraint is the relation $R$ applied to $S_i \in P([n], r)$ (namely, ordered subsets of size $r$) with weight $w_i \in \mathbb{R}^{\geq 0}$, and is denoted by $C_i = w_i \cdot R_{S_i}(x)$, where we understand $R_{S_i}(x) = 1$ if and only if $x_{S_i} \in R$, and is otherwise $0$.\footnote{In general, a well-known reduction allows us to also consider CSPs with constraints that use \emph{different} relations. For such a CSP, we can simply decompose it into independent instances, where each new instance contains all constraints using a single relation. We can then sparsify these instances separately and take their union to get a sparsifier of the original CSP instance. } In general then, for an assignment $x \in D^n$ to the $n$ variables, we say that the \emph{value} taken by the CSP on assignment $x$ is exactly 
\[
\mathrm{sat}_{R,C}(x) := \sum_{i = 1}^m w_i \cdot R_{S_i}(x).
\]
For a choice of $\eps \in (0,1)$, we say that a CSP instance $\widehat{C}$ is a $(1 \pm \eps)$ sparsifier of $C$ if for every assignment $x \in D^n$, 
\[
\mathrm{sat}_{R,\widehat{C}}(x) \in (1 \pm \eps) \cdot \mathrm{sat}_{R,C}(x).
\]
Typically, the goal is to find such (reweighted) sparsifiers while also \emph{minimizing} the number of constraints that $\widehat{C}$ retains.

Due to the flexibility in choosing the relation $R$, CSP sparsification captures a wide variety of natural sparsification tasks. For instance, when $R = \{01, 10 \} \subseteq \zo^2$, CSP sparsification is \emph{equivalent} to cut-sparsification in graphs; when $R = \zo^r - \{0^r, 1^r\}$, CSP sparsification captures hypergraph sparsification for hyperedges of size $r$; and when $R = \{x \in \zo^r: \mathrm{Ham}(x) \mod 2 = 1 \}$, CSP sparsification captures so-called code sparsification over $\mathbb{F}_2$.\footnote{Specifically, this will capture codes whose generating matrices have $\leq r$ $1$'s per row.} 
As the relation $R$ varies more generally, CSP sparsification instead captures a large sub-class of \emph{generalized hypergraph sparsification} problems as introduced in Veldt, Benson, and Kleinberg \cite{VBK22} and which have found numerous applications, for instance, in circuit design \cite{alpert1995recent, lawler1973cutsets}, scientific computing / matrix multiplication \cite{ballard2016hypergraph}, and in many clustering algorithms \cite{agarwal2005beyond, michoel2012alignment, li2017inhomogeneous, li2018submodular}.\footnote{See \cite{VBK22} for further discussion of the applications of sparsification.}
In all of these applications, the running time depends on the \emph{sparsity} of the underlying CSP (hypergraph), and this naturally motivates the study of better sparsification algorithms.  %

However, despite the utility of CSP sparsification, in many senses, our understanding of the sparsifiability of CSPs (as a function of $n, \eps, R$) is far from complete. The works of Filtser and Krauthgamer \cite{FK17} and Butti and {\v Z}ivn{\'y} \cite{BZ20} together gave a complete (and efficient) characterization of the worst-case sparsifiability of binary CSPs (i.e., when $r = 2$). For Boolean ternary CSPs (when $r = 3$ and $D = \zo$), the works of Khanna, Putterman, and Sudan \cite{KPS24, khanna2025efficient} gave an exact characterization of the worst-case sparsifiability, and more broadly, showed that relations which can be expressed as non-zeros of polynomial equations of degree $d$ admit $(1 \pm \eps)$ sparsifiers of size $\widetilde{O}(n^d / \eps^2)$.\footnote{We use $\widetilde{O}(\cdot)$ to hide polylogarithmic factors of $\cdot$.} Most recently, the work of Brakensiek and Guruswami \cite{brakensiek2025redundancy} showed that the worst-case sparsifiability of a CSP with relation $R$ is in fact \emph{equal} (up to $\text{poly}(\log n/\eps)$ factors) to the so-called \emph{non-redundancy} of the relation $R$, a quantity that has received much attention in its own right~\cite{bessiere2020Chain,chen2020BestCase,lagerkvist2020Sparsification,carbonnel2022Redundancy,brakensiek2025Richness}. In doing so, their work provided the most unified picture of the sparsifiability of CSPs to date. Furthermore, such works show that that sparsifiability is intricately related to many other basic questions in the study of CSPs. This includes understanding the space complexity of CSP decidability~\cite{SV26a},  making CSPs as small as possible (kernelization) without changing their decidability~\cite{chen2020BestCase,lagerkvist2020Sparsification,carbonnel2022Redundancy}, and determining the complexity of learning CSPs~\cite{bessiere2020Chain}. As such, we view sparsification as a natural testbed for how to approach these broader questions.

However, this characterization in terms of the non-redundancy is still lacking in many important respects: 
\begin{enumerate}
    \item In general, non-redundancy of arbitrary relations is not known to be efficiently computable, and in fact the asymptotic behavior of the  
non-redundancy of even some concrete small relations remains a mystery (cf. \cite{bessiere2020Chain,brakensiek2025redundancy,brakensiek2025Richness,BGP26,SV26b}). Thus, neither computing the optimal worst-case sparsifier size nor the actual computation of the sparsifier itself are known to be doable in polynomial time.
    \item Non-redundancy is only known to provide a \emph{worst-case} guarantee on the sparsifiability of CSPs with a given relation. This means that for any fixed CSP \emph{instance}, the non-redundancy does not necessarily govern the sparsifiability. In fact, the non-redundancy of a fixed CSP instance is always at least as large as the worst-case sparsifiability of \emph{any sub-instance} (subset of constraints) of that instance.
\end{enumerate}

In practice, many of the CSP instances we encounter are not worst-case, and thus one may hope that such instances are sparsifiable to sizes \emph{smaller} than their worst-case sub-instances, thereby circumventing the non-redundancy type bounds from Brakensiek and Guruswami~\cite{brakensiek2025redundancy}.
In fact, \cite{brakensiek2025redundancy} explicitly poses this question of understanding this typical behavior of the sparsifiability of CSPs.
Towards this end, in this work we consider the following guiding question:

\begin{center}
    \emph{What governs the sparsifiability of random CSP instances?}
\end{center}

Our primary contribution is to give a complete answer to the above question by providing tight bounds on the sparsifiability of random CSPs. In order to bypass the ``worst sub-instance sparsifiability'' lower bound that confounds previous CSP sparsifications \cite{KPS24, khanna2025efficient, brakensiek2025redundancy}, we introduce a new framework of \emph{decomposing} and \emph{sandwiching} CSPs that may be interesting in its own right.

\subsection{Our Results}

To introduce the notion of a random CSP, we first fix the relation $R \subseteq D^r$. There are then two natural ways in which we can define a \emph{random} CSP:

\begin{definition}[Random Uniform Instance]
    A random uniform CSP instance on $n$ variables and $m$ constraints with relation $R \subseteq D^r$ is the result of randomly sampling $S_1, \dots S_m \sim P([n], r)$ (where $P([n], r)$ refers to all ordered subsets of $r$ variables), and including the constraints $R_{S_i}(x): i \in [m]$. 
\end{definition}

\begin{definition}[Random $r$-Partite Instance]
    A random $r$-partite CSP instance on $r$ groups of $n$ variables $V_1, \dots V_r$ and $m$ constraints with relation $R \subseteq D^r$ is the result of randomly sampling $S_1, \dots S_m \sim U[V_1 \times V_2 \times \dots \times V_r]$, and including the constraints $R_{S_i}(x)$.\footnote{In the CSP setting, such multipartite ($r$-partite) instances are also referred to as multisorted ($r$-sorted).} 
\end{definition}

At face value, it may be unclear that in either of these models, random CSPs should be more sparsifiable than their worst-case counterparts; indeed, as the value of $m$ approaches $n^r$, then virtually every constraint will be included in the random CSP instance $C$ that we generate. In particular then, \emph{any worst-case} CSP instance is indeed contained in our CSP $C$!

For such CSPs, existing non-redundancy type characterizations (\cite{brakensiek2025redundancy}, and also the non-redundancy-esque characterizations of \cite{basu2025sparsifying}) will always report a sparsifiability value which is equal to the \emph{worst sparsifiability} of any sub-instance. However, it turns out that the presence of these unsparsifiable sub-instances is \emph{not} an inherent barrier towards sparsification of the global instance. For illustration, one can consider the relation $R = \{01\} \subseteq \zo^2$, which exactly models the directed-cut function in graphs. In this setting, each constraint is a directed edge over vertex set $[n]$, and given a cut $S \subseteq [n]$, an edge $(u,v)$ is crossing the cut $S$ if $u \notin S$ and $v \in S$.
In the worst-case, there is a well known lower bound on the sparsifiability of directed graphs, as one can show that the complete bipartite directed graph $G = (L \sqcup R, E = L \times R)$ requires all $|E| = |L| \times |R| = \Omega(n^2)$ directed edges to be kept (see \cite{CCPS21} for a more detailed argument). But, what then is the sparsifiability of \emph{random} directed graphs?

It turns out that random directed graphs almost always admit $(1 \pm \eps)$ cut-sparsifiers of size $\widetilde{O}(n / \eps^2)$. One way to see this is that random directed graphs are almost Eulerian (with high probability), meaning that their  out-degree minus in-degree is small in absolute value for all/most vertices, and thus can essentially be modeled by undirected graphs, for which we do know near-linear size sparsifiability results (a formal version of this argument appears in \cite{CCPS21}). This argument also shows that the \emph{complete} directed graph on $n$ vertices is sparsifiable to $\widetilde{O}(n / \eps^2)$ many edges. On the other hand, plugging this instance into any non-redundancy based sparsification framework will only guarantee sparsifiability to size $O(n^2)$, as the aforementioned worst sub-instance is the complete bipartite directed graph, which is indeed contained in the complete graph.

\paragraph{Random $r$-Partite CSPs}

As we show, random CSPs admitting smaller size sparsifiers than their worst-case counterparts proves to be a common trend as we consider different relations. 
Not only this, but the smoothed behavior of random CSP instances in fact allows us to derive very tight characterizations of their sparsifiability! 

In order to properly describe our characterizations of random sparsifiability, we must first introduce the notion of restrictions to $\mathbf{AND}$ inside a relation $R$. Intuitively, the relation $\mathbf{AND}_r = \{1^r\}$ is the most \emph{un}-sparsifiable relation there is. Indeed, if one considers the complete $r$-partite CSP instance using the relation $\mathbf{AND}_r$, no sparsification is possible. This is because for any constraint $S_i \in V_1 \times V_2 \times \dots \times V_r$, one can consider the assignment $x$ which assigns $1$ to variables in $S_i$, and $0$'s to all other variables. Under this assignment, the constraint $S_i$ is satisfied, but \emph{every} other constraint is unsatisfied (as each such constraint will have at least one $0$ in this assignment). This is an intrinsic barrier to sparsifiability as one cannot afford to drop the constraint $S_i$: for this assignment $x$, it will then be the case that the number of satisfied constraints goes from $1$ to $0$ (and no re-weighting can compensate for this fact). 

This barrier to sparsification has long been observed, dating back to the works of \cite{FK17, BZ20, KPS24}. In fact, these works show that as long as there is even a \emph{projection} to $\mathbf{AND}_{\ell}$ inside $R$, then there is an $\Omega(n^{\ell})$ lower bound on sparsifiability. Here, given a relation $R \subseteq D^r$, a projection is a mapping of the variables $\pi: \{x_1, \dots x_r\} \rightarrow D \cup \{y_1, \neg y_1 \dots y_{\ell}, \neg y_{\ell} \}$ such that $R(\pi(x_1), \dots \pi(x_r)) = \mathbf{AND}_{\ell}(y_1, \dots y_{\ell})$. This can be thought of as a projection to a (potentially rotated) subcube of $R$, where the relation $R$ now looks exactly like $\mathbf{AND}_{\ell}$. Having a projection to $\mathbf{AND}$ inside $R$ then enables one to effectively simulate the $\mathbf{AND}_{\ell}$ lower bound argument for worst-case sparsification, thereby obtaining the $\Omega(n^{\ell})$ lower bound. It was long conjectured that these ``$\mathbf{AND}$-projections'' characterize worst-case CSP sparsifiability \cite{BZ20, KPS24, khanna2025efficient}  but this was ultimately disproven in \cite{brakensiek2025redundancy}.

In our setting of $r$-partite CSPs, it turns out that projections to CSPs are also not the right notion, but rather we require a slight weakening to the largest ``$\mathbf{AND}$-restriction''. For a relation $R \subseteq D^r$, this is defined as the largest $\ell$ such that there are $D_1, \dots D_r \subseteq D$, $|D_i| \in \{1, 2\}$, with exactly $\ell$ of the $D_i$'s of size $2$ such that $|R \cap (D_1 \times D_2 \times \dots \times D_r)| = 1$. This is effectively a version of $\mathbf{AND}$ projections where rotating the subcubes is \emph{not} allowed. 

With this notion in hand, we then have the following theorem:

\begin{theorem}\label{thm:introrpartite}[Informal]
    Let $R \subseteq D^r$ be a relation, and let $C$ be a random $r$-partite CSP with relation $R$ with $m$ constraints over $n$ variables. Then, for $c$ being the arity of the largest $\mathbf{AND}$ that can be restricted to in $R$, with high probability:
    \begin{enumerate}
        \item For any $\eps > 0$, $C$ admits a $(1 \pm \eps)$ sparsifier of size $\min(m, \widetilde{O}(n^c / \eps^3))$.
        \item There exists a constant $\eps > 0$ for which any $(1 \pm \eps)$ sparsifier of $C$ requires size $\Omega(\min(m, n^c))$.
    \end{enumerate}
\end{theorem}

Thus, in the $r$-partite setting, the sparsifiability of a random CSP undergoes a strict phase transition: when the number of constraints $m$ is smaller than roughly $n^c$, there is no sparsification that can be done. However, as soon as the number of constraints exceeds $n^c$, then the CSP can be sparsified back down to $n^c$ constraints with high probability.
In a strong sense then, this ``presence of an \textbf{AND}'' in $R$ strongly dictates the sparsifiability of random CSPs with the relation $R$. 

As discussed above, the largest $\mathbf{AND}$ that can be \emph{projected to} in a predicate was conjectured to govern worst-case sparsifiability \cite{BZ20, KPS24,khanna2025efficient}, but ultimately proven false in \cite{brakensiek2025redundancy}. Our notion of $\mathbf{AND}$ restriction is slightly different than that of $\mathbf{AND}$ projection, as projection allows for setting variables equal to one another (and to $\{0,1\}$), while restriction only allows setting variables to $\{0,1\}$ (and in particular, the largest $\mathbf{AND}$ under projections is always at least as large as the largest $\mathbf{AND}$ under restrictions). Nevertheless, our theorem formalizes these long-believed connections between sparsifiability of CSPs and the presence of $\mathbf{AND}$'s in the relation.

Another important property of the characterization above is its ``monotonicity'': as one might expect, adding more random constraints makes sparsifying the CSP harder up to a point (roughly $n^c$ constraints), at which point the sparsifier size stagnates. Thus, adding more random constraints \emph{does not} decrease the sparsifier size. 

\paragraph{Random $r$-Partite Valued CSPs}

In addition to our characterizations of the sparsifiability of random CSPs with relations $R \subseteq D^r$, we also consider the sparsifiability of random CSPs with \emph{valued} relations $\bR: D^r \rightarrow \R^{\ge 0}$. Although Butti and {\v Z}ivn{\'y}~\cite{BZ20} make a brief remark concerning the sparsification of valued CSPs (VCSPs), we are the first work to formally do so. Just as in the standard $r$-partite setting with $\{0,1\}$-valued relations, it turns out that the key parameter which governs \emph{most} of the sparsifiability is the presence of $\mathbf{AND}$ in the relation. Note that we now define a restriction to $\mathbf{AND}$ as the largest $\ell$ such that there are $D_1, \dots D_r \subseteq D$, $|D_i| \in \{1, 2\}$, with exactly $\ell$ of the $D_i$'s of size $2$ such that $|(\Supp(\bR)) \cap (D_1 \times D_2 \times \dots \times D_r)| = 1$; in words, this is the largest subcube we can find which has exactly one non-zero value in the relation.
With this notion, for $r$-partite random VCSPs we obtain the following formal characterization:

\begin{theorem}\label{thm:introrpartiteVCSP}[Informal]
    Let $R: D^r \rightarrow \R^{\geq 0}$ be a valued relation, and let $C$ be a random $r$-partite CSP with relation $R$ with $m$ constraints over $nr$ variables. Then, for $c$ being the arity of the largest $\mathbf{AND}$ that can be restricted to in $R$, if every such $\mathbf{AND}_c$ is \textbf{balanced}, let $\widehat{c} = c$, and otherwise let $\widehat{c} = c+1$. For this $\widehat{c}$,
     with high probability:
    \begin{enumerate}
        \item For any $\eps > 0$, $C$ admits a $(1 \pm \eps)$ sparsifier of size $\min(m, \widetilde{O}_{\eps, \bR}(n^{\widehat{c}}))$.
        \item There exists a constant $\eps > 0$ for which any $(1 \pm \eps)$-sparsifier of $C$ requires size $\Omega_{\bR}(\min(m, n^{\widehat{c}}))$.
    \end{enumerate}
\end{theorem}

The key distinction between \Cref{thm:introrpartite} and \Cref{thm:introrpartiteVCSP} is that the size of the largest $\mathbf{AND}$ appearing in the valued relation no longer \emph{solely} governs the sparsifiability of the resulting VCSP instance: but in addition, a lack of \emph{balance}. For example, it is straightforward to compute that any instance of the CSP relation $R = \{0,1\}^r$ can be sparsified to a single constraint (as all constraints behave identically for all assignments). However, one can show that an arbitrary function $\bR : \{0,1\}^r \to \R^{>0}$ where all weights are nonzero but not all equal, must have $\Omega(n)$ constraints in every $(1 \pm \eps)$-sparsifier \emph{even though} this relation has no restriction to $\mathbf{AND}_1$. In general, verifying whether every such $\mathbf{AND}_c$ in a relation is balanced relies on understanding whether every non-zero output in the $\mathbf{AND}_c$'s in fact takes the \emph{same} value.

As another concrete example, we can consider the relation
\[
  \bR(00) = 0,\ \bR(01) = 0,\ \bR(10) = 1,\ \bR(11) = 2.
\]
In this case, $\bR$ has two separate restrictions to $\mathbf{AND}_1$: with $D_1 = \{0,1\}, D_2 = \{0\}$ and $D_1' = \{0,1\}, D_2' = \{1\}$. The key fact however is that for assignment in $D_1 \times D_2$, $\bR$ only takes non-zero value $1$, whereas in $D'_1 \times D'_2$, $\bR$ takes the non-zero value of $2$! This is exactly our definition of imbalance, and hence one can indeed verify that $\bR$ as defined above requires sparsifiers of size $\Omega(n^2)$.

\paragraph{Random Uniform CSPs}
In the \emph{uniform} setting, a similar characterization holds for most relations, in the sense that there is a strict phase transition and that the sparsifier sizes also display this monotonicity property with respect to $m$. However, perhaps surprisingly, \emph{this monotonicity behavior does not hold true for all relations}! 

To start, in the uniform setting, we require an analogous notion of the largest $\mathbf{AND}$ in a relation $R$, which we characterize via so-called \emph{plentiful-ness}. However, the more interesting distinction which arises when compared to the $r$-partite setting is in the characterization of \emph{non-monotonicity}.

To better explain the intuition for this non-monotonicity in the uniform setting, let us consider a relation $R \subseteq D^r$ and suppose that we have a huge number of constraints (approaching $\binom{n}{r} \cdot r!$). In this case, the CSP instance we are dealing with is essentially the \emph{complete} instance. In the uniform model, when we have the complete CSP, this means that we have every constraint $R_S(x)$, for $S \in P([n], r)$. So, for every unordered subset of variables $T \in \binom{[n]}{r}$, we have a constraint for every possible \emph{permutation} of the $r$ variables in $T$. Thus, sparsifying the complete instance with relation $R$ is effectively the same as sparsifying the complete instance with what we call the \emph{symmetrization} of $R$ (essentially, the average over all $r!$ permutations of the variables).

This turns out to have huge ramifications for sparsification: indeed, if we revisit the examples from the $r$-partite setting, we saw that CSPs with so-called ``balanced'' relations like $R = \{0,1\}^r$ are easy to sparsify, while CSPs with ``imbalanced'' relations are harder to sparsify. A key subtlety in the uniform setting is that there are relations which are themselves imbalanced but which, after symmetrization, become balanced! 

As an example, we can consider the predicate $\bR_1: \zo^2 \rightarrow \R_{\geq 0}$ such that
\[
\bR_1(0,0) = 1, \quad \bR_1(0,1) = 1.5, \quad \bR_1(1,0) = 0.5, \quad \bR_1(1,1) = 1.
\]
The predicate $\bR_1$ on its own is not ``balanced'', as it alternates between non-zero values of $0.5, 1, 1.5$. Importantly however the \emph{symmetrization} of $\bR_1$ now \emph{is} balanced: indeed, the only non-trivial permutation one can apply to $\bR_1$ yields the relation $\bR_1'$ such that 
\[
\bR'_1(0,0) = 1, \quad \bR'_1(0,1) = 0.5, \quad \bR'_1(1,0) = 1.5, \quad \bR'_1(1,1) = 1.
\]
Thus, the symmetrization of $\bR$ is exactly 
\[
\mathrm{Sym}(\bR_1) = \frac{\bR_1 + \bR'_1}{2},
\]
or equivalently,
\[
\mathrm{Sym}(\bR_1)(0,0) = 1, \quad \mathrm{Sym}(\bR_1)(0,1) = 1, \quad \mathrm{Sym}(\bR_1)(1,0) = 1, \quad \mathrm{Sym}(\bR_1)(1,1) = 1.
\]
This relation $\mathrm{Sym}(\bR_1)$ is now \emph{extremely easy} to sparsify as it is the trivial relation - much moreso than the original predicate $\bR_1$!

The upshot for random CSP sparsification is that for denser random CSPs, sparsification becomes easier, as, after symmetrization, we are potentially dealing with balanced relations, while for less dense random CSPs, sparsification is more difficult as we are dealing with imbalanced relations. The effect is of sufficient magnitude that it leads to a \emph{non-monotone} relationship between the density of the random CSP and the size of the optimal sparsifier.  We present our characterization theorem as follows.  %

\begin{theorem}\label{thm:introuniform}[Informal - see \cref{sec:completeArbitrary} for formal definitions.]
    Let $R \subseteq D^r$ be a relation, and let $C$ be a random uniform CSP with relation $R$ with $m$ constraints over $n$ variables. Then, with high probability:
    \begin{enumerate}
        \item If $R$ is \textbf{precisely} $k$\textbf{-plentiful}, and the \textbf{symmetrization} of $R$ is \textbf{not marginally balanced}, then (1) $C$ is $(1 \pm \eps)$-sparsifiable to $\widetilde{O}_{\bR}(\min(m, n^{r-k+1} / \eps^2))$ constraints, and (2) for a sufficiently small, constant $\eps$, any $(1 \pm \eps)$-sparsifier of $C$ requires $\Omega_{\bR}(\min(m, n^{r-k+1}))$ constraints. 
        \item If $R$ is \textbf{precisely} $k$\textbf{-plentiful}, and $R$ is \textbf{marginally balanced}, then (1) $C$ is $(1 \pm \eps)$-sparsifiable to $\widetilde{O}_{\bR}(\min(m, n^{r-k} / \eps^3))$ constraints, and (2) for a sufficiently small, constant $\eps$, any $(1 \pm \eps)$-sparsifier of $C$ requires $\Omega_{\bR}(\min(m, n^{r-k}))$ constraints. 
        \item If $R$ is \textbf{precisely} $k$\textbf{-plentiful}, the \textbf{symmetrization} of $R$ is \textbf{marginally balanced}, but $R$ itself is \textbf{not marginally balanced}, then (1)  if $m \geq n^{r-k+1}$, $C$ is $(1 \pm \eps)$-sparsifiable to $\widetilde{O}_{\bR}(n^{r-k} / \eps^3)$, but (2) if $m \leq n^{r-k+1}$, there is a small, constant $\eps$, such that any $(1 \pm \eps)$-sparsifier of $C$ requires $\Omega_{\bR}(m)$ constraints. 
    \end{enumerate}
\end{theorem}

In the first two cases, the CSP behaves in a similar manner to \cref{thm:introrpartite}, as there is a single threshold at which the sparsifiability begins to stay the same as one adds in more constraints. However, of key importance is the distinction in the final case above. Here, for any $m \leq n^{r-k+1}$ constraints, $C$ is essentially unsparsifiable, as any sparsifier requires $\Omega_{\bR}(m)$ constraints. However, once $m$ reaches roughly $n^{r-k+1}$, the CSP suddenly becomes \emph{more sparsifiable}, with sparsifiers being computable of size roughly $n^{r-k}$. Note that this is not simply a figment of our terminology, and there are in fact concrete (even non-valued) relations $R \subseteq D^r$ of constant size for which this behavior appears!

In many senses our results present a very strict \emph{phase transition} in the sparsifiability of random CSPs: for instance, in the $r$-partite model, our results indicate that there is a constant $c$ which depends on the relation at hand such that (with high probability) any random CSP on $n$ variables with $\leq n^c$ constraints \emph{cannot} be sparsified, while any CSP with more than $n^c$ constraints can be sparsified, with the number of constraints being reduced all the way down to roughly $n^c$. In this manner, our results can be seen as an extension of a long line of works studying phase transitions in random CSPs (see for instance \cite{friedgut1996every}). Perhaps more exciting though, our results also indicate a slight deviation from the typical setting of phase transitions. Indeed, random CSPs often portray monotone properties, where as one increases the number of constraints, a property (such as say, satisfiability \cite{ding2015proof}) becomes strictly less likely. In our setting however, we show that there are some CSPs where the property is non-monotone, as adding more constraints makes the problem of sparsification harder up until a point, where there is then a phase transition, subsequent to which the problem becomes significantly easier.

For uniform random VCSPs, the sparsifiability classification mirrors exactly that of \cref{thm:introuniform} (see \cref{thm:UniformGeneral}). In fact, in the uniform model the abstraction of CSPs as VCSPs greatly aids in studying the classification. 

\paragraph{Some General Remarks}

In all of our main results, it is worth noting that the characterizations of sparsifiability depend only on the relation $R$, and so are computable in constant time (as we deal only with relations $R$ of arbitrary constant size). This, in combination with the fact that our cases are exhaustive, yields an efficient, complete characterization of the sparsifiability of (V)CSPs in both of our studied models. 

Furthermore, and perhaps surprisingly, in both the random uniform and random $r$-partite cases, the random sparsifiability of a CSP on $n$ variables always behaves as $n^c$ where $c$ (which depends on the case and the relation) is \emph{integral}.\footnote{Of course, ignoring cases where no sparsification is possible, and $m$ itself happens to be $n^{\gamma}$, for $\gamma$ non-integral.} This stands in stark contrast to the sparsifiability of \emph{worst-case} instances, where it had been \emph{conjectured} that the sparsifiability exponent was always integral \cite{FK17, BZ20, KPS24, khanna2025efficient}, until the work of \cite{brakensiek2025redundancy}, who constructed explicit CSP instances whose sparsifiability behaves as $n^{\gamma}$ for $\gamma \notin \mathbb{Z}$. Very recently, it was even shown that every rational $\gamma \in \mathbb Q \cap [1,\infty)$ can appear as the exponent for some concrete CSP relation~\cite{brakensiek2025Richness}.

\subsection{Technical Overview}\label{sec:techOverview}

\subsubsection{Intuition}

In order to derive all of the above theorems, our results rely on several new definitions, techniques, and decompositions for CSPs. As mentioned above, the best known characterizations of CSP sparsification like \cite{brakensiek2025redundancy} (and for other classes of sparsifiers more generally, like \cite{basu2025sparsifying}) struggle when dealing with \emph{instance-optimal} sparsification, as these methods can only guarantee sparsifier sizes \emph{equal} to the sparsifiability of the worst-case sub-instance. For instance, when invoking the main result of \cite{brakensiek2025redundancy} on a complete directed graph, the sparsifier size will only be guaranteed to be $O(n^2)$, whereas we know from the previous discussion that sparsifiers of size $\widetilde{O}(n)$ exist. It turns out that this same phenomenon presents us with difficulties in our current setting as well. Indeed, one of the main barriers to optimal sparsification of random CSPs is that there can be too many assignments who satisfy a small number of constraints, (and thus random sampling + a naive Chernoff and union bound does not work for sparsification) as a result of a small subset of ``unluckily'' generated random constraints. 

However, for a random CSP instance $C$ with relation $R$, we provide a general structural theorem that shows that this excess of assignments satisfying a small number of constraints can be remedied by instead considering a ``sandwich CSP'', whereby we replace the relation $R$ with two relations $R_0 \subseteq R \subseteq R_1$ of a specific form.\footnote{Note that these types of sandwiches are also studied phenomena in graph theory, under the label of graph sandwich problems \cite{golumbic1995graph}.} We show that these new CSPs $C_0$, $C_1$ that result from replacing $R$ with $R_0$ and $R_1$ now \emph{are sparsifiable} by a random sample to size equal to our classification theorems above. Thus, for a given assignment, if it is the case that $C_0$ and $C_1$ have approximately the same number of satisfied constraints, then the simple fact that $R$ is sandwiched (i.e., $R_0 \subseteq R \subseteq R_1$) implies that the original CSP $C$ is also tightly sandwiched in between $C_0$ and $C_1$, and so a random sample of $C$ will also be a sparsifier of $C$ with high probability. Otherwise, if $C_0$ and $C_1$ have a vastly differing number of satisfied constraints for a given assignment, we show that (using the structure of our specific sandwiching relations $R_0, R_1$) this is because $C$ itself must have \emph{many} satisfied constraints for this assignment. In this latter case, we are no longer dealing with the ``low-weight'' regime, and can now afford a Chernoff bound and union bound argument.

Designing these sandwiching predicates itself relies on a delicate analysis of the influence of specific variables in our relation $R$. In the $r$-partite setting, this relies on notions of relevant and irrelevant coordinates, as we discuss in the deeper technical overview (\cref{sec:techOverviewIrrelevant}), and in the more difficult arbitrary domain, uniform CSP setting, relies on more involved notions of ``plentifulness'' of the relation $R$, which we omit from the discussion here for brevity. 

Nevertheless, these methods provide a concrete framework for circumventing the unsparsifiability of sub-instances in a CSP by analyzing CSPs with carefully chosen, close relations where these sub-instances are no longer unsparsifiable. We view it as a very promising future direction to use these techniques and this broader paradigm to derive ``instance-optimal'' characterizations of sparsifiability; i.e., to say for a fixed CSP instance $C$ what the \emph{optimal} sparsifier size is.

In the following section, we explain these techniques in more detail, by deriving (part of) the characterization in the significantly simpler Boolean setting. 

\subsubsection{A Crude Sparsifier}

As an introduction to the techniques we use to establish the sparsifiability of random CSPs, in this section we present an informal proof of \cref{thm:introrpartite} in the specific setting where the relation $R$ is \emph{Boolean}, i.e., $R \subseteq \{0,1\}^r$.

To start, instead of considering a \emph{random} instance with $m$ constraints and $n$ variables, we first consider the \emph{complete} $r$-partite instance $C$. Recall that here, there are $nr$ variables, broken up into $r$ groups of $n$ variables (denoted $V_1, V_2, \dots V_r$), and there is a constraint for every subset of $V_1 \times \dots \times V_r$. Ultimately, our goal will be to show that a uniformly random subset of (re-weighted) constraints from $C$ will indeed be a $(1 \pm \eps)$-sparsifier of $C$, and thus we will argue that a random sample of constraints has essentially the same ``behavior'' as the complete $r$-partite instance. Understanding the sparsifiability of $C$ will then completely determine the sparsifiability of such a random instance. Note that while this plan works for our current setting of Boolean, $r$-partite sparsification, it does not work for arbitrary domain, uniform CSP sparsification, thus leading to the counterintuitive non-monotonicity in sparsifier size witnessed in \cref{thm:introuniform}.

Next, we present some crude bounds on the sparsifiability of the complete $r$-partite instance $C$, for a relation $R \subseteq \{0,1\}^r$. We let $c$ denote the size of the largest $\mathbf{AND}$ that can be restricted to in the relation $R$, i.e., $c$ is the largest integer such that there are $D_1, \dots D_r \subseteq \{0,1\}$, with $|D_i| \in \{1,2\}$, and exactly $c$ of the $D_i$'s of size $2$ such that 
\[
|R \cap (D_1 \times D_2 \times \dots \times D_r)| =1.
\]
While ultimately we will show that the optimal sparsifier size of $C$ is approximately $n^c$, we start by instead showing that the optimal sparsifier size of $C$ lies in the range $[n^c, n^{c+1}]$.

To see the lower bound, we consider the sets $D_1, \dots D_r$ exactly realizing the $\mathbf{AND}$ of arity $c$, and we let $a_1a_2\dots a_r$ be the unique satisfying assignment in $R \cap (D_1 \times D_2 \times \dots \times D_r)$. We let $T \subseteq [r]$ denote the indices where the $D_i$'s are of size $2$, i.e., $T = \{i \in [r]: |D_i| = 2 \}$. Now, we consider the set of assignments $X \subseteq \zo^{nr}$ which, for indices $i \in [r] - T$ always take value $a_i$ for variables in $V_i$, and for indices $i \in T$ take value $a_i$ in \emph{exactly one variable} in $V_i$, which we call the distinguished variable, and are otherwise given value $1 - a_i$.

Now intuitively, for an assignment $\psi \in X$, the only satisfied constraints will be exactly those that contain \emph{all} of the distinguished variables for that assignment. For each $V_i: i \in T$ there are $n$ choices of the distinguished variable, so this effectively creates a partition of the constraints into $n^c$ disjoint classes where no assignment $\psi \in X$ simultaneously satisfies constraints from \emph{two different classes}. Thus, for any sparsifier to ensure that every assignment in $X$ still satisfies at least one constraint, it must be the case that the sparsifier retains at least one constraint from each class, thus yielding an $n^c$ lower bound. 

On the upper bound front, our primary building block is the following claim:

\begin{claim}\label{clm:crudeintroub}
    Let $\psi \in \{0,1\}^{nr}$ be any assignment such that at least one constraint in $C$ is satisfied. Then, $\psi$ satisfies $\Omega(n^{r-c})$ constraints in $C$. 
\end{claim}

If we assume the above claim is true, then creating a sparsifier which preserves $O(n^{c+1} / \eps^2)$ constraints follows almost immediately by random sampling. Indeed, if we sample the constraints of $C$ at rate roughly $\frac{1}{\eps^2 \cdot n^{r-c-1}}$, then for every assignment $\psi$ such that at least one assignment is satisfied, we should expect at least 
\[
\Omega(n^{r-c}) \cdot \frac{1}{\eps^2 \cdot n^{r-c-1}} = \Omega \left ( \frac{n}{\eps^2} \right )
\]
constraints to survive. By a Chernoff bound then, a random subsampling will preserve the fraction of satisfying assignments for assignment $\psi$ with probability $1 - 2^{- \Omega(n)}$, which is sufficiently large to survive a union bound over all $2^{O_r(n)}$ possible assignments $\psi$. Thus, our (reweighted) subsampling at rate roughly $\frac{1}{\eps^2 \cdot n^{r-c-1}}$ will constitute a sparsifier for $C$ with high probability. 

To see why the claim above is true, we start by considering any assignment $\psi$ which satisfies at least one assignment, and we assume that in each block of $n$ variables $\psi$ has at least one $1$ and at least one $0$, i.e., 
\[
|\psi^{-1}(0) \cap V_i| \geq 1, \quad \quad |\psi^{-1}(1) \cap V_i| \geq 1.
\]
Now, within each block of variables, let $b_i$ denote the more common symbol in $\{0,1\}$ which appears: $b_i = \mathrm{argmax}_{b \in \{0,1\}} |\psi^{-1}(b) \cap V_i|$.

Because the relation $R$ does not have a restriction to $\mathbf{AND}_{c+1}$, this means that there must be some tuple $t \in R$ such that $\Delta(t, b_1b_2\dots b_r) \leq c$. For this tuple $t \in \{0,1\}^r$, we can then lower bound the number of constraints $S \in V_1 \times V_2 \times \dots \times V_r$ for which $\psi_S = t$: For an index $i \in [r]$, when $t_i = b_i$ (which is true for $r-c$ indices), we then know that there are $\geq n/2$ choices of index $j \in V_i$ for which $\psi_j = b_i$. On the other hand, when $t_i \neq b_i$, we know there is at least one choice, and so altogether, there are at least $(n/2)^{r-c}$ satisfied constraints.

However, in order to shave off the extra $+1$ in the exponent of the sparsifier size, we need to take a more careful look at the structure of the satisfied constraints.

\subsubsection{Attempts at Improving the Upper Bound}

When creating sparsifiers, a common technique for reducing their size is to create \emph{smooth counting bounds}, as pioneered in the work of Karger \cite{DBLP:conf/soda/Karger93}. In our context, where our target is sparsifiers of size roughly $n^c$, our goal would be to show that we can in fact sample \emph{more aggressively}, namely at rate $\frac{1}{n^{r-c}}$, while still preserving the number of satisfied constraints for each assignment, thus yielding sparsifiers of size essentially $n^c$.

To show that such a sampling scheme would work in our context, a counting bound analogous to \cite{DBLP:conf/soda/Karger93}'s would require that the number of assignments $\psi \in \{0,1\}^{nr}$ for which the number of satisfied constraints is positive and at most  $\lambda \cdot n^{r-c}$ is at most $n^{O(\lambda)}$. Unfortunately, it turns out that such a counting bound is not true. Indeed, we can consider the simple relation $R = \{ 000, 001\} \subseteq \{0,1\}^3$, where the largest arity $\mathbf{AND}$ is of size $c = 2$. Now, consider the family of assignments $X = \{\psi \in \{0,1\}^{3n}: \psi_{1:2n} = 0 \circ 1^{n-1} \circ 0 \circ 1^{n-1} \}$. For each such assignment, exactly $n$ constraints will be satisfied, corresponding to the constraints which contain variables $x_1$ and $x_{n+1}$, and any choice of variable in $V_3$. However, because we allow the assignments in $X$ to vary arbitrarily in $V_3$, the size of $|X|$ is $2^n$, and so we have an \emph{exponential} number of assignments satisfying $n$ constraints, thereby providing a strong refutation to our smooth counting bound. 

At the same time, one can see that the relation is slightly contrived, as the variables in the third coordinate are essentially irrelevant; i.e., it does not matter if a variable in $V_3$ is set to $0$ or $1$, as $000$ and $001$ are the only satisfying assignments, and are symmetric with respect to $0$ vs. $1$ in the third coordinate. In fact, as we will see, this notion of \emph{irrelevance} turns out to be fundamental in deriving improved sparsifiability bounds.

\subsubsection{Decomposing Relations Through Irrelevant Coordinates}\label{sec:techOverviewIrrelevant}

To formalize this notion of coordinates being irrelevant, we first require the notion of an ``extreme'' tuple. Specifically, we say that a tuple $t \in \{0,1\}^r$ is extreme if the distance from $t$ to $R$ is exactly $c$. Note that this distance cannot be $\geq c+1$, as then this would imply the existence of an $\mathbf{AND}$ of arity $\geq c+1$. So an extreme tuple is one whose distance is the maximum possible.

Going forward, we restrict our attention to assignments $\psi \in \zo^{nr}$ which are $0^r$-dominant, meaning that for every $i \in [r]$, $|\psi^{-1}(0) \cap V_i| \geq n/2$. We will assume that $0^r$ is one such extreme tuple for the relation $R$. Note that if $0^r$ were not extreme, then this means that $0^r$ is distance $\leq c-1$ from the relation $R$. So, for any assignment $\psi$ such that for every $i \in [r]$, $|\psi^{-1}(0) \cap V_i| \geq n/2$ (i.e., an assignment where in each part, the most common value is a $0$), such an assignment would actually satisfy at least $(n/2)^{r-c+1}$ constraints, a factor $n$ improvement over our crude analysis in \cref{clm:crudeintroub}. Thus, the more interesting case is when a tuple (like $0^r$) is distance \emph{exactly} $c$ from $R$.

Now, we look at the entire set of tuples in $R$ which are at distance exactly $c$ from $t = 0^r$: formally, we consider \emph{all} possible sets $S \in \binom{[r]}{c}$ such that $0^r \oplus e_S = e_S \in R$ (corresponding to which bits must be flipped). For each index $i \in [r]$, we look at whether there exists an $S$ such that $e_S \in R$ and $i \in S$. If no such set exists, we say that the coordinate $i$ is \emph{irrelevant}, as intuitively, this coordinate never acts as a witness to the tuple $0^r$ being distance $c$ from $R$. 

In fact, there is an even stronger sense in which so-called irrelevant coordinates are irrelevant:

\begin{proposition}
Let $t = 0^r$ be extreme for $R$. Let $I \subseteq [r]$ be the set of all irrelevant coordinates for $t$. Then, for any $S$ such that $|S| = c$, and $e_S \in R$, and any $T \subseteq I$, we have that $e_{S \cup T} \in R$.
\end{proposition}

That is to say, for a set of $c$ coordinates that can be flipped to make a satisfying assignment, we can also arbitrarily flip irrelevant coordinates while still ensuring that the resulting tuple is satisfying. While we ultimately prove this claim formally by induction in \cref{sec:rpartiteBoolean}, it is illustrative to consider the case where $T = \{i\}$ is just a single irrelevant coordinate. To start, we consider the tuples $t' = t \oplus e_S = e_S$, and $\hat{t} = t \oplus e_{\{i\}} = e_{\{i\}}$. Because $i \notin S$, these tuples are at distance $c+1$ from one another, and so there must be some tuple $t'' \in R$ \emph{between} $t'$ and $\hat{t}$, in order to avoid an $\mathbf{AND}$ of arity $c+1$. For this tuple $t'' = t \oplus e_{S''} = e_{S''}$ however, it must still be at distance $\geq c$ from $0^r$ as $0^r$ is an extreme tuple for $R$. If $S'' = S \cup \{i\}$, then we are immediately done, as this implies that $0^r \oplus e_{S \cup T} \in R$. Otherwise, we know that $S'' \subsetneq S \cup \{i\}$ and $|S''| = c$, which means that $i$ \emph{must} be in $S''$. This yields a contradiction however, as this would mean $i$ is in fact a \emph{relevant} coordinate for $0^r$. So, the only possibility is that $0^r \oplus e_{S \cup T} \in R$. An inductive argument with a generalization of the above logic extends the proof to arbitrary subsets of irrelevant coordinates.

As further motivation for the above proposition, we introduce the notion of a \emph{decomposable} relation. Again, we consider the case when $t = 0^r$ (assuming it is extreme), and we let $I$ denote the set of irrelevant coordinates for $t$.  We say that $R$ is decomposable if $R = Q \times \{0,1\}^I$, where $I$ is the set of irrelevant coordinates and $Q \subseteq \{0,1\}^{\bar{I}}$. The benefit of this characterization is that any decomposable predicate satisfies a smooth counting bound of our desired form, when we consider the ``codeword'' perspective of the CSP: i.e., for an assignment $\psi \in \zo^{nr}$, we associate it with the corresponding vector in $\zo^{m}$ (where $m$ is the number of constraints) which is the indicator vector of the satisfied constraints under $\psi$. Morally, this perspective treats two assignments $\psi_1$ and $\psi_2$ as being the same if the \emph{sets of constraints} satisfied by $\psi_1$ and $\psi_2$ are the same.

\begin{claim}
    Let $X$ contain all assignments such that for every $i \in [r]$, $|\psi^{-1}(0) \cap V_i| \geq n/2$  and suppose that $0^r$ is an extreme tuple for $R$, with irrelevant coordinates $I$ and that $R$ is decomposable with $R = Q \times \{0,1\}^I$. Then, the complete $r$-partite instance has at most $n^{O(\lambda)}$ distinct codewords of weight $\leq \lambda \cdot n^{r-c}$, for assignments in $X$.
\end{claim}

To see why this holds, we consider a coordinate $i \in [r]$ which is relevant (not in $I$) under $0^r$. For such a coordinate, by definition there must be a tuple $s$ in $R$ such that $s_i = 1$. Now, for an assignment $\psi$, the number of constraints satisfied by $\psi$ is at least the number of constraints where the assignment $\psi$ evaluates to $s$, i.e.:
\[
\mathrm{sat}_R(\psi) \geq \prod_{i \in [r]: s_i = 0} |\psi^{-1}(0) \cap V_i| \cdot \prod_{i \in [r]: s_i = 1} |\psi^{-1}(1) \cap V_i|.
\]
Using our assumption that $|\psi^{-1}(0) \cap V_i| \geq n/2$, this then evaluates to $\mathrm{sat}_R(\psi) \geq (n/2)^{r-c} \cdot |\psi^{-1}(1) \cap V_i|$. In particular, if $\psi$ satisfies fewer than $\lambda \cdot n^{r-c}$ constraints, then it must be the case that $|\psi^{-1}(1) \cap V_i| = O_r({\lambda})$, and so there are at most $\binom{n}{O_r({\lambda})} = n^{O_r(\lambda)}$ choices for how to choose the positions of $1$'s in the $i$th part of the assignment $\psi$. 

Finally, for the irrelevant coordinates $I$, because of the decomposability of the predicate, we know that the satisfaction of a constraint effectively \emph{ignores} whether the coordinates in $I$ are $0$ or $1$. Thus, it suffices to consider the version of the assignment $\psi$ where these coordinates are all set to be $0$, as the ``codeword'' of satisfied constraints will be unchanged. This does not alter the counting bound established above for the number of possible assignments $\psi$ that satisfy fewer than $\lambda \cdot n^{r-c}$ constraints.

\subsubsection{Sandwiching to Create Sparsifiers}

In the previous section we established a counting bound (and thus sparsifiability) for so-called \emph{decomposable} relations. However, in general there is no guarantee that a relation $R$ forms this type of product structure with the irrelevant coordinates, as the above proposition only guarantees that \emph{extreme tuples} can be arbitrarily extended with irrelevant coordinates while maintaining satisfaction. So, it is possible for instance, that assignments $\psi_1$ and $\psi_2$ do agree on constraints satisfied by extreme tuples, but nevertheless satisfy different sets of constraints overall due to the ``lower order'' contributions from tuples of weight $c+1$ in $R$. How then can we still guarantee the sparsifiability of such CSPs?

To approach this, we introduce the notion of \emph{sandwiching}. We consider the relations $R_0 \subseteq R \subseteq R_1$, where $R_0$ and $ R_1$ are the maximal and minimal \emph{decomposable} relations that are contained in or containing $R$ respectively. We can show that $R_0$ and $R_1$ still satisfy the counting bound from the previous subsection, and it follows then that the complete $r$-partite CSPs with $R_0$ and $R_1$ are both sparsifiable to $\widetilde{O}(n^{c} / \eps^2)$ constraints. But, this fact alone does not suffice for sparsification: indeed, every relation $R$ trivially contains the empty relation, and is contained in the full relation $\zo^r$, which themselves are both sparsifiable to constant size. 

Instead, for each assignment $\psi \in \zo^{nr}$, we consider \emph{how different} the number of satisfied constraints with $R_0$ (denoted $\mathrm{sat}_{R_0}(\psi)$) is when compared to the number of satisfied constraints with $R_1$ (denoted $\mathrm{sat}_{R_1}(\psi)$). If say, $\mathrm{sat}_{R_1}(\psi) \leq (1 + \eps / 3) \cdot \mathrm{sat}_{R_0}(\psi)$, then we can show that any set of constraints which is a $(1 \pm \eps/3)$ sparsifier for the complete $r$-partite CSP on both $R_0$ and $R_1$ is \emph{also} a $(1 \pm \eps)$ sparsifier for the complete $r$-partite CSP over $R$. If, for an assignment $\psi$, this does not hold true, then instead we show that such an assignment $\psi$ must satisfy many more constraints; i.e., on the order of at least $n^{r-c + 1}$ constraints. For such assignments, we trivially can sample at rate $1 / n^{r-c}$, as we have an extra factor of $n$ compared to \cref{clm:crudeintroub} (just as was the case with the non-extreme tuples).  In this way, either an assignment $\psi$ is sparsified by \emph{sandwiching} between two sparsifiable, decomposable relations, or $\psi$ must satisfy so many constraints that sampling at rate $\approx \frac{1}{n^{r-c}}$ works with a trivial union bound. In either case, this allows us to sparsify complete $r$-partite CSP with $R$ to roughly $n^{r-c}$ constraints.

While these techniques ultimately suffice for deriving a characterization of the sparsifiability of $r$-partite Boolean CSPs, we dedicate the majority of this work towards the challenges that arise in \emph{arbitrary domains}, valued CSPs, and the \emph{uniform} setting. For the sake of brevity, we postpone this discussion to the main body of the work.

\subsection{Open Questions}

Despite the generality of our results, there are still several directions of future study that remain open:

\begin{enumerate}
\item  Can the sandwiching and decomposability techniques be used to derive a general instance-optimal characterization of sparsifiability? I.e., to characterize for any specific CSP instance $C$ what the \emph{optimal} sparsifier size is for $C$. Note that, as mentioned before, existing characterizations \cite{brakensiek2025redundancy} are always lower bounded by the worst-case sparsifiability of any sub-instance of $C$ which is not in general equal to the sparsifiability of $C$ as a whole.
    \item What is the sparsifiability of random CSPs using a relation $R \subseteq D^r$ when $r$ \emph{grows} with the number of variables $n$ (that is to say, when $r$ is no longer constant)? Note that in our results, our sparsifier sizes depend (often exponentially) on the number of variables $r$ and the size of the domain $|D|$. Thus, even for relations $R \subseteq \{0,1\}^{\mathrm{polylog}(n)}$, our characterizations lose their tightness. 
    \item What is the \emph{spectral} sparsifiability of random CSPs (as defined in \cite{KPS25spectral})? Here, instead of considering assignments in $\{0,1\}^r$, one instead considers continuous-valued assignments in $[0,1]^r$. For a single constraint $c$ in a CSP, the value of $c$ on assignment $x \in [0,1]^r$ is then defined to be $\Pr_{\theta \sim [0,1]}[c(x^{(\theta)}) =1]^2$, where $x^{(\theta)}$ refers to the rounded version of $x$, where every entry below $\theta$ gets sent to $0$, and every entry larger than $\theta$ gets sent to $1$. As shown in \cite{KPS25spectral}, this notion generalizes spectral sparsification in graphs and hypergraphs, and thus one can naturally ask whether our characterizations hold for this stronger notion of sparsification. 
    \item Can we understand the sparsifiability of CSPs in \emph{semi-random} models? As mentioned above, one of our primary contributions is to give a framework for reasoning about how the worst-case sparsifiability of a sub-instance of a given CSP affects the global sparsifiability of the \emph{entire} CSP. One way to further test this framework is to instead consider a model where $m$ constraints are generated at random, and then for some fraction $p \in (0,1)$, an additional $p \cdot m$ constraints are chosen by an adversary. As a function of $p, R, m ,n$, what then governs the sparsifiability of the resulting instance? Intuitively, this question asks how large an adversarial worst-case sub-instance must be before it ``dominates'' the asymptotics of the optimal sparsifier size.
\end{enumerate}

\subsection*{Outline}

In \Cref{sec:prelim}, we outline the basic notation used in this paper as well as some concentration inequalities. In \Cref{sec:uniform-boolean}, we prove the classification of Boolean CSPs in the uniform model. In \Cref{sec:rpartiteBoolean}, we prove the classification of Boolean CSPs in the $r$-partite model. In \Cref{sec:rpartiteGeneral}, we prove the classification of general CSPs in the $r$-partite model. 
In \Cref{sec:rpartiteVCSP}, we prove the classification of general VCSPs in the $r$-partite model. 
In \Cref{sec:completeArbitrary}, we establish the classification of general (V)CSPs in the uniform model.
In \Cref{sec:random}, we prove that the sparsifiability of the complete instance closely matches the sparsifiability of random instances (in either model).
The sections containing the most general results for both $r$-partite and uniform cases can be read by themselves, but reading the special cases first is recommended to build intuition before the technicalities and complexities of the most general treatment surface.

\subsection*{AI Disclosure} No AI was used in the discovery of our results, nor in the general exposition of this manuscript. An LLM tool for theoretical computer science researchers\footnote{\url{https://acm-stoc.org/stoc2026/stoc2026-LLM_feedback.html}} was used to catch a missing case in the proofs of \cref{prop:irrelevant} and  \cref{prop:irrelevantGeneral} as well as a few other typographical errors. No other LLM assistance was used in the production of this manuscript.

\subsection*{Acknowledgments} We thank anonymous conference referees for various comments which have improved the exposition of this manuscript.

\section{Preliminaries}\label{sec:prelim}

\subsection{Notation}

We begin with some notation that we will use throughout the paper. We will use $R \subseteq D^r$ to denote the \emph{relation} used in our CSP (also referred to as the predicate), where we understand $R$ to be the \emph{satisfying} assignments to each constraint. We will let $n$ denote the number of variables, and let $x_1, \dots x_n$ denote the variables themselves, where each $x_i \in D$, for $D$ being our domain. When $|D|=2$ we have the Boolean CSP setting. We will only focus on the case when the \emph{arity} $r$ of the relation is bounded by a constant, i.e., $r = O(1)$. 

We refer to vectors in $D^n$ as \emph{assignments}, and refer to vectors in $D^r$ as \emph{tuples}. For any vector $v$ over a domain $D$, and a symbol $d \in D$, we will also use $\#_d(v)$ to denote the number of occurrences of $d$ in the vector $v$. For two tuples $x, y \in \zo^r$, we say that $x \leq y$ if for every $i \in [r]$ $x_i \leq y_i$. Additionally, when $x, y \in D^r$, we say that $z$ is between $x, y$ if $z_i \in \{x_i, y_i\}$. 

Note that throughout this work, we will often refer to a CSP $C$, with the understanding that we are referring to a CSP \emph{instance} $C$. For a CSP $C$ and an assignment $x$, we will use the notation $\mathrm{sat}_C(x)$ to denote the number of satisfied constraints in $C$ under the assignment $x$.
Additionally, we will often use the notion of \emph{codewords} for an assignment $x$:

\begin{definition}\label{def:codeword}
    For a CSP instance $C$ with $m$ constraints and an assignment $x$, we say that the \emph{codeword} for $C$ on assignment $x$ is the indicator vector in $\zo^m$ of the constraints that are satisfied by $x$. In general, the codewords of $C$ is then the multiset of the codewords for all possible assignments.
\end{definition}

\subsection{Concentration Bounds}

We will often make use of the following concentration bounds:

\begin{claim}\cite{SY19}\label{clm:concentrationBound2}
    Let $X_1, \dots X_{\ell}$ be independent random variables bounded by the interval $[0, W]$. 
    Then, for $X = X_1 + \dots + X_{\ell}$, $\mu = \E[X]$, and $\eps \in (0,1)$, we have 
    \[
    \Pr[|X - \mu| \geq \eps \mu] \leq 2 e^{- \left (\frac{\eps^2 \mu}{3W} \right )}.
    \]
\end{claim}

As a corollary of the above claim, we immediately have the following:

\begin{corollary}\label{cor:concentration}
    Let $x \in [0,W]^m$ be an arbitrary vector. Now, let $X_1, \dots X_{\ell}$ be random variables, where each $X_i$ samples a uniformly random index $q \in [m]$, and then takes on value $x_q \cdot \frac{m}{\ell}$. Then, for $\eps > 0$,
    \[
    \Pr\left [\sum_{j = 1}^{\ell} X_j  \in (1 \pm \eps) \wt(x)\right ] \geq 
    1 - 2 e^{- \left (\frac{\eps^2 \ell \cdot \wt(x)}{3Wm} \right )} \ .
    \]
\end{corollary}

The above corollary will typically be invoked with $W=1$ and $m \approx n^r$.

\section{Sparsifiability of the Complete, Uniform Instance with Boolean Relations}\label{sec:uniform-boolean}

To start, we analyze the sparsifiability of the complete uniform instance with Boolean relations. Specifically, we fix $R \subseteq \zo^r$ to be our relation (for $r = O(1)$), and consider the CSP instance $C$ over $n$ variables which consists of $R$ applied to all possible ordered subsets of $r$ variables (so there are $\binom{n}{r} \cdot r!$ constraints). We prove the following theorem in this section:

\begin{theorem}\label{thm:BooleanUniform}
Let $C$ be the complete, uniform instance of a Boolean relation $R\subseteq \zo^r$, let $a_{\min}$ denote the $r$-tuple of minimum Hamming weight in $R$, and let $a_{\max}$ denote the $r$-tuple of maximum Hamming weight in $R$.
For $c = \max(\wt(a_{\min}), r - \wt(a_{\max}))$, and for any $\eps > 0$, we have the following characterization:
\begin{enumerate}
    \item If $c \geq 1$, there exists a $(1 \pm \eps)$ sparsifier of $C$ which preserves $\widetilde{O}(n^{c} / \eps^2)$ constraints. \label{BooleanUniformItem1}
    \item If $c \geq 1$, any $(1 \pm \eps)$ sparsifier of $C$ must preserve $\Omega(n^c)$ constraints. \label{BooleanUniformItem2}
    \item If $c = 0$, and $R = \zo^r$, then $C$ is sparsifiable to $O(1)$ constraints. Otherwise if $c = 0$ and $R \neq \zo^r$, then $C$ is sparsifiable to $\widetilde{O}(n / \eps^2)$ constraints. \label{BooleanUniformItem3}
    \item If $c = 0$ and $R \neq \zo^r$, then for small enough (constant) $\eps'$, any $(1 \pm \eps')$ sparsifier must have $\widetilde{\Omega}(n)$ constraints.\label{BooleanUniformItem4}
\end{enumerate}
\end{theorem}

Before we begin proving the items of \cref{thm:BooleanUniform}, we make the following simplifying observation: 

\begin{claim}\label{clm:onlyWeightMatters}
    Let $C$ be the complete CSP instance for a relation $R$. Let $x, y \in \zo^n$ be such that $\wt(x) = \wt(y)$. Then, $\sat_C(x) = \sat_C(y)$ (in the sense that both have the same number of satisfying assignments). 
\end{claim}

\begin{proof}
    Because $x$ and $y$ have the same number of $1$'s, there is a permutation $\pi: [n] \rightarrow [n]$ such that $\pi(x) = y$. In particular, this lets us make a bijection between the satisfied constraints in $\sat_C(x)$ and the satisfied constraints in $\sat_C(y)$. For any ordered set $S \subseteq [n]$ of size $r$, $R_S(y)$ is satisfied if and only if $R_{\pi^{-1}(S)}(x)$ is satisfied. 
\end{proof}

This immediately lets us conclude the following: 

\begin{claim}\label{clm:numSatisfiedBooleanUniform}
    Let $x \in \zo^n$, let $\ell$ denote $\max_{b \in \zo} \#_b(x)$, let $c$ be as defined in \cref{thm:BooleanUniform}, and suppose $c \geq 1$. Then, $\sat_C(x) = \Omega_{r}(\max(1,(n-\ell)) \cdot n^{r-c})$.
\end{claim}

\begin{proof}
    Let $b \in \zo$ denote the symbol which achieves $\max_{b \in \zo} \#_b(x)$ (so $b$ is the more frequent symbol in $x$). By \cref{clm:onlyWeightMatters}, we can assume WLOG that $x$ starts with $\ell$ $b$'s, and the final $n-\ell$ coordinates are all $(1-b)$'s. Now, by definition, recall that $c = \max(\wt(a_{\min}), r - \wt(a_{\max}))$. So, for the symbol $b$, if we let
    \[
    a^* = \mathrm{argmax}_{a \in R^{-1}(1)} |\{j \in [r]: a_j = b\}|,
    \]
    (i.e., $a^*$ is the tuple in $R^{-1}(1)$ which has the most symbols which are $b$), then if $b = 0$, then $a^*$ is the lowest weight tuple in $R^{-1}(1)$, denoted $a_{\min}$, and if $b = 1$, then $a^*$ is the largest weight tuple, denoted $a_{\max}$. In either case, we can observe that 
    \[
    |\{j \in [r]: a^*_j = b\}| \geq r -c,
    \]
    as $c = \max(\wt(a_{\min}), r - \wt(a_{\max}))$ and therefore measures how many symbols are \emph{not equal} to $b$.

    Finally, for our assignment $x \in \zo^n$, we now lower bound the number of satisfied constraints. We let $k = |\{j \in [r]: a^*_j = b\}| \geq r -c$, let $S_b = \{i: x_i = b\}$ and let $S_{1-b} = \{i: x_i = 1-b\}$. Now, let $T \subseteq [n]$ denote any subset of $r$ coordinates which is formed by taking $k$ coordinates from $S_b$ and $r-k$ coordinates from $S_{1-b}$. Then, observe that there is some ordering of the set $T$ such that $R_T(x) = 1$ (as this ordering of $x_T$ will be exactly $a^*$, and therefore a satisfying assignment). This means that for our assignment $x$, there must be at least
    \[
    \binom{\ell}{k} \cdot \binom{n - \ell}{r-k} \geq \binom{n/2}{k} \cdot \binom{n - \ell}{r-k}
    \]
    satisfying assignments, as each distinct choice of the set $T$ will yield a satisfying assignment. Because $k \geq r- c$, this means that there are 
    \[
    \Omega_{r}(n^{r-c} \cdot \max(1,(n-\ell)))
    \]
    satisfied constraints for our assignment $x$.  
\end{proof}

With these preliminary claims established, we now prove the items of \cref{thm:BooleanUniform} one at a time, as the arguments have relatively small overlap. We begin with the upper bound for $c \geq 1$:

\begin{proof}[Proof of \cref{thm:BooleanUniform}, \cref{BooleanUniformItem1}.]
Let $c \geq 1$, let $x \in \zo^n$ and let $\ell$ denote $\max_{b \in \zo} \#_b(x)$. Note that for a fixed value of $\ell \ge n/2$, the number of possible assignments $x$ with this value of $\ell$ is at most $2 \cdot \binom{n}{n - \ell}$ (i.e., choosing whether $b = 0$ or $1$, and then the location of the $\ell$ positions that are not $b$). Additionally, we can observe that $n - \ell \leq n/2$.

    By \cref{clm:numSatisfiedBooleanUniform}, we also know that any such assignment satisfies $\Omega_r(n^{r-c} \cdot \max(1,(n-\ell)))$ constraints. 
    Now, let us consider selecting $w = \frac{\kappa n^c \log(n)}{\eps^2}$ constraints as per \cref{cor:concentration}, and giving weight $\binom{n}{r} \cdot r! / w$ to each kept constraint. 
    
    As per \cref{cor:concentration} (with $m = \binom{n}{r} \cdot r!$), this will preserve the weight of the satisfied constraints for $x$ to a factor of $(1 \pm \eps)$ with probability $1 - 2^{-10(n - \ell)\log(n)}$ (by choosing $\kappa$ to be a sufficiently large constant). This ensures that the probability the value of the CSP on $x$ is preserved to a $(1 \pm \eps)$ factor is at least $1 - 1 / n^{10 \max(1, n - \ell)}$, so we can then take a union bound over all $2 \cdot \binom{n}{\ell}$ assignments $x$ for which $\max_{b \in \zo} \#_b(x) = \ell$. Thus, we obtain that 
    \[
    \Pr[\exists x \in \zo^n: \sat_C(x) \text{ not preserved to }(1 \pm \eps)] \leq \sum_{n-\ell =0}^{n/2} \frac{2 \binom{n}{\ell}}{n^{10 \max(1, n - \ell)}} \leq \sum_{n-\ell =0}^{n/2} \frac{2 n^{n- \ell}}{n^{10 \max(1, n - \ell)}} \leq 1/n^2.
    \]
    Finally, it remains only to observe that with high probability the \emph{sparsity} (i.e., number of surviving constraints) obtained by our sparsification is indeed $\widetilde{O}(n^c / \eps^2)$. Thus, with probability $1 - 1/n^2$, the resulting sample contains $\widetilde{O}(n^c / \eps^2)$ constraints, and is a $(1 \pm \eps)$ sparsifier of the original CSP instance $C$; this yields the lemma. 
\end{proof}

\begin{proof}[Proof of \cref{thm:BooleanUniform}, \cref{BooleanUniformItem2}.]
As above, let $c = \max(\wt(a_{\min}), r - \wt(a_{\max}))$, and suppose $c \geq 1$. Now, if $c = \wt(a_{\min})$, then let $b = 1$, and otherwise, let $b = 0$. For our lower bound, we consider the set of assignments 
\[
X = \{x \in \zo^n: \#_{b}(x) = c\}.
\]
Next, we consider any constraint in our instance $C$, and we suppose that such a constraint operates on a subset of variables $S \subseteq [n], |S| = r$ (we denote this constraint by $R_S$). For any assignment $x \in X$, observe that the only way for $R_S(x)$ to be $1$ is if \emph{all} $c$ of the entries of $x$ which are $b$ are contained in $S$; i.e., if 
\[
\{i: x_i = b \} \subseteq S. 
\]
This is because if there is any entry $i$ such that $x_i = b$ and $i \notin S$, then this means that $|S \cap \{i: x_i = b \}| \leq c-1$. But, there cannot be any tuple in $R^{-1}(1)$ which has fewer than $c-1$ symbols that are equal to $b$; if $b = 1$, 
this implies that there is a satisfying tuple with weight $\wt(a_{\min})-1$, and if $b=0$, 
this implies that there is a satisfying tuple with weight $\wt(a_{\max})+1$, in either case yielding a contradiction.

Finally, because $R_S(x)$ can only be $1$ if $\{i: x_i = b \} \subseteq S$, and $|\{i: x_i = b \}| = c$, we can observe that $R_S$ can only be satisfied for $\leq \binom{r}{c} = O_r(1)$ different assignments $x$ (the assignment $x$ must place all $c$ of its coordinates such that $x_i = b$ within the set $S$ that $R_S$ acts on). 

Finally, we see that there are $\binom{n}{c}$ assignments $x \in X$. Any sparsifier of the complete instance $C$ must ensure that every assignment $x \in X$ receives at least one satisfied constraint. However, any single constraint can only be satisfied for $O_r(1)$ different assignments $x \in X$, and thus our sparsifier must retain 
\[
\Omega\left ( \frac{\binom{n}{c}}{O_r(1)}\right )= \Omega_{r}(n^c)
\]
different constraints. 
\end{proof}

Now, we prove \cref{BooleanUniformItem3}:

\begin{proof}[Proof of \cref{thm:BooleanUniform}, \cref{BooleanUniformItem3}]
To start, observe that if $R = \zo^r$, then every constraint is always satisfied. So, for every $x \in \zo^r$, $\sat_C(x) = \binom{n}{r} \cdot r!$. Thus, to create a sparsifier of $C$ it suffices to retain one constraint with weight $\binom{n}{r} \cdot r!$.

Otherwise, let us assume only that $c = 0$. In particular, this means that both $0^r$ and $1^r$ are in $R$. Thus, for any assignment $x \in \zo^r$, $\sat_C(x) \geq \binom{n/2}{r} \cdot r!$, as there are at least $n/2$ $0$'s or at least $n/2$ $1$'s in $x$. 

Thus, if we consider sampling $w$ constraints in $C$, for $w = \kappa_r \cdot n \log(n) / \eps^2$ (for a sufficiently large constant $\kappa_r$), and giving weight $\binom{n}{r} \cdot r! / w$ to kept constraints, we see that by \cref{cor:concentration}, every assignment $x$ will have $\sat_C(x)$ preserved to a $(1 \pm \eps)$ factor with probability $1 - 2^{-2n}$. Taking a union bound over all $2^n$ assignments then yields the claim. 
\end{proof}

Finally, we prove a stronger lower bound in the setting where $c = 0$, but $R \neq \zo^r$.

\begin{proof}[Proof of \cref{thm:BooleanUniform}, \cref{BooleanUniformItem4}]

Indeed, because $c = 0$ but $R \neq \zo^r$, this means that there must be some $t \in \zo^r$ such that $t \notin R$. 

Now, for an assignment $x \in \zo^n$, observe that we can write 
\[
\sat_C(x) \leq r! \cdot \binom{n}{r} - \sum_{t \notin R} \binom{\mathrm{Ham}(x)}{\mathrm{Ham}(t)} \cdot \binom{n-\mathrm{Ham}(x)}{r-\mathrm{Ham}(t)},
\]
as there are $r! \cdot \binom{n}{r}$ constraints total, and for each unsatisfying tuple $t \notin R$, there are $\binom{\mathrm{Ham}(x)}{\mathrm{Ham}(t)} \cdot \binom{n-\mathrm{Ham}(x)}{r-\mathrm{Ham}(t)}$ ways to choose constraints where the restriction of $x$ becomes $t$ (and are thus unsatisfied). In particular, this means that for an assignment $x \in \{0,1\}^n$ such that $\mathrm{Ham}(x) = n/2$, 
    \[
    \sat_C(x) \leq r! \cdot \binom{n}{r} - \sum_{t \notin R} \binom{n/2}{\mathrm{Ham}(t)} \cdot \binom{n/2}{r-\mathrm{Ham}(t)} \leq n^r - \frac{n^r}{2^r \cdot \lfloor r/2\rfloor! \lceil r/2\rceil!}.
    \]

At the same time, for an assignment $x \in \{0,1\}^n$ such that $\mathrm{Ham}(x) = n/\log(n)$,  
 \[
  \mathrm{sat}_R(x) \geq r! \cdot \binom{(1 - 1 /\log(n))\cdot n}{r} = (n(1 - 1 / \log(n) )\cdot \dots (n(1 - 1 / \log(n)) -(r-1) ) 
  \]
  \[
  = n^r \cdot (1 - 1 / \log(n))^r - O(n^{r-1}).
 \]
In particular, if we let $\mathrm{sat}_R(\ell)$ denote the number of satisfied constraints when using an assignment of weight $\ell$, we see that 
\begin{align}
\frac{\mathrm{sat}_R(n/\log(n))}{\mathrm{sat}_R(n/2)} \geq \frac{(1 - 1 / \log(n))^r - O(1/n)}{1 - \frac{1}{2^r \lfloor r/2\rfloor! \lceil r/2\rceil!}} \geq 1 + \delta, \label{eq:relateHamweights}
\end{align}
for some constant $\delta > 0$ (since $r$ is constant, and we take $n \rightarrow \infty$).

Next, consider any CSP instance with $<n / (r\log(n))$ constraints, and denote this instance by $\widehat{C}$. This means there are at most $n / \log(n)$ variables $x_i: i \in [n]$ which participate in the constraints. We call this set of variables $V_1 \subseteq [n]$. The remaining variables $V_2 \subseteq [n]$ are those which do not participate in any constraints. We have that $|V_1| \leq n / \log(n)$, and $|V_2| \geq n - n / \log(n)$.

In particular, for any assignments $x, x'$, provided that $x|_{V_1} = x'|_{V_1}$, it is the case that $\mathrm{sat}_{\widehat{C}, R}(x) = \mathrm{sat}_{\widehat{C}, R}(x')$. Now, we consider $x$ such that $x|_{V_1} = 1^{|V_1|}$, and is $0$ otherwise (to ensure $\mathrm{Ham}(x) = n/\log(n)$), and $x'$ such that $x'|_{V_1} = 1^{|V_1|}$, and the remaining coordinates are balanced such that $\mathrm{Ham}(x') = n/2$. 

Clearly, $\mathrm{sat}_{\widehat{C}, R}(x) = \mathrm{sat}_{\widehat{C}, R}(x')$. But, if we suppose for the sake of contradiction that $\widehat{C}$ is a $(1 \pm \delta/3)$ sparsifier of $C$, then it must also be the case that 
    \[
    (1 - \delta/3)\mathrm{sat}_{C, R}(\psi) \leq \mathrm{sat}_{\widehat{C}, R}(\psi) = \mathrm{sat}_{\widehat{C}, R}(\psi') \leq (1 + \delta/3) \mathrm{sat}_{C, R}(\psi'),
    \]
    which means that 
    \[
    \frac{\mathrm{sat}_{C, R}(\psi)}{\mathrm{sat}_{C, R}(\psi')} \leq \frac{1 + \delta / 3}{1 - \delta/3} ,
    \]
    and therefore
    \[ \frac{\mathrm{sat}_{C, R}(n/\log(n))}{\mathrm{sat}_{C, R}(n/2)} \leq \frac{1 + \delta / 3}{1 - \delta/3} < 1+\delta.
    \]
    But, this is a contradiction with \cref{eq:relateHamweights}.

    Hence, there can be no such $(1 \pm \delta / 3)$ sparsifier $\widehat{C}$ with $\leq n / r \log(n)$ constraints.
\end{proof}

\section{Sparsifiability of the Complete, $r$-Partite Instance with Boolean Relations}\label{sec:rpartiteBoolean}

In this section, we consider CSPs where each constraint uses a relation $R\subseteq \zo^r$. We assume that $r = O(1)$, and that there are $nr$ global variables, split into $r$ sets $V_1, \dots V_r$, each of size $n$. We denote a variable in $V_i$ by $x^{(i)}_j$, where $i \in [r]$ and $j \in [n]$. We work with the \emph{complete $r$-partite instance}, which contains a constraint for every $r$-tuple $S \in V_1 \times V_2 \times \dots \times V_r$.

Now, we have the following definition for such a relation $R$:

\begin{definition}
    We say a relation $R\subseteq \zo^r$ has a restriction to $\mathbf{AND}_k$ if there exist sets $D_1, \dots D_r \subseteq \zo$, such that $1 \leq |D_i| \leq 2$, exactly $k$ of the $D_i$'s are of size $2$, and all together, $|R \cap D_1 \times D_2 \dots \times D_r| = 1$.
\end{definition}

With this, we present below our characterization of the sparsifiability of complete, $r$-partite, Boolean CSPs. Note that although the notion of non-redundancy can be defined for the complete instance $C$, non-redundancy only governs the worst sparsifiability of any \emph{sub-instance} of $C$. In general, by including \emph{all constraints}, we can actually get smaller size sparsifiers:
\begin{theorem}\label{thm:BooleanRpartite}
    Let $C$ be the complete, $r$-partite instance of a Boolean relation $R \subseteq \zo^r$, and let $c = \max \{k: R \text{ has a restriction to } \mathbf{AND}_k \} $. Then:
\begin{enumerate}
    \item For any $\eps > 0$, there exists a $(1 \pm \eps)$ sparsifier of $C$ which preserves $\widetilde{O}(n^{c} / \eps^3)$ constraints. \label{item:BooleanRpartite1}
    \item For any $\eps > 0$, a $(1 \pm \eps)$ sparsifier of $C$ must preserve $\Omega(n^c)$ constraints. \label{item:BooleanRpartite2}
\end{enumerate}
\end{theorem}

\subsection{Proof of Lower Bound}

We start by proving the lower bound: 

\begin{proof}[Proof of \cref{thm:BooleanRpartite}, \cref{item:BooleanRpartite2}]
    Let $c$ denote the maximum size of an $\mathbf{AND}$ that can be restricted to in the relation $R$, and let $D_1, \dots D_r$ denote the sets which realize this $\mathbf{AND}_c$. In particular, we know that $|R \cap D_1 \times D_2 \dots \times D_r| = 1$, and so we let this single satisfying assignment in the intersection be denoted by $a_1 a_2 \dots a_r$.
    
    Additionally, among these sets $D_1, \dots D_r$, exactly $c$ of them will be of size $2$. We let $T \subseteq [r]$ be exactly these sets, i.e., $T = \{i \in [r]: |D_i| = 2 \}$. With this, we can describe our set of assignments $X$ that we will use to derive the lower bound. To start, for each assignment, for every $i \in T$, we choose a special \emph{distinguished vertex} $x^{(i)}_{j^*_i} \in V_i$. The assignment $x \in \zo^{nr}$ is then given by setting $x^{(i)}_{j} = a_i$ for every $i \in [r] -T, j \in [n]$. For the remaining $i \in T$, we set $x_{j^*_i}^{(i)} = a_i$, and for all other $j \in V_i - \{j^*\}$, we set $x_j^{(i)} = 1 - a_i$. In this way, the assignment $x$ is completely determined by our choices of $j^*_i: i \in T$. In fact, we can immediately see that the number of possible assignments we generate is exactly $n^c$ (since $|T| = c$, and for each set in $T$, there are $n$ options for $j^*_i$). Going forward, we parameterize the assignments $x$ by tuples $\tau \in [n]^{T}$.
    
    To conclude our sparsifier bound, we show that the assignment $x(\tau)$ satisfies exactly the constraints in $\tau \times [n]^{[r] \setminus T}$. To see this, observe that for any constraint $c \in V_1 \times \dots \times V_r$, the only way for $c(x(\tau))$ to be satisfied is if for every $j \in T$, $c_j = \tau_j$, as otherwise the $j$th variable $c$ operates on will be $1 - a_j$, for which there is no satisfying assignment in our restriction of the relation. Now, because these constraints that are satisfied by $x(\tau)$ are exactly those in $\tau \times [n]^{[r] \setminus T}$, we have the additional property that the sets of satisfied constraints are \emph{disjoint}. The lower bound then follows because there are $\Omega(n^c)$ distinct assignments we create, each satisfying at least one constraint, and there is no constraint which is satisfied by \emph{two} different assignments. Thus, in order to preserve a non-zero number of satisfied constraints for every assignment $x(\tau)$, we must keep $\Omega(n^c)$ distinct constraints. 
\end{proof}

\subsection{Proof of Upper Bound}

Now, we continue to a proof of the upper bound. For this we require several auxiliary claims and definitions. To start, we can make the following, simple observation regarding the number of satisfying assignments. 

\begin{claim}
    Let $x \in \zo^{nr}$ be any assignment which satisfies at least one constraint in the complete $r$-partite CSP instance $C$ for relation $R$, and let $c$ be the size of the maximum arity $\mathbf{AND}$ in the relation $R$. Then, $x$ must satisfy $\Omega(n^{r-c})$ constraints. 
\end{claim}

\begin{proof}
    As before, we break up the variables in the assignment $x$ into their constituent groups $x^{(1)}, \dots x^{(r)}$ (each containing $n$ variables). In each such group of variables $x^{(i)}$, there must be some symbol $b_i \in \zo$ which is taken by at least $1/2$ of the variables. Additionally, for every index $i \in [r]$ such that every variable in $x^{(i)}$ is the same, we include $i$ in the set $P$.

    Now, we claim that there must be some tuple $(a_1, \dots a_r) \in R$ such that $(a_1, \dots a_r)$ and $(b_1, \dots b_r)$ differ in at most $c$ coordinates, and such that for every index $i \in P$, $a_i = b_i$. To see why, we design a restriction to $\mathbf{AND}$. To start, for every index $i \in P$, we set $D_i = \{b_i\}$. Under this restriction of the variables, there must still be a satisfying assignment in the relation $R$, as we are promised that our initial assignment $x$ satisfies some constraint (and satisfies this restriction). Next, let us suppose for the sake of contradiction that every satisfying tuple $(a_1, \dots a_r) \in R$ (that also satisfies that $a_i = b_i$ for $i \in P$) is distance at least $c+1$ away from $(b_1, \dots b_r)$, and let $(a^*_1, \dots a^*_r)$ denote one such tuple of minimum distance from $(b_1, \dots b_r)$. Let $T \subseteq [r], |T| \geq c+1$ denote the coordinates for which $(b_1, \dots b_r)$ and $(a^*_1, \dots a^*_r)$ disagree. 

    For the restriction, for every index $i \in [r] - T$, we set $D_i = \{b_i\} = \{a_i\}$. For every index $i \in T$, we set $D_i = \{ 0,1\}$. It follows then that 
    \[
    |R \cap D_1 \times \dots \times D_r| = 1,
    \]
    as if there were any other satisfying assignment besides $(a^*_1, \dots a^*_r)$ in $R \cap D_1 \times \dots \times D_r$, it would contradict the minimality of the distance between $(b_1, \dots b_r)$ and  $(a^*_1, \dots a^*_r)$. However, this then exactly satisfies the condition of being a projection to an $\mathbf{AND}$ of arity $|T| \geq c+1$, which contradicts our assumption that $R$ only has a projection to an $\mathbf{AND}$ of arity $c$. 

    To conclude then, we have established that there must be a tuple $(a^*_1, \dots a^*_r) \in R$ such that $a^*_i = b_i$ for $i \in P$, and that among the remaining coordinates, there are $\leq c$ coordinates on which $a^*$ and $b$ disagree. This means that there must be $\geq r - c$ coordinates on which $a^*$ and $b$ do agree. For each such coordinate $i \in [r]$ such that $a^*_i = b_i$, we know that there are $\Omega(n)$ choices of $j_i \in [n]$ such that $x^{(i)}_{j_i} = b_i$. For the coordinates $i$ such that $a^*_i \neq b_i$, we know that there is at least one choice of $j_i \in [n]$ for which $x^{(i)}_{j_i} = a^{*}_i$ (because these disagreement coordinates must be in parts $x^{(i)}$ for $i \notin P$). Thus, there will be $\Omega(n^{r-c})$ tuples in $x^{(1)} \times x^{(2)} \times \dots \times x^{(r)}$ which equal $(a^*_1, \dots a^*_r)$, and thus there must be $\Omega(n^{r-c})$ satisfied constraints in the CSP. 
\end{proof}

However, the above claim does not suffice for sparsification. Random sampling at rate $\approx \frac{1}{n^{r-c}}$ ensures that every assignment has $\Omega(1)$ surviving satisfied constraints in expectation, but importantly, this does not ensure that we can take a union bound over all possible assignments. To overcome this, we introduce a more detailed breakdown of the number of satisfied constraints. To do this, we require some new definitions.

\begin{definition}
    Let $t \in \zo^r$. We say that $t$ is \emph{extreme} for $R$ if the Hamming distance from $t$ to $R$ is exactly $c$ (where $c$ is defined as the maximum arity $\mathbf{AND}$ that $R$ has a projection to). 
\end{definition}

Note that $t$ cannot have distance $> c$ from $R$, as otherwise this would allow us to project to an $\mathbf{AND}$ of arity $> c+1$.

\begin{definition}
    Now, let $t \in \zo^r$ be extreme for $R$. We let $\mathcal{S}_t$ be the set of all $S \in \binom{[r]}{c}$ such that $t \oplus e_{S} \in R$. For an index $i \in [r]$, we say that $i$ is \emph{covered} by $t$ if there exists a set $S \in \mathcal{S}_t$ such that $i \in S$, and otherwise we say that $i$ is \emph{irrelevant} for $t$. 
\end{definition}

Irrelevant coordinates derive their name because of the following key proposition:

\begin{proposition}\label{prop:irrelevant}
Let $t \in \{0,1\}^r$ be extreme for $R$. Let $I \subseteq [r]$ be the set of all irrelevant coordinates for $t$. Then, for any $S \in \mathcal{S}_t$ and any $T \subseteq I$, we have that $t \oplus e_{S \cup T} \in R$.
\end{proposition}

\begin{proof}
We prove this claim by induction on $|T|$. The base case of $|T| = 0$ follows from the definition of $\mathcal{S}_t$. Now, assume the claim holds for any $T$ with $|T| \le \ell$. Pick $T$ such that $|T| = \ell+1$ and pick an arbitrary $T' \subsetneq T$ of size $\ell$ with $i_0 \in T \setminus T'$. Note that $t' := t \oplus e_{S\cup T'} \in R$ by the induction hypothesis.

Define $\bar{t} := t \oplus e_T$. First, we observe that $\bar{t} \notin R$. Indeed, if $\bar{t} \in R$ and $|T| < c$, then this violates the extremality of $t$. If $|T| = c$, then this violates the irrelevance of the coordinates in $T$. Otherwise, if $|T| > c$, then $t \notin R$, but $\bar{t} := t \oplus e_T \in R$, and so there must be some tuple strictly between $t$ and $\bar{t}$ which is in $R$ to avoid a copy of $\mathbf{AND}_{c+1}$. But this yields some new tuple $\bar{t}' = t \oplus e_{T''} \in R$, for $T'' \subsetneq T$, and so we can repeat this argument until we either contradict extremality of $T$ or irrelvance of coordinates in $T$.

Now, we know that $\bar{t}$ has distance $c+1$ from $t'$ and that $\bar{t} \notin R$. Thus, to avoid a copy of $\mathbf{AND}_{c+1}$ in $R$, there must be some $t'' \in R$ between $\bar{t}$ and $t'$. In other words, there exists $S' \subsetneq S \cup \{i_0\}$ such that $\bar{t} \oplus e_{S'} \in R$. We now have a few cases based on the composition of $S'$.

If $S' = S$, then $\bar{t} \oplus e_{S'} = t \oplus e_{S \cup T} \in R$, as desired. Otherwise, $S' \cap S \subsetneq S$. In particular, consider $\bar{t} \oplus e_{S'} = t \oplus e_{S' \oplus T} \in R$. Note that if $|S' \oplus T| < c$, we contradict the fact that $t$ is extreme. If $|S' \oplus T| = c$, then $S' \oplus T \in \mathcal{S}_t$. But, since $S' \oplus T \neq S$, there must be some irrelevant coordinate in $S' \oplus T$. This contradicts the fact that the coordinate is irrelevant. Finally, assume that $|S' \oplus T| > c$. Then, to avoid a copy of $\AND_{c+1}$ in $R$, there must be $S'' \subseteq S' \oplus T$ of size $c$ with $t \oplus e_{S''} \in R$, so $S'' \in \mathcal{S}_t$. Again, $S'' \cap S \subseteq S' \cap S \subsetneq S$, so $S''$ must contain an irrelevant coordinate, a contradiction.

The only possible outcome is $t \oplus e_{S \cup T} \in R$, as desired.
\end{proof}

Now, let us consider an assignment $\psi \in \zo^{nr}$. As before, we use the convention that the $nr$ variables are broken up into groups $V_1, \dots V_r$, where each $V_i$ is of size $n$.

\begin{definition}
    For a tuple $t \in \zo^r$, we say that an assignment $\psi \in \zo^{nr}$ is $t$-dominant if for every $i \in [r]$
    \[
    |\psi^{-1}(t_i) \cap V_i | \geq |\psi^{-1}(\bar{t_i}) \cap V_i |.
    \]
\end{definition}

Going forward, we will focus only on assignments that are $0^r$ dominant. Our reasoning will likewise extend for any other dominant tuples, and we can ultimately afford to take a union bound over all these $(2^r)$ different dominant tuples separately. We let $a_i = |\psi^{-1}(0) \cap V_i |$, and let $b_i =|\psi^{-1}(1) \cap V_i |$. Because $0^r$ is dominant, we know that $a_i \geq b_i$. Additionally, we can assume that $b_i \geq 1$. Otherwise, we can simply restrict to the relation where $x_i  = 0$, and repeat our argument for this sub-relation. This restriction cannot increase the size of the largest existing $\mathbf{AND}$, and we can therefore simply take a union bound over all these possible restrictions (of which there are $O_r(1)$).

Now, we have the following claim:

\begin{claim}\label{clm:notExtreme}
    Suppose that $t = 0^r$ is \emph{not} an extreme tuple for the relation $R$. Then, any assignment $\psi \in \zo^{nr}$ which is $t$-dominant has $\Omega(n^{r-c +1})$ satisfied constraints.
\end{claim}

\begin{proof}
Let $\mathrm{sat}_{R}(\psi)$ denote the number of satisfied constraints in the complete $r$-partite instance with relation $R$ by the assignment $\psi$. Because $t$ is not extreme, it must be the case that there is some assignment $t' \in \zo^r$ such that $d_{\mathrm{Ham}}(t, t') \leq c - 1$ but $t' \in R$. Now, we can observe that 
\[
\mathrm{sat}_{R}(\psi) \geq \prod_{i = 1}^r |\psi^{-1}(t'_i) \cap V_i| \geq \left ( \prod_{i: t'_i = 1} b_i  \right ) \cdot \left ( \prod_{i: t'_i = 0} a_i  \right ) \geq \left ( \prod_{i: t'_i = 0} a_i  \right ) \geq \left ( \frac{n}{2}\right )^{r-c+1}.
\]
Here, we have used that every $b_i \geq 1$ (without loss of generality), that every $a_i \geq n/2$ by our assumption that $0^r$ is dominant, and that because $d_{\mathrm{Ham}}(t, t') \leq c - 1$, there must be $\geq r - c + 1$ indices such that $t'_i = 0$.
\end{proof}

Thus, in the rest of this section we focus on the case where $0^r$ \emph{is} an extreme tuple for the relation $R$. Recall then that by \cref{prop:irrelevant}, we can identify a set of \emph{irrelevant} coordinates $I \subseteq  [r]$ such that for any $t \in R$ with $\mathrm{Ham}(t)  = c$ and $t|_I = 0^{|I|}$, we have that $t|_{\bar{I}} \times \zo^I \subseteq R$ (i.e., flipping any of these irrelevant coordinates maintains a satisfying assignment). With this, we introduce the notion of decomposability:

\begin{definition}
    We say that a relation $R$ is \emph{decomposable} if there exists $Q \subseteq \zo^{\bar{I}}$ such that $R = Q \times \zo^{I}$.
\end{definition}

In particular, we immediately obtain the following for decomposable relations:

\begin{claim}\label{clm:countingBound}
    Let $R = Q \times \zo^{I}$ be a decomposable predicate. Then, for any $\lambda \in \Z^+$, the number of distinct codewords of $C$ (in the sense of \cref{def:codeword}) of weight $\leq  \lambda \cdot (n/2)^{r-c}$, generated by $0^r$ dominant assignments is at most $n^{O_r(\lambda)}$.
\end{claim}

\begin{proof}
    First, we can observe that because $R = Q \times \zo^{I}$, the assignments to the coordinates in $I$ do not matter, as setting them to be $0$ vs $1$ does not change the satisfaction of any given constraint. Thus, for an assignment $\psi$, the corresponding codeword (i.e., indicator vector of which constraints are satisfied) is only determined by the values $\psi$ takes on coordinates in $\bigcup_{i \in \bar{I}} V_i$. We say that any two assignments that are exactly the same on $\bigcup_{i \in \bar{I}} V_i$ are \emph{equivalent}.

    Next, recall that our goal is to bound the number of assignments with $\leq \lambda n^{r-c}$ satisfied constraints. To do this, by \cref{prop:irrelevant}, we know that for every $j \in \bar{I}$, there exists a tuple $t \in R$ such that $\mathrm{Ham}(t) = c$ and $t_j = 1$. Thus, any $\psi$ for which $\mathrm{sat}_{R}(\psi) \leq\lambda n^{r-c} $ must satisfy 
    \[
    \lambda n^{r-c} \geq \mathrm{sat}_{R}(\psi) \geq \prod_{i \in \supp(t)} b_i \prod_{i \notin \supp(t)} a_i \geq b_j \cdot \left ( \frac{n}{2}\right )^{|\bar{I}|-c} \cdot n^{|I|}.
    \]
    This means that $b_j \leq O_r(\lambda)$. In particular, the number of possible distinct codewords such that $b_j \leq O_r(\lambda)$ for every $j \in \bar{I}$ is at most 
    \[
    \binom{n}{O_r(\lambda)}^r = n^{O_r(\lambda)}.
    \]
    Thus, we get that the number of distinct codewords of weight $\leq  \lambda \cdot n^{r-c}$ is at most $n^{O_r(\lambda)}$, as we desire.
\end{proof}

Thus, the only case that remains to be analyzed is the case where $R$ is \emph{not} decomposable. In this case, we introduce a final definition:

\begin{definition}\label{def:nonDecompose}
    Let $R$ be a non-decomposable relation. Then, we let $Q_0, Q_1$ be relations on $\zo^{\bar{I}}$ such that 
    \[
    R_0 = Q_0 \times \zo^I \subseteq R \subseteq Q_1 \times \zo^I = R_1,
    \]
    with $Q_0$ being maximal, and $Q_1$ being minimal. 
\end{definition}

The following facts are immediate regarding $Q_0, Q_1$:

\begin{fact}\label{fact:covered}
    If $t \in R$ satisfies $\mathrm{Ham}(t) = c$, then $t \in R_0$ by virtue of \cref{prop:irrelevant}. This means that every covered index $i \in \bar{I}$ is also covered in $R_0$.
\end{fact}

\begin{fact}\label{fact:alsoUnion}
    By \cref{fact:covered} and \cref{clm:countingBound}, the complete $r$-partite instance with relation $R_0$ also satisfies the counting bound of \cref{clm:countingBound} (when restricting attention to $0^r$ dominant assignments). Since $R_0 \subseteq R_1$ (and has the same set of irrelevant coordinates), we also have that the complete $r$-partite instance with relation $R_1$ satisfies the counting bound (again when restricting our attention to $0^r$ dominant assignments). More clearly, for any assignment $\psi$, we have that $\mathrm{sat}_{R_0}(\psi) \leq \mathrm{sat}_{R_1}(\psi)$. By  \cref{clm:countingBound}, there are at most $n^{O_r(\lambda)}$ $0^r$-dominant non-equivalent assignments $\psi$ such that $\mathrm{sat}_{R_0}(\psi) \leq \lambda \cdot (n/2)^{r-c}$. Since $\mathrm{sat}_{R_0}(\psi) \leq \mathrm{sat}_{R_1}(\psi)$, there can only be fewer such assignments for the relation $R_1$.
\end{fact}

\begin{fact}\label{fact:weightLB}
    Consider any $t \in R \setminus R_0$. By \cref{prop:irrelevant} and \cref{fact:covered}, we have that $\mathrm{Ham}(t) \geq c+1$. In order to avoid a projection to $\mathbf{AND}_{c+1}$, there must be some other tuple $s \leq t$ such that $\mathrm{Ham}(s) = c$ and $s \in R$ ($s$ cannot have weight less than $c$ as we are assuming that $0^r$ is extreme). But, by \cref{fact:covered}, we must have that $s \in R_0$, and so it must be the case that $t \notin s_{\bar{I}} \times \zo^I$, i.e., $t$ must differ from $s$ in one of the \emph{non-irrelevant} coordinates. So, $\mathrm{Ham}(t_{\bar{I}}) \geq c + 1$. In general, \emph{every} $t \in Q_1 \setminus Q_0$ has weight $\geq c + 1$.
\end{fact}

With this, we are now able to establish the following key claim:

\begin{claim}\label{clm:CaseAnalysis}
    For every $0^r$ dominant assignment $\psi \in \zo^{nr}$, any $\eps \in (0,1)$, and any $R$ for which $0^r$ is extreme, at least one of the following holds:
    \begin{enumerate}
        \item $(1 + \eps) \mathrm{sat}_{R_0}(\psi) \geq \mathrm{sat}_{R_1}(\psi)$.
        \item $\mathrm{sat}_{R_0}(\psi) \geq \eps \cdot n^{r-c+ 1}$.
    \end{enumerate}
\end{claim}

\begin{proof}
Assume that (1) is false. This means that there must be some $t \in R_1 \setminus R_0$ such that $\mathrm{sat}_{t}(\psi) \ge \Omega_r(\eps \cdot \mathrm{sat}_{R_0}(\psi))$. By Fact~\ref{fact:weightLB}, we have that $\mathrm{Ham}(t|_{\bar{I}}) \ge c+1$. Furthermore, since $a_i \ge b_i$ for all $i \in [r]$, we may assume WLOG that $t|_I = 0^I$. Together, this means that
\begin{align}
    \mathrm{sat}_{t} (\psi) = \prod_{i \in \supp(t)} b_i \prod_{i \in [r]\setminus \supp(t)} a_i > \Omega_r(\eps\cdot  \mathrm{sat}_{R_0} \psi).\label{eq:sat-t}
\end{align}

Now, note that since $\mathrm{Ham}(t) \ge c+1$ but the largest arity $\mathbf{AND}$ that we can project to in $R$ is of size $c$, there must exist $0 \le s \le t$ with $\mathrm{Ham}(s) \leq c$ and $s \in R$. In fact, because $0^r$ is extreme for $R$, it must be the case that there is such an $s$ with $\mathrm{Ham}(s) = c$. For this $s$, as in \cref{fact:weightLB}, $s \in R_0$, so
\begin{align}
    \mathrm{sat}_{R_0} \psi \ge \mathrm{sat}_{s} \psi = \prod_{i \in \supp(s)} b_i \prod_{i \in [r]\setminus \supp(s)} a_i.\label{eq:sat-s}
\end{align}
Now putting together (\ref{eq:sat-t}) and (\ref{eq:sat-s}) we have
\[
\prod_{i \in \supp(t) \setminus \supp(s)} b_i \ge \eps \prod_{i \in \supp(t) \setminus \supp(s)} a_i = \Omega_r(\eps n^{|\supp(t) \setminus \supp(s)|}).
\]
In particular, since each $b_i \le n$, we must have that $b_i = \Omega_r(\eps n)$ for all $i \in \supp(t) \setminus \supp(s)$. 

Further observe that since $\mathrm{Ham}(t|_{\bar{I}}) \ge c+1$, there must exist some $j \in \supp(t) \setminus (\supp(s) \cup I)$. Thus, by definition of $I$, there is a $t' \in R$ with $\mathrm{Ham}(t') = c$ (and therefore also in $R_0$) such that $j \in \supp(t')$. For this $t'$, we have  
\[
    \mathrm{sat}_{R_0} (\psi) \ge \mathrm{sat}_{t'} (\psi) = \prod_{i\in \supp(t')}b_i \prod_{i \in [r] \setminus \supp(t')} a_i \ge \Omega_r\left(b_j n^{r - |\supp(t')|}\right) \ge \Omega_r(\eps n^{r-c+1}),
\]
as desired.
\end{proof}

Now, it remains only to piece these claims together to the proof of our upper bound in \cref{thm:BooleanRpartite}:

\begin{proof}[Proof of \cref{thm:BooleanRpartite}, \cref{item:BooleanRpartite1}]
The actual procedure for constructing the sparsifier is simple: we simply sample $w = \frac{\kappa \cdot n^c \log(n)}{\eps^3}$ constraints, where $\kappa$ is a large constant that depends on $r$.
In the resulting sample, every surviving constraint is given weight $n^r / w$. Thus, the sampling procedure is unbiased (i.e., the expected number of satisfied constraints for each assignment is preserved). 

We denote the resulting sparsifier by $\widehat{C}$. All that remains to be shown is that with high probability \emph{for every} assignment $\psi \in \zo^{nr}$, 
\[
\mathrm{sat}_{R, \widehat{C}}(\psi) \in (1 \pm \eps)\mathrm{sat}_{R}(\psi),
\]
where we use $\mathrm{sat}_{R, \widehat{C}}(\psi)$ to denote the weight of satisfied constraints for assignment $\psi$ in the instance $\widehat{C}$ (and when there is no subscript, we use the complete instance). We assume here that in every $V_i$, $\psi$ has at least $1$ coordinate equal to $1$ and at least one coordinate equal to $0$. If this does not hold true, we can instead restrict $R$ to the case where the $i$th coordinate is either $0$ or $1$, and instead apply our analysis on that sub-predicate. In our final union bound, we will pay this factor of $\leq 2^{2r}$ (the number of possible restrictions). We have several cases.
\begin{enumerate}
    \item If the relation $R$ is a decomposable relation, then recall that by \cref{cor:concentration}, for every $\lambda \in \mathbb{Z}^+$, the number of distinct codewords of weight $\leq \lambda \cdot n^{r-c}$ is at most $n^{O_r(\lambda)}$. In particular, we know that by \cref{cor:concentration}, for any codeword $x$ of weight $[(\lambda / 2)\cdot  (n/2)^{r-c}, \lambda \cdot (n/2)^{r-c})]$, the probability that $x$ has its weight preserved to a $(1 \pm \eps)$ factor under our sampling procedure is at least 
\[
1 - 2e^{-0.33 \eps^2 (\lambda / 2)\cdot  (n/2)^{r-c} \cdot w / n^r} \geq 1 - 2e^{-(0.33/2^{r+1}) \log(n) \cdot \kappa(r)} = 1 - n^{-0.33\kappa(r)\lambda / 2^{r+1}}.
\]
By setting $\kappa(r)$ to be sufficiently large, we can ensure that this probability is sufficiently large to survive a union bound. I.e., we have, 
\[
\Pr[\exists \psi: \mathrm{sat}_{R, \widehat{C}}(\psi) \notin (1 \pm \eps)\mathrm{sat}_{R}(\psi)] \leq \sum_{\psi \in \zo^{nr}} \Pr[\mathrm{sat}_{R, \widehat{C}}(\psi) \notin (1 \pm \eps)\mathrm{sat}_{R}(\psi)]
\]
\[
\leq \sum_{\lambda \in \Z^+} \sum_{\psi: \mathrm{Ham}(\psi) \in [(\lambda / 2)\cdot  (n/2)^{r-c}, \lambda \cdot (n/2)^{r-c})]}\Pr[\mathrm{sat}_{R, \widehat{C}}(\psi) \notin (1 \pm \eps)\mathrm{sat}_{R}(\psi)]
\]
\begin{align}
    \leq \sum_{\lambda \in \Z^+} n^{O_r(\lambda)} \cdot n^{-0.33\kappa(r)\lambda / 2^{r+1}} \leq \frac{1}{2^{3r} \cdot \mathrm{poly}(n)} \label{eq:decompUnionBound}, 
\end{align}
for a sufficiently large choice of $\kappa(r)$. Thus, sampling $w$ many constraints will indeed yield a sparsifier. 
\item Otherwise, let us suppose that $R$ is not decomposable. As before, let $t \in \zo^r$ denote the dominant assignment in the assignment $\psi$. 
I.e., 
\[
t_i = \mathrm{argmax}_{b \in \zo} |\psi^{-1}(b) \cap V_i|.
\]

As in \cref{clm:notExtreme}, if $t$ is not an extreme tuple for the relation $R$, then we immediately know that $\mathrm{sat}_{R}(\psi)$ is $\Omega_r(n^{r - c + 1})$. In particular, by sampling $w$ constraints, \cref{cor:concentration} guarantees that
\[
\Pr\left[\mathrm{sat}_{\widehat{C}, R}(\psi) \in (1 \pm \eps ) \cdot \mathrm{sat}_{R}(\psi) \right] \geq 1 - 2^{-2nr},
\]
where we use that $\kappa$ is a sufficiently large constant depending on $r$. Thus, every assignment has its weight preserved with probability $\geq 1 - 2^{-nr}$.
\item So, the only remaining case for us to analyze is when $t$ is extreme and $R$ is not decomposable. Here, we instead invoke \cref{clm:CaseAnalysis}. In particular, we either have that \begin{align}
(1 + \eps/3) \mathrm{sat}_{R_0}(\psi) \geq\mathrm{sat}_{R_1}(\psi), \label{eq:sandwich}
\end{align}
for $R_0, R_1$ as defined in \cref{def:nonDecompose},
or that $\mathrm{sat}_{R}(\psi) \geq \mathrm{sat}_{R_0}(\psi) \geq \eps/3 \cdot n^{r-c+1}$. In the second case, we can immediately see (again via \cref{cor:concentration}), that 
\[
\Pr\left[\mathrm{sat}_{\widehat{C}, R}(\psi) \in (1 \pm \eps ) \cdot \mathrm{sat}_{R}(\psi) \right] \geq 1 - 2^{-2nr},
\]
and taking a union bound over all $2^{nr}$ such possible strings yields a sparsifier with probability $1 - 2^{-nr}$.

Otherwise, we let $C_0(\psi), C(\psi),$ and $C_1(\psi)$ denote the satisfied constraints for relation $R_0, R$ and $R_1$ respectively on assignment $\psi$. We have that 
\[
C_0(\psi) \subseteq C(\psi) \subseteq C_1(\psi).
\]
Simultaneously, we know that $R_0$ and $R_1$ are both decomposable and that by \cref{fact:alsoUnion}, both $R_0$ and $R_1$ satisfy the counting bound of \cref{clm:countingBound} for assignments that are $t$-dominant. In particular, we can use the exact same union bound of \cref{eq:decompUnionBound} to show that with probability $\geq 1 - \frac{1}{2^{3r} \cdot \mathrm{poly}(n)}$, both $C_0$ and $C_1$ are $(1 \pm \eps/3)$ sparsified when sampling at rate $p$. Now, because 
\[
C_0(\psi) \subseteq C(\psi) \subseteq C_1(\psi),
\]
if we let $\widehat{C}$ denote the sampled constraints, we have 
\[
\mathrm{sat}_{R_0, \widehat{C}}(\psi) \leq \mathrm{sat}_{R, \widehat{C}}(\psi) \leq \mathrm{sat}_{R_1, \widehat{C}}(\psi).
\]
Simultaneously, because $\widehat{C}$ yields a $(1 \pm \eps/3)$ sparsifier of $C_0$ (i.e., $C$ with relation $R_0$) and $C_1$ (i.e., $C$ with relation $R_1$), we see that 
\[
(1 - \eps / 3) \cdot \mathrm{sat}_{R_0, C}(\psi) \leq \mathrm{sat}_{R, \widehat{C}}(\psi) \leq (1 + \eps / 3) \cdot \mathrm{sat}_{R_1, C}(\psi).
\]
Now, we can plug in the inequality from \cref{eq:sandwich}, namely that $ \mathrm{sat}_{R_0}(\psi) \geq \frac{\mathrm{sat}_{R_1}(\psi)}{(1 + \eps/3)} \geq \frac{\mathrm{sat}_{R}(\psi)}{(1 + \eps/3)}$ and that $\mathrm{sat}_{R_1}(\psi) \leq (1 + \eps/3) \cdot \mathrm{sat}_{R_0}(\psi) \leq (1 + \eps/3) \cdot \mathrm{sat}_{R}(\psi)$ to conclude that
\[
(1 - \eps)\mathrm{sat}_{R, C}(\psi) \leq  \mathrm{sat}_{R, \widehat{C}}(\psi)\leq (1 + \eps)\mathrm{sat}_{R, C}(\psi),
\]
as we desire.

In particular, in this case we see that every assignment with dominant tuple $t$ has its weight preserved with probability $\geq \min(1 - \frac{1}{2^{3r} \cdot \mathrm{poly}(n)}, 1 - 2^{-nr})$.
\end{enumerate}

Finally, we take a union bound over all $2^r$ choices of dominant tuple $t$, and all $2^{2r}$ choices of possible projections and conclude that $\widehat{C}$ is indeed a $(1 \pm \eps)$ sparsifier of $C$ with probability $\geq 9/10$, as we desire.
\end{proof}

\section{Sparsifiability of the Complete, $r$-Partite Instance with Relations over Arbitrary Domains}\label{sec:rpartiteGeneral}

In this section, we extend the argument from \cref{sec:rpartiteBoolean} to relations $\bR \subseteq D_1 \times D_2 \times \dots D_r$. Note that we still assume $r = O(1)$ and further assume that $|D_i| = O(1)$. As in \cref{sec:rpartiteBoolean}, our characterization of the sparsifiability relies on the largest arity \textbf{AND} that we can restrict to in our relation:

\begin{definition}
    We say a relation $\bR\subseteq D_1 \times \dots \times D_r$ has a restriction to $\mathbf{AND}_k$ if there exist sets $E_1 \subseteq D_1, \dots E_r \subseteq D_r$, such that $1 \leq |E_i| \leq 2$, exactly $k$ of the $E_i$'s are of size $2$, and all together, $|\bR \cap E_1 \times E_2 \dots \times E_r| = 1$.
\end{definition}

With this, we can present our characterization of the sparsifiability of complete $r$-partite CSPs with a relation $\bR \subseteq D_1 \times \dots \times D_r$.
\begin{theorem}\label{thm:RpartiteArbDomain}
    Let $C$ be the complete, $r$-partite instance of a relation $\bR \subseteq D_1 \times \dots \times D_r$, with $n$ variables in each part, and let $c(\bR) = \max \{k: \bR \text{ has a restriction to } \mathbf{AND}_k \} $. Then:
\begin{enumerate}
    \item For any $\eps > 0$, there exists a $(1 \pm \eps)$ sparsifier of $C$ which preserves $\widetilde{O}_{\bR}(n^{c(\bR)} / \eps^3)$ constraints. \label{item:BooleanRpartiteArbDom1}
    \item For any $\eps > 0$, a $(1 \pm \eps)$ sparsifier of $C$ must preserve $\Omega_{\bR}(n^{c(\bR)})$ constraints. \label{item:BooleanRpartiteArbDom2}
\end{enumerate}
\end{theorem}

\subsection{Proof of the Lower Bound}

We start with a proof of the lower bound, which is nearly identical to the previous proof from \cref{sec:rpartiteBoolean}. Nevertheless, we include the proof here for completeness.

\begin{proof}[Proof of \cref{thm:RpartiteArbDomain}, \cref{item:BooleanRpartiteArbDom2}]
Let $c=c(\bR)$ denote the maximum size of an $\mathbf{AND}$ that can be projected to in the relation $\bR$, and let $E_1, \dots E_r$ denote the sets which realize this $\mathbf{AND}_{c}$. In particular, we know that $|\bR \cap (E_1 \times E_2 \dots \times E_r)| = 1$, and so we let this single satisfying assignment in the intersection be denoted by $a_1 a_2 \dots a_r$. When $|E_i| = 2$, we let $b_i$ be the single element in $E_i - \{a_i \}$.
    
    Additionally, among these sets $E_1, \dots E_r$, exactly $c$ of them will be of size $2$. We let $T \subseteq [r]$ be exactly these sets, i.e., $T = \{i \in [r]: |E_i| = 2 \}$. With this, we can describe our set of assignments $X$ that we will use to derive the lower bound. To start, for each assignment, for every $i \in T$, we choose a special \emph{distinguished vertex} $x^{(i)}_{j^*_i} \in V_i$. The assignment $x \in D_1^n \circ D_2^{n} \dots \times D_r^n$ is then given by setting $x^{(i)}_{j} = a_i$ for every $i \in [r] -T, j \in [n]$. For the remaining $i \in T$, we set $x_{j^*_i}^{(i)} = a_i$, and for all other $j \in V_i - \{j^*\}$, we set $x_j^{(i)} = b_i$. In this way, the assignment $x$ is completely determined by our choices of $j^*_i: i \in T$. In fact, we can immediately see that the number of possible assignments we generate is exactly $n^c$ (since $|T| = c$, and for each set in $T$, there are $n$ options for $j^*_i$).

    To conclude our sparsifier lower bound, we show that for any two assignments $x, x'$ generated in the above manner, there is no constraint $\bR_S$ which is satisfied by \emph{both} $x$ and $x'$. To see this, recall that any constraint $\bR_S$ operates on a set of variables $S \in V_1 \times V_2 \times \dots \times V_r$ (we denote this set by $(j_1, j_2, \dots j_r)$). For any assignments $x, x'$, by construction, we enforce that for each $i \in [r]$, and for each $j \in [n]$, $x^{(i)}_j \in E_i$ (and likewise for $x'$). In particular, this means that the only way for $\bR_S(x)$ to be satisfied is if $x^{(1)}_{j_1} = a_1, x^{(2)}_{j_2} = a_2, \dots x^{(r)}_{j_r} = a_r$. However, by construction of our assignment $x$, for any $i \in T$, there is only one choice of index $j_i \in [n]$ for which $x^{(i)}_{j_i} = a_i$, namely when $j_i = j^*_i$ (the distinguished index). For every $i \in T$, this forces $j_i^* = j_i$. However, this yields a contradiction, as this would imply that $x, x'$ both have the same set of distinguished indices, and would therefore be the same assignment.

    The lower bound then follows because there are $\Omega(n^c)$ distinct assignments we create. No single constraint can be satisfied by two assignments, and so in order to get positive weight for each of the $\Omega(n^c)$ distinct assignments, we must retain $\Omega(n^c)$ distinct constraints. 
\end{proof}

\subsection{Proof of the Upper Bound}

To prove our upper bound, we require generalizations of many of the notions discussed in \cref{sec:rpartiteBoolean}.

\begin{definition}
    Let $s$ be a tuple in the domain of $\bR$ (i.e., in $D_1 \times \dots \times D_r$). We say that a tuple $s$ is \emph{extreme} for $\bR$ if the Hamming distance from $s$ to $\bR$ is exactly $c(\bR)$. 
\end{definition}

\begin{definition}
    For an extreme tuple $s \in D_1 \times \dots \times D_r$, we let $\mathcal{S}_s = \{t \in \bR: \mathrm{Ham}(s, t) = c(\bR) \}$.
\end{definition}

Going forward, we will often work (WLOG) using $s = 0^r$ under the assumption that $0^r$ is extreme for $\bR$. The argument can then be generalized to other arbitrary extreme tuples $s$. 
Now, we also have the following generalization of \emph{irrelevance}:

\begin{definition}
    For a relation $R \subseteq D_1 \times \dots \times D_r$, $i \in [r], d \in D_i$ with $0^r$ as an extreme tuple, we say that $(i, d)$ is \emph{irrelevant} for $0^r$, if there is no tuple $t \in \bR$ with $\mathrm{Ham}(t) = c(\bR)$ for which $t_i = d$. We let $I \subseteq \{(i, j): j \in D_i, i \in [r] \}$ denote the set of all irrelevant coordinates / symbols (for $0^r$). Note that we always include $(i,0)$ in $I$ by default, as the purpose of $I$ is to look at irrelevant alternatives of $0^r$, which includes $0^r$ itself.
\end{definition}

With this, we have the following proposition:

\begin{proposition}\label{prop:irrelevantGeneral}
    Let $0^r$ be an extreme tuple for $\bR$, and let $t \in D_1 \times \dots \times D_r$ satisfy $\mathrm{Ham}(t) =c(\bR)$ and $t \in \bR$. Then, for every $s \in D_1 \times \dots \times D_r$ satisfying for every $i \in [r]$:
    \begin{enumerate}
        \item if $t_i \neq 0$, then $s_i = t_i$
        \item if $t_i = 0$, then either $s_i = 0$, or $(i, s_i) \in I$,
    \end{enumerate}
    then $s$ is in $\bR$. 
\end{proposition}

\begin{proof}
    We prove this via induction on the Hamming weight of $s$, and for ease of notation we set $c = c(\bR)$. For the base case, if $\mathrm{Ham}(s) = c$, then the only way to satisfy the above condition is for $s = t$, and thus $s \in \bR$.

    Let us assume inductively that the above proposition is true for tuples $s$ of weight $\leq i$. We will show that it holds true for tuples of weight $i + 1$. So let $s$ be a tuple of the above form such that $\mathrm{Ham}(s) = i+1$. We let $T \subseteq [r]$ denote the set of coordinates in $[r]$ for which $s, t$ disagree. Note that $T \subseteq I$ and that $T$ must be a subset of $[r] \setminus \supp(t)$. Now, let us pick an arbitrary set $T' \subseteq T$ such that $|T'| = |T| -1$, and let $i_0 \in [r]$ be the singular index in $T \setminus T'$. We let $t' \in D_1 \times \dots \times D_r$ be the tuple such that $t'_{i_0} = 0$, but otherwise $t'_i = s_i$. By induction, this means that $t' \in \bR$, as $\mathrm{Ham}(t') = \mathrm{Ham}(s) -1 = i$, and $i_0 \in I$, so setting coordinate $i_0$ to be $0$ ensures $t'$ satisfies our conditions above. 

    Next, we define $\bar{t}$ as $\bar{t}_i = 0$ for $i \in \bar{T}$, and otherwise $\bar{t}_i = s_i$. As in \cref{prop:irrelevant}, we first claim that $\bar t \notin \bR$. Indeed, if $\bar{t} \in \bR$, we have a few cases: (1) if $|T| < c$, then this violates the extremality of $0^r$, (2) if $|T| = c$, then this violates the irrelevance of the coordinate, value pairs in $(i, \bar{t}_i): i \in T $, or (3) if $|T| > c$, because $0^r \notin \bR$ and $\bar{t} \in \bR$, then in order to not violate the maximum arity $\mathbf{AND}$ being $c$, there must be a tuple $\bar{t}'$ \emph{strictly between} $0^r$ and $\bar{t}$ such that $\bar{t}' \in \bR$. Importantly, $\bar{t}'$ is only non-zero on irrelevant coordinates, and so we repeat the above argument with cases (1) and (2), eventually reaching a contradiction. Thus, $\bar{t} \notin \bR$.
    
    Now, we have that $\mathrm{Ham}(\bar{t}, t') = c+1$. So, because $\bar{t}\notin \bR$, in order to ensure that there is no $\mathbf{AND}_{c+1}$ in $\bR$, it must be the case that there is some $t'' \in R$ which is between $\bar{t}$ and $t'$ (here, we use between to mean that $t''_i \in \{t'_i, \bar{t}_i\}$). Now, we let $S' \subsetneq \supp(t) \cup \{i_0\}$ denote the coordinates where $\bar{t}$ and $t''$ disagree. We have a few cases depending on $S$.

    First, if $S' = \supp(t)$, then we are done, as this implies that $t''_i = \bar{t}_i = s_i$ for all $i \in [r] - \supp(t)$, and $t''_i = t'_i = s_i$ for all $i \in \supp(t)$, and thus $t'' = s \in \bR$. Otherwise, it must be the case that $S' \cap \supp(t) \neq \supp(t)$. Now, observe that $t''$ is non-zero exactly in the coordinates $T \oplus S'$. Thus, we have several cases depending on the cardinality of $T \oplus S'$:
    \begin{enumerate}
        \item If $|T \oplus S'| < c$, this contradicts the extremality of $0^r$, as we have a satisfying assignment of weight $\leq c-1$.
        \item If $|T \oplus S'| = c$, then the tuple $t''$ is of Hamming weight exactly $c$. Importantly, this means that every coordinate symbol pair in $t''$ is relevant. But, we know that for $i \in T'$, $t''_i = s_i$, and thus this contradicts the irrelevance of $(i, s_i)$ for $i \in T'$.
        \item If $|T \oplus S'| > c$, then there must be a satisfying tuple between $0^r$ and $t''$, whose support is a strict subset of $\supp(t'') = T \oplus S'$. We denote this tuple by $t'''$, and its support by $S'''$. $S'''$ satisfies $S''' \cap \supp(t) \neq \supp(t)$, as $S''' \subset T \oplus S'$ and $T$ and $\supp(t)$ are disjoint. Thus, we again land in case 1 or case 2. 
    \end{enumerate}

    Thus, the only possible outcome is that $S' = \supp(t)$, in which case $s \in \bR$, as we desire.
\end{proof}

Now, let us consider a global assignment $\psi \in D_1^n \circ \dots \circ D_r^n$. As before, we use the convention that the $nr$ variables are broken up into groups of size $n$, which we denote by $V_1, \dots V_r$.

\begin{definition}
    For a tuple $t \in D_1 \times \dots \times D_r$, we say that an assignment $\psi \in D_1^n \circ \dots \circ D_r^n$ is $t$-dominant if for every $i \in [r]$
    \[
    |\psi^{-1}(t_i) \cap V_i | \geq \max_{d \in D_i - \{t_i\}}|\psi^{-1}(d) \cap V_i |.
    \]
\end{definition}

Going forward, we will focus only on assignments that are $0^r$ dominant. Our reasoning will likewise extend for any other dominant tuples, and we can ultimately afford to take a union bound over all these $(|D_1| \cdot \dots \cdot |D_r|) = O_{\bR}(1)$ different dominant tuples separately. For $i \in [r]$ and $d \in D_i$, we let $a_{i,d} = |\psi^{-1}(d) \cap V_i |$. Because $0^r$ is dominant, we know that $a_{i,0} \geq a_{i,d}$ for all other $d \neq 0$. Additionally, we will assume that $a_{i,d} \geq 1$. Otherwise, we can simply restrict to the sub-relation where the $i$th variable is restricted to never be $d$, and repeat our argument for this sub-relation. 
This restriction cannot increase the size of the largest existing $\mathbf{AND}$, and we can therefore simply take a union bound over all these possible restrictions (of which there are $O_{\bR|}(1)$).

Now, we have the following claim:

\begin{claim}\label{clm:notExtremeGeneral}
    Suppose that $t = 0^r$ is \emph{not} an extreme tuple for the relation $\bR$, and let $c = c(\bR)$. Then, any assignment $ \psi \in D_1^n \circ \dots \circ D_r^n$ which is $t$-dominant has $\Omega_{\bR}(n^{r-c +1})$ satisfied constraints.
\end{claim}

\begin{proof}
Let $\mathrm{sat}_{\bR}(\psi)$ denote the number of satisfied constraints in the complete $r$-partite instance with relation $\bR$ by the assignment $\psi$. Because $t$ is not extreme, it must be the case that there is some assignment $t' \in D_1 \times \dots \times D_r$ such that $d_{\mathrm{Ham}}(t, t') \leq c - 1$ but $t' \in \bR$. Now, we can observe that 
\[
\mathrm{sat}_{R}(\psi) \geq \prod_{i = 1}^r |\psi^{-1}(t'_i) \cap V_i| \geq \left ( \prod_{i: t'_i \neq 0} a_{i, t_i}  \right ) \cdot \left ( \prod_{i: t'_i = 0} a_{i,0}  \right ) \geq \left ( \prod_{i: t'_i = 0} a_{i,0}  \right ) \geq \left ( \frac{n}{\max_{i \in [r]}|D_i|}\right )^{r-c+1}.
\]
Here, we have used that every $a_{i,d} \geq 1$ (without loss of generality), that every $a_{i,0} \geq n/|D_i|$ by our assumption that $0^r$ is dominant, and that because $d_{\mathrm{Ham}}(t, t') \leq c - 1$, there must be $\geq r - c + 1$ indices such that $t'_i = 0$.
\end{proof}

Thus, in the rest of this section we focus on the case where $0^r$ \emph{is} an extreme tuple for the relation $\bR$. Next, we introduce a notion of decomposability:

\begin{definition}
    For a relation $\bR \subseteq D_1 \times \dots D_r$, we let $I$ denote the set of irrelevant coordinates / symbols for $0^r$. We say that $\bR$ is \emph{decomposable } if for every tuple $s \in D_1 \times \dots D_r$, we have $s \in \bR$ if and only if the tuple $s'$ such that for $i \in [r]$
    \begin{enumerate}
        \item if $(i, s_i) \in I$, $s'_i = 0$
        \item if $(i, s_i) \notin I$, $s'_i = s_i$
    \end{enumerate}
    is also in $\bR$. 
\end{definition}

The following claim summarizes the utility of decomposable predicates.

\begin{claim}\label{clm:countingBoundGeneral}
    Let $\bR$ be a decomposable relation over $D_1 \times \dots D_r$ such that its largest arity $\mathbf{AND}$ is of arity $c = c(\bR)$. Then, the number of distinct codewords of weight $\leq \lambda \cdot \left ( \frac{n}{\max_{i \in [r]}|D_i|}\right )^{r-c}$ is at most $n^{O_{\bR}(\lambda)}$.
\end{claim}

\begin{proof}
    Consider an index $i$ and all values $L_i \subseteq D_i$ such that $(i, d) \in \bar{I}$ for every $d \in L_i$. We know that for every choice of such $(i,d)$, there is a tuple $t \in \bR$ such that $t_i = d$ and $\mathrm{Ham}(t) = c$, as these are exactly the \emph{relevant} coordinates for $0^r$. We denote such a tuple by $t^{(i, d)}$, and let $T_{i}$ denote the set of all such tuples.  In particular, we can also observe that $\mathrm{sat}_{\bR}(\psi) = \sum_{t \in \bR} \mathrm{sat}_{t}(\psi)$.

    Thus, any $\psi$ which satisfies $\mathrm{sat}_{\bR}(\psi) \leq \lambda \cdot \left ( \frac{n}{\max_{i \in [r]}|D_i|}\right )^{r-c}$ also satisfies for a fixed $i \in [r]$:
    \[
    \lambda \cdot \left ( \frac{n}{\max_{i \in [r]}|D_i|}\right )^{r-c} \geq \mathrm{sat}_R(\psi) \geq \sum_{t^{(i,d)} \in T_i} \mathrm{sat}_{t^{(i,d)}}(\psi) \geq \sum_{d \in L_i} a_{i, d} \cdot \left ( \frac{n}{\max_{i \in [r]}|D_i|}\right )^{r-c}= 
    \]
    \[
    \left ( \sum_{d \in L_i} a_{i,d} \right ) \cdot \left ( \frac{n}{\max_{i \in [r]}|D_i|}\right )^{r-c},
    \]
    and therefore $\sum_{d \in D_i} a_{i,d}\leq \lambda$. Note that the final inequality relies on the fact that each $t^{(i,d)}$ has Hamming weight $c$, and so is $0$ in $r-c$ of the coordinates (hence leading to the extra factor of $\left ( \frac{n}{\max_{i \in [r]}|D_i|}\right )^{r-c}$).

    Now, for a fixed choice of $i \in [r]$ and $a_{i,d}: d \in L_i$, observe that there are at most $\binom{n}{\lambda} \cdot |D_i|^{\lambda} = n^{O_{\bR}(\lambda)}$ choices of assignments for the variables in $V_i$ that are in $L_i$ in the assignment $\psi$, provided $\mathrm{sat}_R(\psi) \leq \lambda \cdot \left ( \frac{n}{\max_i|D_i|}\right )^{r-c}$. Across all $r$ parts $V_1, \dots V_r$, this means that there are at most $n^{O_{\bR}(\lambda)}$ choices for all such variables.

    Finally, let $\psi, \psi'$ be any two distinct assignments such that for all $i \in [r]$, and for every $d \in L_i$, if $j \in V_i$, then $\psi_j = d$ if and only if $\psi'_j = d$ (i.e., the two assignments exactly match on all relevant coordinates). Then, $\psi$ and $\psi'$ satisfy exactly the same constraints, and therefore define exactly the same codeword. To see this, any constraint operates on $r$ variables, denote these $y_1, \dots y_r$ corresponding to the subset of $\psi$, and $y'_1, \dots y'_r$ corresponding to $\psi'$. If any $y_i \in L_i$ or $y'_i \in L_i$, then $y_i = y_i'$. Thus, the only way for $y_i$ and $y_i'$ to differ is if neither of $y_i$ and $y_i'$ are in $L_i$. But this means that both must take on \emph{irrelevant} values in their $i$th position, and $\bR$ being decomposable ensures that these assignments are either both satisfiable or both unsatisfiable. 

    In total then, we see that there are at most $n^{O_{\bR}(\lambda)}$ distinct codewords of weight $\leq \lambda \cdot \left ( \frac{n}{\max_i|D_i|}\right )^{r-c}$, as the satisfied constraints are uniquely determined by the positions of the relevant values. 
\end{proof}

Now, we define our sandwiching relations:

\begin{definition}\label{def:nonDecomposeGeneral}
    Let $\bR \subseteq D_1 \times \dots D_r$ be a non-decomposable relation. Let $\bR_0$ be a maximal decomposable relation such that $\bR_0 \subseteq \bR$, and let $\bR_1$ be a minimal decomposable relation such that $\bR \subseteq \bR_1$.
\end{definition}

The following facts are immediate regarding $\bR_0, \bR_1$:

\begin{fact}\label{fact:coveredGeneral}
If $t \in \bR$ satisfies $\mathrm{Ham}(t) = c(\bR)$, then $t \in \bR_0$. This follows directly from \cref{prop:irrelevantGeneral}. This means every covered tuple $(i,d)$ in $\bR$ is also covered in $\bR_0$.
\end{fact}

\begin{fact}\label{fact:alsoUnionGeneral}
    By \cref{fact:coveredGeneral} and \cref{prop:irrelevantGeneral}, the complete $r$-partite instance with relation $\bR_0$ also satisfies the counting bound of \cref{clm:countingBoundGeneral}, when restricting attention to $0^r$ dominant assignments. Since $\bR_0 \subseteq \bR_1$, this means the same must be true for the complete instance with relation $\bR_1$. More clearly, for any assignment $\psi$, we have that $\mathrm{sat}_{\bR_0}(\psi) \leq \mathrm{sat}_{\bR_1}(\psi)$. By  \cref{clm:countingBound}, there are at most $n^{O_{\bR}(\lambda)}$ $0^r$-dominant non-equivalent assignments $\psi$ such that $\mathrm{sat}_{\bR_0}(\psi) \leq \lambda \cdot (n/\max_i |D_i|)^{r-c}$. Since $\mathrm{sat}_{\bR_0}(\psi) \leq \mathrm{sat}_{\bR_1}(\psi)$, there can only be fewer such assignments for the relation $\bR_1$.
\end{fact}

\begin{fact}\label{fact:weightLBGeneral}
    Consider any $t \in \bR \setminus \bR_0$. By \cref{fact:coveredGeneral}, then $\mathrm{Ham}(t) \geq c+1$. In order to avoid a projection to $\mathbf{AND}_{c(\bR)}$, there must also be some other tuple $s$ between $0^r$ and $t$ such that $\mathrm{Ham}(s) = c(\bR)$ and $s \in \bR$. But, by \cref{fact:coveredGeneral}, this would mean that $s \in \bR_0$. In order for $t \notin \bR_0$ then, this must mean that $t$ differs from $s$ in a \emph{relevant} coordinate, and thus $t$ has at least $c(\bR)+1$ relevant (non-zero) coordinates. Because $\bR_1$ is the closure of $\bR$ with respect to the irrelevant coordinates / values, it must be the case that every $t \in \bR_1 \setminus \bR_0$ satisfies $\mathrm{Ham}(t) \geq c(\bR)+1$.
\end{fact}

With this, we are now able to establish the following key claim:

\begin{claim}\label{clm:CaseAnalysisGeneral}
    For every $0^r$ dominant assignment $\psi \in D_1^n \circ \dots \circ D_r^n$, any $\eps \in (0,1)$, and any $\bR$ for which $0^r$ is extreme, at least one of the following holds:
    \begin{enumerate}
        \item $(1 + \eps) \mathrm{sat}_{\bR_0}(\psi) \geq \mathrm{sat}_{\bR_1}(\psi)$.
        \item $\mathrm{sat}_{\bR_0}(\psi) = \Omega_{\bR}( \eps \cdot n^{r-c(\bR)+ 1})$.
    \end{enumerate}
\end{claim}

\begin{proof}
	We let $c = c(\bR)$.
    Assume that (1) is false. This means that there must be some $t \in \bR_1 \setminus \bR_0$ such that $\mathrm{sat}_{t}\psi \ge \Omega_{|D|, r}(\eps \cdot \mathrm{sat}_{\bR_0}(\psi))$. By \cref{fact:weightLBGeneral}, we have that $t$ is non-zero in at least $c+1$ relevant coordinates. Likewise, in all irrelevant coordinates, because $\bR_1$ is decomposable, we can assume WLOG that $t$ is $0$ in these coordinates. Thus, we see that
\begin{align}
    \mathrm{sat}_{t} (\psi) = \prod_{i \in [r]: (i, t_i) \in I \vee t_i =0} a_{i, 0} \cdot \prod_{i \in [r]: (i, t_i) \in \bar{I}} a_{i, t_i} \geq \Omega_{\bR}(\eps\cdot  \mathrm{sat}_{\bR_0} \psi).\label{eq:sat-tgeneral}
\end{align}

Now, note that since $\mathrm{Ham}(t) \ge c+1$ but the largest arity $\mathbf{AND}$ that we can project to in $\bR$ is of size $c$, there must exist $0 \le s \le t$ with $\mathrm{Ham}(s) \leq c$ and $s \in \bR$. In fact, because $0^r$ is extreme for $\bR$, it must be the case that there is such an $s$ with $\mathrm{Ham}(s) = c$. For this $s$, as in \cref{fact:weightLBGeneral}, $s \in \bR_0$, so
\begin{align}
    \mathrm{sat}_{\bR_0} \psi \ge \mathrm{sat}_{s} \psi = \prod_{i \in [r]: (i, s_i) \in I \vee s_i = 0} a_{i, 0} \cdot \prod_{i \in [r]: (i, s_i) \in \bar{I}} a_{i, s_i}.\label{eq:sat-sgeneral}
\end{align}

We let the set of \emph{relevant} coordinates for $s$ be denoted by $K_s$ and the relevant coordinates for $t$ be denoted by $K_t$. Because $0 \leq s \leq t$, note that $K_s \subsetneq K_t$. We let $K' = K_t \setminus K_s$ denote the relevant coordinates for $t$ that are not relevant in $s$. 

Now, dividing (\ref{eq:sat-tgeneral}) by (\ref{eq:sat-sgeneral}) we have
\[
\prod_{i \in K'} a_{i, t_i} \ge \eps \prod_{i \in K'} a_{i,0} = \Omega_{\bR}(\eps n^{|K'|}).
\]
In particular, since for each $i \in K'$, $a_{i,t_i} \le n$, we must have that $a_{i,t_i} = \Omega_{\bR}(\eps n)$ for all $i \in K'$. 

Additionally, since $t$ is non-zero in at least $c+1$ relevant coordinates, and $\mathrm{Ham}(s) = c$, there must exist some coordinate $j \in K'$ such that $(j, t_j)$ is relevant. Since $(j, t_j)$ is relevant, this implies that there is a string $t' \in \bR$ such that $\mathrm{Ham}(t') = k$ and $t'_j = t_j$. For this $t'$, we have that 
\[
\mathrm{sat}_{\bR_0}(\psi) \geq \mathrm{sat}_{t'}(\psi) = \prod_{i \in [r]} a_{i,t'_i} \geq a_{i, t'_i} \cdot \prod_{i \in [r]: t'_i = 0} a_{i,0} = \Omega_{\bR}(\eps n) \cdot \left ( \frac{n}{|D|}\right )^{r-c} = \Omega_{\bR}(\eps n^{r-c+1}).
\]
This concludes the proof. 
\end{proof}

Finally, we use this structural trade-off to derive our sparsification result.

\begin{proof}[Proof of \cref{thm:RpartiteArbDomain}, \cref{item:BooleanRpartiteArbDom1}]
	Let $c = c(\bR)$.
    The actual procedure for constructing the sparsifier is simple: we sample $w$ constraints uniformly at random, where $w = \frac{\kappa(r, |D|) \log(n) \cdot n^c}{\eps^3}$, where $\kappa$ is a large constant that depends on $r$ and $|D|$ (we use $|D| = |D_1 \cup D_2 \dots \cup D_r|$ as a crude upper bound). In the resulting sample, every surviving constraint is given weight $\frac{n^r}{w}$. Thus, the sampling procedure is unbiased (i.e., the expected number of satisfied constraints for each assignment is preserved). 

We denote the resulting sparsifier by $\widehat{C}$. All that remains to be shown is that with high probability \emph{for every} assignment $\psi \in D_1^n \circ \dots \circ D_r^n$, 
\[
\mathrm{sat}_{\bR, \widehat{C}}(\psi) \in (1 \pm \eps)\mathrm{sat}_{\bR}(\psi),
\]
where we use $\mathrm{sat}_{\bR, \widehat{C}}(\psi)$ to denote the weight of satisfied constraints for assignment $\psi$ in the instance $\widehat{C}$ (and when there is no subscript, we use the complete instance). We assume here that in every $V_i$, $\psi$ has at least $1$ coordinate equal to $d$ for each $d \in D_i$. If this does not hold true, we can instead restrict $\bR$ to the case where the $i$th coordinate is never equal to $d$, and instead apply our analysis on this restricted sub-relation, as our claims hold for any multi-sorted relation. In our final union bound, we will pay this factor of $\leq 2^{|D|r}$ (the number of possible restrictions). We have several cases.
\begin{enumerate}
    \item If the relation $\bR$ is a decomposable relation, then recall that by \cref{clm:countingBoundGeneral}, for every $\lambda \in \mathbb{Z}^+$, the number of distinct codewords of weight $\leq \lambda \cdot n^{r-c}$ is at most $n^{O_{\bR}(\lambda)}$. In particular, we know that by \cref{cor:concentration}, for any codeword $x$ of weight $[(\lambda / 2)\cdot n^{r-c}, \lambda \cdot n^{r-c})]$, the probability that $x$ has its weight preserved to a $(1 \pm \eps)$ factor under our sampling procedure is at least 
\[
1 - 2e^{-0.33 \eps^2 (\lambda / 2)\cdot  n^{r-c} \cdot w / n^r} \geq 1 - 2e^{-0.16 \lambda \cdot \log(n) \kappa(r, |D|)} = 1 - n^{-0.16 \lambda \cdot \kappa(r, |D|)}.
\]
By setting $\kappa(r, |D|)$ to be sufficiently large, we can ensure that this probability is sufficiently large to survive a union bound. I.e., we have, 
\[
\Pr[\exists \psi: \mathrm{sat}_{\bR, \widehat{C}}(\psi) \notin (1 \pm \eps)\mathrm{sat}_{\bR}(\psi)] \leq \sum_{\psi \in \zo^{nr}} \Pr[\mathrm{sat}_{\bR, \widehat{C}}(\psi) \notin (1 \pm \eps)\mathrm{sat}_{\bR}(\psi)]
\]
\[
\leq \sum_{\lambda \in \Z^+} \sum_{\psi: \mathrm{Ham}(\psi) \in [(\lambda / 2)\cdot  n^{r-c}, \lambda \cdot n^{r-c})]}\Pr[\mathrm{sat}_{\bR, \widehat{C}}(\psi) \notin (1 \pm \eps)\mathrm{sat}_{\bR}(\psi)]
\]
\begin{align}
    \leq \sum_{\lambda \in \Z^+} n^{O_r(\lambda)} \cdot n^{-0.16 \lambda \cdot \kappa(r, |D|)} \leq \frac{1}{2^{3|D|r} \cdot \mathrm{poly}(n)} \label{eq:decompUnionBoundGeneral}, 
\end{align}
for a sufficiently large choice of $\kappa(r, |D|)$. Thus, sampling these $w$ constraints will indeed yield a sparsifier. 
\item Otherwise, let us suppose that $\bR$ is not decomposable. As before, let $t \in D_1 \times \dots \times D_r$ denote the dominant assignment in the assignment $\psi$. 
I.e., 
\[
t_i = \mathrm{argmax}_{b \in \zo} |\psi^{-1}(b) \cap V_i|.
\]

As in \cref{clm:notExtremeGeneral}, if $t$ is not an extreme tuple for the relation $\bR$, then we immediately know that $\mathrm{sat}_{\bR}(\psi)$ is $\Omega_{\bR}(n^{r - c + 1})$. In particular, upon sampling $w$ constraints (and re-weighing), \cref{cor:concentration} guarantees that
\[
\Pr\left[\mathrm{sat}_{\widehat{C}, \bR}(\psi) \in (1 \pm \eps ) \cdot \mathrm{sat}_{\bR}(\psi) \right] \geq 1 - 2^{-4n|D|r},
\]
where we use that $\kappa$ is a sufficiently large constant depending on $r, |D|$. Thus, taking a union bound over all possible $\leq |D|^{nr}$ such assignments, every such assignment has its weight preserved with probability $\geq 1 - 2^{-3n|D|r}$.
\item So, the only remaining case for us to analyze is when the dominant tuple $t$ is extreme and $\bR$ is not decomposable. Here, we instead invoke \cref{clm:CaseAnalysisGeneral}. In particular, we either have that \begin{align}
(1 + \eps/3) \mathrm{sat}_{\bR_0}(\psi) \geq\mathrm{sat}_{\bR_1}(\psi), \label{eq:sandwichGeneral}
\end{align}
for $\bR_0, \bR_1$ as defined in \cref{def:nonDecomposeGeneral},
or that $\mathrm{sat}_{\bR}(\psi) \geq \mathrm{sat}_{\bR_0}(\psi) \geq \Omega_{\bR}(\eps/3 \cdot n^{r-c+1})$. In the second case, we can immediately see (again via \cref{cor:concentration}), that 
\[
\Pr\left[\mathrm{sat}_{\widehat{C}, \bR}(\psi) \in (1 \pm \eps ) \cdot \mathrm{sat}_{\bR}(\psi) \right] \geq 1 - 2^{-4n|D|r},
\]
and taking a union bound over all $\leq |D|^{nr}$ such possible strings yields a sparsifier with probability $1 - 2^{-3n|D|r}$.

Otherwise, we let $C_0(\psi), C(\psi),$ and $C_1(\psi)$ denote the satisfied constraints for relation $\bR_0, \bR$ and $\bR_1$ respectively on assignment $\psi$. We have that 
\[
C_0(\psi) \subseteq C(\psi) \subseteq C_1(\psi).
\]
Simultaneously, we know that $\bR_0$ and $\bR_1$ are both decomposable and that by \cref{fact:alsoUnionGeneral}, both $\bR_0$ and $\bR_1$ satisfy the counting bound of \cref{clm:countingBoundGeneral} for assignments that are $t$-dominant. In particular, we can use the exact same union bound of \cref{eq:decompUnionBoundGeneral} to show that with probability $\geq 1 - \frac{1}{2^{3|D|r} \cdot \mathrm{poly}(n)}$, both $C_0$ and $C_1$ are $(1 \pm \eps/3)$ sparsified when sampling $w$ constraints.
Now, because 
\[
C_0(\psi) \subseteq C(\psi) \subseteq C_1(\psi),
\]
if we let $\widehat{C}$ denote the sampled constraints, we have 
\[
\mathrm{sat}_{\bR_0, \widehat{C}}(\psi) \leq \mathrm{sat}_{\bR, \widehat{C}}(\psi) \leq \mathrm{sat}_{\bR_1, \widehat{C}}(\psi).
\]
Simultaneously, because $\widehat{C}$ yields a $(1 \pm \eps/3)$ sparsifier of $C_0$ (i.e., $C$ with relation $\bR_0$) and $C_1$ (i.e., $C$ with relation $\bR_1$), we see that 
\[
(1 - \eps / 3) \cdot \mathrm{sat}_{\bR_0, C}(\psi) \leq \mathrm{sat}_{\bR, \widehat{C}}(\psi) \leq (1 + \eps / 3) \cdot \mathrm{sat}_{\bR_1, C}(\psi).
\]
Now, we can plug in the inequality from \cref{eq:sandwich}, namely that $ \mathrm{sat}_{\bR_0}(\psi) \geq \frac{\mathrm{sat}_{\bR_1}(\psi)}{(1 + \eps/3)} \geq \frac{\mathrm{sat}_{\bR}(\psi)}{(1 + \eps/3)}$ and that $\mathrm{sat}_{\bR_1}(\psi) \leq (1 + \eps/3) \cdot \mathrm{sat}_{\bR_0}(\psi) \leq (1 + \eps/3) \cdot \mathrm{sat}_{\bR}(\psi)$ to conclude that
\[
(1 - \eps)\mathrm{sat}_{\bR, C}(\psi) \leq  \mathrm{sat}_{\bR, \widehat{C}}(\psi)\leq (1 + \eps)\mathrm{sat}_{\bR, C}(\psi),
\]
as we desire.

In particular, in this case we see that every assignment with dominant tuple $t$ has its weight preserved with probability $\geq \min(1 - \frac{1}{2^{3|D|r} \cdot \mathrm{poly}(n)}, 1 - 2^{-3n|D|r})$.
\end{enumerate}

Finally, we take a union bound over all $\leq |D|^r$ choices of dominant tuple $t$, and all $\leq 2^{|D|r}$ choices of possible restrictions and conclude that $\widehat{C}$ is indeed a $(1 \pm \eps)$ sparsifier of $C$ with probability $\geq 9/10$.
\end{proof}

\section{Sparsifiability of the Complete, $r$-Partite Instance for Valued Relations}\label{sec:rpartiteVCSP}

The goal of this section is to extend \cref{thm:RpartiteArbDomain} to arbitrary \emph{valued} CSPs (VCSP). That is, instead of viewing our relation $R$ as a subset of the space of local assignments $D^r$, we instead view $\bR$ as a \emph{valued relation} (VR)--that is as a function from $D^r$ to the non-negative integers. In many cases, the sparsifiability of the support of a VR is a good proxy for the sparsifiability of the VR itself. However, greatly complicating the classification, there are notable situations in which the two classifications diverge. Such complications also occur in arbitrary classification of complete uniform instance for ``ordinary'' CSPs in \cref{sec:completeArbitrary}, so we present the $r$-partite setting first.

\subsection{VCSP Notation}\label{subsec:vcsp}

We begin by formally defining some notation for valued CSPs. Formally a \emph{valued relation} (VR) of arity $r$ over domains $D$ is any map $\bR : D^r \to \mathbb Z^{\ge 0}$. We use boldface to distinguish valued relations from ordinary relations. We define $\supp \bR$ to be the set of $t \in D^r$ for which $\bR(t) \neq 0$. As in previous sections, we assume that $r, D$ and every value $\bR(t)$ for $t \in D^r$ is bounded by some constant.

In general, an instance $C$ of $\VCSP(\bR)$ consists $n$ variables and $m$ clauses, where for each $i \in [m]$, the $i$th clause is defined by an $r$-tuple of variables $S_i \in [n]^r$ and has a corresponding weight $w_i \in \mathbb R^{> 0}$. 
Given an assignment $x \in D^n$, we define its value to be
\[
  \sat_{\bR,C}(x) = \sum_{i=1}^m w_i \cdot \bR_{S_i}(x).
\]
If $\bR$ consists of a single element $t$, then we sometimes write $\sat_{t,C}(x)$ instead of $\sat_{\bR,C}(x)$.

As before, for any $\eps \in (0,1)$, we say that another instance $\widehat{C}$ is a $1 \pm \eps$ sparsifier of $C$ if for every $x \in D^n$, we have that
\[
  \sat_{\bR,\widehat{C}}(x) \in (1 \pm \eps) \cdot \sat_{\bR,C}(x).
\]

In this section, we consider the complete $r$-partite instance $C$ to be the instance on $nr$ variables, which we view as the disjoint union $r$ sets $V_1, \hdots, V_r$ of size $n$. The clauses of the instance consist of the $n^r$ possible $r$-tuples $S \in V_1 \times \cdots \times V_r$, each having the same weight of $1$. Since the variable sets applied to each coordinate of $\bR$ are disjoint, we consider the more general $r$-sorted setting in which the valued relation is of the form $\bR : D_1 \times \cdots \times D_r \to \Z^{\ge 0}$, with the global assignments being of the form $x \in D_1^n \times \cdots \times D_r^n$.

A primary (but not only) factor governing the sparsifiability of $C$ is the presence of $\AND_k$, which we define below for $r$-sorted valued relations.

\begin{definition}\label{def:vr-and}
    We say an $r$-sorted valued relation $\bR : D_1 \times \cdots \times D_r \to \mathbb Z^{\ge 0}$ has a restriction to $\AND_k$ if there exist sets $E_1 \subseteq D_1, \hdots, E_r \subseteq D_r$ such that $1 \leq |E_i| \leq 2$, exactly $k$ of the $E_i$'s are of size $2$, and all together, $|\supp \bR \cap E_1 \times E_2 \dots \times E_r| = 1$.
\end{definition}

However, unlike \cref{thm:RpartiteArbDomain}, the largest $\AND_k$ in $\bR$ does not suffice to understand the size of the sparsifier. To see why, consider the following example. Let $D = \{0,1\}$ and $r = 2$ and let $\bR$ be
\[
  \bR(00) = 0,\ \bR(01) = 0,\ \bR(10) = 1,\ \bR(11) = 2.
\]
One can show that $\bR$ has a restriction to $\AND_1$ but not $\AND_2$. Thus, one might expect that the sparsifiability of the complete $r$-partite instance to be on the order of $\widetilde{\Theta}_{\eps}(n)$. However, for reasons similar to that of \cref{thm:BooleanUniform}, \cref{BooleanUniformItem4}, any sparsifier of $\bR$ must have size at least $\Omega_{\eps}(n^2)$. In short, for every $i \in V_1$, we need to preserve at least $\Omega_{\eps}(n)$ edges of the form $(i,j)$ with $j \in V_2$ in order to correctly estimate the total contribution of $\bR(10)$ and $\bR(11)$.

\begin{remark}
One can also show from this example that non-redundancy does not govern the sparsifiability of valued CSPs, which is in stark contrast to the result of Brakensiek and Guruswami~\cite{brakensiek2025redundancy} for ordinary CSPs. 
\end{remark}

\subsection{Balanced Marginals and Imbalanced \texorpdfstring{$\AND$}{AND}}

For succinctness, we define $\vD^r := D_1 \times \cdots \times D_r$. Given a valued relation $\bR : \vD^r \to \Z^{\ge 0}$ and a relation $R_0 \subseteq \vD^r$, we say that $\bR \upharpoonright R_0$ is \emph{balanced} if for all $t \in R_0$, $\bR(t)$ has the same value. This motivates the definition of a \emph{generalized} $\AND_k$.

\begin{definition}\label{def:generalized-AND}
We say that an $r$-sorted valued relation $\bR : \vD^r \to \Z^{\ge 0}$ has a restriction to a generalized $\AND_k$ if there exists $E_1 \subseteq D_1, \hdots, E_r \subseteq D_r$ as well as a set of $k$ distinguished coordinates $S \subseteq [r]$ with the following properties.
\begin{itemize}
\item For every $i \in S$, we have that $|E_i| = 2$, where $E_i = \{d_i, e_i\}$. 
\item For every $t \in \vE^r$, we have that $\bR(t) \neq 0$ if and only if $t_i = e_i$ for all $i \in S$.
\end{itemize}
We say that $\bR \upharpoonright \vE^r$ is a \emph{balanced} $\AND_k$ if every nonzero output has the same value. Otherwise, we say that $\bR \upharpoonright \vE^r$ is an \emph{imbalanced} $\AND_k$. 
\end{definition}

Note that $\bR$ has a restriction to at least one generalized $\AND_k$ if and only if $\bR$ has a restriction to $\AND_k$ (as in \cref{def:vr-and}).

To give some intuition, recall our previous example of

\[
  \bR(00) = 0,\ \bR(01) = 0,\ \bR(10) = 1,\ \bR(11) = 2.
\]
In this case, $\bR \upharpoonright D^2$ is an imbalanced $\AND_1$ because the value of the first coordinate suffices to distinguish between zero and non-zero but the second coordinate still affects the value of the output. We can now state our classification.

\begin{theorem}\label{thm:RpartiteVCSP}
    Let $C$ be the complete, $r$-partite instance of an $r$-sorted valued relation $\bR : \vD^r \to \Z^{\ge 0}$ with $n$ variables in each part, and let $c = \max \{k: \bR \text{ has a restriction to } \mathbf{AND}_k \} $. If every generalized $\AND_c$ of $\bR$ is balanced, let $\widehat{c} = c$. Otherwise, let $\widehat{c} = c+1$. Then:
\begin{enumerate}
    \item For any $\eps > 0$, there exists a $(1 \pm \eps)$ sparsifier of $C$ which preserves $\widetilde{O}_{\bR}(n^{\widehat{c}} / \eps^3)$ constraints. \label{item:rpartiteVCSP1}
    \item For any $\eps \in (0, O_{\bR}(1))$, a $(1 \pm \eps)$ sparsifier of $C$ must preserve $\Omega_{\eps}(n^{\widehat{c}})$ constraints. \label{item:rpartiteVCSP2} 
\end{enumerate}
\end{theorem}

\subsection{Proof of the Lower Bound}

As before, we begin with a proof of the lower bound (\cref{thm:RpartiteVCSP}, \cref{item:rpartiteVCSP2}). We note this directly follows by proving the following two lemmas.

\begin{lemma}\label{lem:hatc=c}
If $\bR$ has a restriction to $\AND_k$, then every $(1 \pm \eps)$ sparsifier of $C$ must preserve $\Omega(n^k)$ constraints.
\end{lemma}

\begin{lemma}\label{lem:hatc=c+1}
If $\bR$ has a restriction to an imbalanced $\AND_k$, then every $(1 \pm \eps)$ sparsifier of $C$ must preserve $\Omega(n^{k+1})$ constraints if $\eps$ is a sufficiently small constant depending on the relation $\bR$. 
\end{lemma}

In particular, \cref{lem:hatc=c} handles the case in which $\widehat{c} = c$, and \cref{lem:hatc=c+1} handles the case in which $\widehat{c} = c+1$. We observe that the proof of \cref{lem:hatc=c} is identical to that of \cref{thm:RpartiteArbDomain}, \cref{item:BooleanRpartiteArbDom2} as the lower bound argument only looks at a single tuple in the relation. We omit further details. We thus focus on proving \cref{lem:hatc=c+1}. The main idea is similar to the proof of \cref{thm:BooleanUniform}, \cref{BooleanUniformItem4}.

\begin{proof}[Proof of \cref{lem:hatc=c+1}]
Consider $E_1 \subseteq D_1, \hdots, E_r \subseteq D_r$ as well as a set of $k$ distinguished coordinates $S \subseteq [r]$ such that $\bR \upharpoonright \vE^r$ which is an imbalanced generalized $\AND_k$. Without loss of generality, we may assume that $S = [k]$ and $E_1 = \cdots = E_k = \{0,1\}$ so that if $\bR(t) > 0$ for some $t \in \vE^r$ then $(t_1, \hdots, t_k) = (1,\hdots, 1)$.

Since $\bR \upharpoonright \vE^r$ is imbalanced, there exists $s,t \in \vE^r$ with $0 < \bR(s) < \bR(t)$. Since $\vE^r$ is a product set, we can construct a chain of tuples $s^0 = s, \hdots, s^{\ell} = t \in \vE^r$ such that $\Ham(s^i, s^{i+1}) = 1$ throughout the chain and $s^i_j = 1$ whenever $j \in [k]$. Since $\bR(s) \neq \bR(t)$, there must be two consecutive terms in this sequence  which differ in value. Thus, we may assume without loss of generality that $\Ham(s,t) = 1$. Furthermore, we may assume that $s_{k+1} \neq t_{k+1}$ and $\bR(s) < \bR(t)$.

For $i \in [k]$, fix $x_i \in V_i$. Let $\Phi_x$ be the family of assignments $\psi \in D_1^n \times \cdots \times D_r^n$ with the following properties.
\begin{itemize}
\item For $i \in [k]$, we have that $\psi_{x} = \one[x = x_i]$ for all $x \in V_i$.
\item For all $x \in V_{k+1}$, $\psi_x \in \{s_{k+1}, t_{k+1}\}$. 
\item For all $i \in \{k+2, \hdots, r\}$, for all $x \in V_{i}$, $\psi_x = s_i$.
\end{itemize}
We claim that any $(1\pm \eps)$ sparsifier of $C$ must keep at least $n/2$ constraints $y \in V_1 \times \cdots \times V_r$ with $y_i = x_i$ for all $i \in [k]$. This immediately implies an $n^{k+1}/2$ lower bound on the sparsifier size, since the choice of $x_i \in V_i$ for $i \in [k]$ was arbitrary.

 Since $\eps$ is sufficiently small, we may assume that $1 + 10\eps < \bR(t) / \bR(s)$. Assume we have a $1\pm \eps$ sparsifier which only preserved less than $n/2$ constraints with $y_i = x_i$ for all $i \in [k]$. Let $V'_{k+1} \subseteq V_{k+1}$ be the vertices which are contained in at least one constraint. By assmption, we have $|V'_{k+1}| < n/2$. Now look at two assignments $\psi, \psi' \in \Phi_x$, where
\begin{itemize}
\item $\psi_x = s_{k+1}$ for all $x \in V_{k+1}$.
\item $\psi'_x = s_{k+1}$ for all $x \in V'_{k+1}$.
\item $\psi'_x = t_{k+1}$ for all $x \in V_{k+1} \setminus V'_{k+1}$.
\end{itemize}

Then, we can compute that
\begin{align*}
  \sat_{\bR,C}(\psi) &= n^{r-k}\bR(s)\\ 
  \sat_{\bR,C}(\psi') &= n^{r-k-1}(|V'_{k+1}|\bR(s) + |V_{k+1} \setminus V'_{k+1}|\bR(t))\\ 
\end{align*}
Thus, \begin{align*}
\frac{\sat_{\bR,C}(\psi')}{\sat_{\bR,C}(\psi)} = \frac{|V'_{k+1}|}{n} + \frac{|V_{k+1} \setminus V'_{k+1}|}{n} \frac{\bR(t)}{\bR(s)} > 1 + 5\eps.
\end{align*}
However, since $\psi$ and $\psi'$ have the same weight with respect to the $(1\pm \eps)$ sparsifier, we have that
\[
  \frac{1+\eps}{1-\eps} > \frac{\sat_{\bR,C}(\psi')}{\sat_{\bR,C}(\psi)},
\]
which is a contradiction.
\end{proof}

\subsection{Proof of the Upper Bound}

Let $W := \max \{\bR(t) : t \in \vD^r\} = O_{\bR}(1)$. We first show a simple $n^{c+1}$ upper bound, which suffices to handle the case that $\widehat{c} = c+1$.

\begin{lemma}\label{lem:c+1-ub}
 For any $\eps > 0$, there exists a $(1 \pm \eps)$ sparsifier of $C$ which preserves $O_{\bR}(n^{c+1} / \eps^2)$ constraints.
\end{lemma}

\begin{proof}
As in previous constructions, we sample $\ell$ constraints uniformly at random (with replacement) from $V_1 \times \cdots \times V_r$, where $\ell = \frac{\kappa(\bR) \cdot n^{c+1}}{\eps^2}$ for a sufficiently large finite constant $\kappa(\bR)$. Every surviving constraint is given weight $n^{r}/\ell$, so the sampler is unbiased. We denote the resulting sparsifier by $\widehat{C}$. We now show that with high probability \emph{for every} assignment $\psi \in D_1^{n} \times \cdots \times D_r^n$, 
\[
\mathrm{sat}_{\bR, \widehat{C}}(\psi) \in (1 \pm \eps)\mathrm{sat}_{\bR,C}(\psi) \ .
\]
If $\sat_{\bR,C}(\psi) = 0$, then we automatically have that $\sat_{\bR, \widehat{C}}(\psi) = 0$ as $\widehat{C}$ is a reweighting of $C$. Henceforth, we assume that $\sat_{\bR,C}(\psi) > 0$. For each $i \in [r]$, and $d \in D_i$, let $a_{i,d}$ denote the number of $x \in V_i$ for which $\psi_x = d$. Let $d_i \in D_i$ be such that $a_{i,d_i}$ is maximal. In particular, $a_{i,d_i} \ge n/|D_i|$.

Since $\sat_{\bR,C}(\psi) > 0$, there exists $e_1 \in D_1, \hdots, e_r \in D_r$ such that $(e_1, \hdots, e_r) \in \supp \bR$ and $a_{i,e_i} \ge 1$ for all $i \in [r]$. Of all such choices of $(e_1, \hdots, e_r)$, pick the one which maximizes the number of $i \in [r]$ for which $e_i = d_i$. Now for all $i \in [r]$, let $E_i := \{d_i, e_i\}.$ By the maximality condition on the choice of $(e_1, \hdots, e_r)$, we have that $\bR \upharpoonright \vE^r$ is a copy of $\AND_{k}$ for some $k \le c$. Therefore,
\[
  \mathrm{sat}_{\bR,C}(\psi) \ge \prod_{i=1}^r a_{i,e_i} \ge \prod_{\substack{i \in [r]\\d_i=e_i}} \frac{n}{|D_i|} \ge \frac{n^{r-k}}{\eta(\bR)} \ge \frac{n^{r-c}}{\eta(\bR)},
\]
where $\eta(\bR)$ is a value depending only on $\bR$.

We now apply \cref{cor:concentration} with $x \in [0,W]^{n^r}$ being the vector of $\psi$'s clause values, $\ell = \frac{\kappa(\bR) \cdot n^{c+1}}{\eps^2}$, and $X_1, \hdots, X_\ell$ be the clause values in $\widehat{C}$ to see that
\begin{align*}
  \Pr_{\widehat{C}}\left [\sat_{\bR,\widehat{C}}(\psi)  \in (1 \pm \eps) \sat_{\bR,C}(\psi)\right ] &\geq 1 - 2 \exp\left(- \frac{\eps^2 \ell \cdot \wt(x)}{3Wm} \right )\\
&= 1 - 2 \exp\left(- \frac{\kappa(\bR)\cdot n^{c+1} \cdot n^{r-c}/\eta(\bR)}{3W n^r} \right )\\
&= 1 - 2\exp\left(-\frac{\kappa(\bR)}{3\eta(\bR)W}n\right). 
\end{align*}
In particular, for $\kappa(\bR) := 3\eta(\bR)W \log(3|\vD^r|)$, we have that
\[
  \Pr_{\widehat{C}}\left [\sat_{\bR,\widehat{C}}(\psi)  \in (1 \pm \eps) \sat_{\bR,C}(\psi)\right ] \ge 1 - \frac{2}{3|\vD^r|^n},
\]
so by a union bound over all $|\vD^r|^n$ assignments $\psi$, we have that a $1\pm \eps$ sparsifier $\widehat{C}$ of $C$ exists with $O_{\bR}(n^{c+1}/\eps^2)$ clauses.
\end{proof}

Moving forward, we assume that $\widehat{c} = c$. That is, every generalized $\AND_{c}$ appearing in $\bR$ is balanced. With this assumption, we now proceed to prove a stronger upper bound of $\widetilde{O}_{\bR,\eps}(n^c)$ on the size of the sparsifier using the methods of \cref{sec:rpartiteGeneral}. More precisely, let $R := \supp \bR$ and note that $c = \max \{k: R \text{ has a restriction to } \mathbf{AND}_k \}$. Therefore, we can use many results of \cref{sec:rpartiteGeneral} verbatim.

Now, we consider an assignment $\psi \in D_1^n \times \cdots \times D_r^n$ and let $a_{i,d}$ be the number of times $\psi(x) = d$ for $x \in V_i$ i.e., $a_{i,d} = |\psi^{-1}(d) \cap V_i|$. WLOG, we assume that $\psi$ is $0^r$-dominant, meaning that $a_{i,0} \geq n / |D_i|$. Ultimately, we will take a union bound over all $|D|^r$ possible dominant tuples, and by a simple relabeling of the elements, every case is isomorphic to the case when $0^r$ is dominant. 

Next, recall, that $0^r \in \vD^r$ is \emph{extreme} if $\Ham(0^r, R) = c$ and that $\mathcal S_{0^r} := \{t \in R : \Ham(0^r,t) = c\}.$  As in \cref{sec:rpartiteGeneral}, if $0^r$ is \emph{not extreme}, then we directly get a lower bound of $\Omega_{\bR}(n^{r-c+1})$ satisfied (positive value) constraints. I.e., by invoking \cref{clm:notExtremeGeneral}, we have that 
\[
\sat_{\bR,C} \ge \sat_{R,C} (\psi) \ge \Omega_{\bR}(n^{r-c+1}),
\]
which is sufficient for the union bound, so we turn to the case in which $0^r$ is extreme.

Thus, we focus WLOG on the case where $0^r$ is extreme.
Now, assuming $0^r$ is extreme, by Proposition~\ref{prop:irrelevantGeneral}, there a set $I \subseteq \bigcup_{i=1}^r \{i\} \times D_i$ of \emph{irrelevant coordinates} with the following properties.

\begin{itemize}
\item If $i \in [r]$ and $d \in D_i$ such that $(i,d) \not\in I$, then there exists $t \in \cS_{0^r}$ with $t_i = d$.
\item For all $t \in \cS_{0^r}$ and $s \in \vD^r$, if for all $i \in [r]$, we have that $s_i = t_i$ or $t_i = 0$ and $(i,s_i) \in I$, then $s \in R$. 
\end{itemize}
As in \cref{prop:irrelevantGeneral}, we also use the convention that $(i,0)$ is in $I$ by default, as the purpose of $I$ is to look at irrelevant alternatives of $0^r$, which includes $0^r$ itself.

At this point, \cref{sec:rpartiteGeneral} considered the notion of \emph{decomposability}, which we generalize to valued relations.

\begin{definition}
    For a valued relation $\bR : \vD^r \to \Z^{\ge 0}$, and a set of irrelevant coordinates / symbols $I \subseteq \bigcup_{i=1}^r \{i\} \times D_i$, we say that $\bR$ is \emph{decomposable} if for every tuple $s \in D^r$, we have $s \in \supp \bR$ if and only if the tuple $s'$ such that for $i \in [r]$
    \begin{enumerate}
        \item if $(i, s_i) \in I$, $s'_i = 0$
        \item if $(i, s_i) \notin I$, $s'_i = s_i$
    \end{enumerate}
    is also in $\supp \bR$. Furthermore, we assert that $\bR(s) = \bR(s').$ 
\end{definition}

We also generalize the concept of \emph{codewords}. Given $\psi \in D_1^n \times \cdots \times D_r^n$, we let $C_{\bR}(\psi) \in \Z^{n^r}_{\ge 0}$ denote the value of $\psi$ on each of the $n^r$ clauses of $C$ with respect to $\bR$. Note when analyzing the performance of a sparsifier, it suffices to take the union bound over all possible codewords rather than all possible assignments. This naturally leads to the following generalization of \cref{clm:countingBoundGeneral}.

\begin{claim}\label{clm:countingBoundVCSP}
    Let $\bR : \vD^r \to \Z^{\ge 0}$ be a decomposable relation such that $\Ham(0^r, \supp \bR) \le c$. Then, the number of distinct $0^r$-dominant codewords of weight $\leq \lambda \cdot \left ( \frac{n}{|\vD|}\right )^{r-c}$ is at most $n^{O_{\bR}(\lambda)}$, where $|\vD| := \max_{i \in [r]}|D_i|$. 
\end{claim}

\begin{proof}
    Fix $i \in [r]$ and define $E_i \subseteq D_i$ such that $(i, d) \in \bar{I} \subseteq \bigcup_{i=1}^r \{i\} \times D_i$ if and only if $d \in E_i$. We know that for every choice of such $(i,d)$, there is a tuple $t^{(i,d)} \in \cS_{0^r}$ such that $t^{(i,d)}_i = d$. Thus, any $0^r$-dominant $\psi \in D_1^n \times \cdots \times D_r^n$ obeying $\mathrm{sat}_{\bR}(\psi) \leq \lambda \cdot \left ( \frac{n}{|\vD|}\right )^{r-c}$ also satisfies the following inequalities for every $i \in [r]$
    \[
    \lambda \cdot \left ( \frac{n}{|\vD|}\right )^{r-c} \geq \mathrm{sat}_{\bR, C}(\psi) \geq \sum_{d \in E_i} \sat_{t^{(i,d)},C}(\psi) \ge \sum_{d \in E_i} a_{i,d} \cdot \left ( \frac{n}{|\vD|}\right )^{r-c},
    \]
 therefore yielding $\sum_{d \in E_i} a_{i,d}\leq \lambda$.

    As such, for any $i \in [r]$, observe that there are at most $\binom{n}{\lambda} \cdot |\vec D|^{\lambda} = n^{O_{\bR}(\lambda)}$ choices of assignments for the variables in $V_i$ which the assignment $\psi$ maps to $E_i$, provided $\mathrm{sat}_R(\psi) \leq \lambda \cdot \left ( \frac{n}{|D|}\right )^{r-c}$. Across all $r$ parts $V_1, \dots V_r$, this means that there are at most $n^{O_{\bR}(\lambda)}$ choices for all such variables.

    Finally, let $\psi, \psi'$ be any two distinct assignments such that for all $i \in [r]$, and for every $d \in E_i$, if $j \in V_i$, then $\psi_j = d$ if and only if $\psi'_j = d$ (i.e., the two assignments exactly match on all relevant coordinates). Then, $\psi$ and $\psi'$ satisfy exactly the same constraints, and therefore define exactly the same codeword. To see this, any constraint operates on $r$ variables $x_1 \in V_1, \hdots, x_r \in V_r$. For all $i \in [r]$, let $y_i = \psi_{x_i}$ and $y'_i = \psi'_{x_i}$. For any $i \in [r]$, if either $y_i \in E_i$ or $y'_i \in E_i$, then $y_i = y_i'$. Thus, the only way for $y_i$ and $y_i'$ to differ is if neither of $y_i$ and $y_i'$ are in $E_i$. As such, for every $i \in [r]$, they must take on \emph{irrelevant} values in their $i$th position. Since $R$ is decomposable, we thus have that there is $t \in \cS_{0^r}$ such that $\bR(y) = \bR(t) = \bR(y')$. In other words, $C_{\bR}(\psi) = C_{\bR}(\psi')$.

    In summary then, we see that there are at most $n^{O(\lambda)}$ distinct codewords of weight $\leq \lambda \cdot \left ( \frac{n}{|D|}\right )^{r-c}$, as the satisfied constraints are uniquely determined by the positions of the relevant values. 
\end{proof}

As we shall soon see, \cref{clm:countingBoundVCSP} suffices for the eventual union bound when $\bR$ is decomposable. We now turn to the case in which $\bR$ is not decomposable. Given two relations $\bR, \bS : \vD^r \to \Z_{\ge 0}$, we say that $\bR \le \bS$ if $\bR(t) \le \bS(t)$ for all $t \in \vD^R$. We say that $(\bR_0, \bR_1)$ \emph{sandwiches} $\bR$ if $\bR_0 \le \bR \le \bR_1$.

Define an equivalence relation $\sim_{I}$ on $\vD^r$ where $s \sim_I t$ if for all $i \in [r]$, if $(i,s_i) \in I$ then $(i,t_i) \in I$ and if $(i,s_i) \not\in I$, then $s_i = t_i$. The fact that $\sim_I$ is an equivalence relation was implicitly used in \cref{clm:countingBoundVCSP}. Now define for all $t \in \vD^r$,
\begin{align}
\bR_0(t) &:= \min_{s \sim_I t} \bR(s) \nonumber \\
\bR_1(t) &:= \max_{s \sim_I t} \bR(s) \ . \label{eq:relation-sandwich}
\end{align}

It is straightforward to see that $\bR_0$ and $\bR_1$ are decomposable and $(\bR_0, \bR_1)$ sandwich $\bR$. Furthermore, the choice of $\bR_0$ is maximal and the choice of $\bR_1$ is minimal. We now show that $\bR_0$ respects the ``$0^r$-dominant'' structure of $\bR$. In particular, we generalize \cref{fact:coveredGeneral} and \cref{fact:weightLBGeneral}.

\begin{claim}\label{claim:Ham-c-VCSP}
For all $t \in \vD^r$ with $\Ham(t) = c$, we have that $\bR_0(t) = \bR(t) = \bR_1(t)$.
\end{claim}
\begin{proof}
Fix $t \in \vD^r$ with $\Ham(t) = c$. Note that we already have $\bR_0(t) \le \bR(t) \le \bR_1(t)$. First, assume for sake of contradiction that $\bR_0(t) < \bR(t)$. Thus, there is $s \sim_I t$ such that $\bR(s) < \bR(t)$. In particular, $t \in \supp \bR$. Now, for all $i \in [r],$ define $E_i \subseteq D_i$ such that
\begin{itemize}
\item If $t_i \neq 0$, then $E_i = \{0, t_i\}$. Note then that $s_i = t_i$.
\item Otherwise, $E_i = \{s_i, t_i\} = \{0, s_i\}$. Note then that $(i,s_i) \in I$.
\end{itemize}
By \cref{prop:irrelevantGeneral}, we have that $t' \in E_1 \times \cdots \times E_r$ satisfies $t' \in \supp \bR$ if and only if $t'_i = t_i$ whenever $t_i \neq 0$.  Thus, $\bR \upharpoonright \vE^r$ is a generalized $\AND_c$ (\cref{def:generalized-AND}) since $\Ham(t) = c$. Since we assume that every generalized $\AND_c$ is balanced, we must have that $\bR(s) = \bR(t)$, a contradiction.

Second, assume for the sake of contradiction that $\bR(t) < \bR_1(t)$. Thus, there is $s \sim_I t$ such that $\bR(s) > \bR(t)$. In particular, $s \in \supp \bR$. If $t \in \supp \bR$ as well, then we can get a contradiction similar to the previous argument, so assume that $t \not\in \supp \bR$.  Note that $\Ham(s) \ge c+1$ as $s$ and $t$ can only differ in coordinates $i \in [r]$ for which $t_i = 0$.  Since $\bR$ lacks $\AND_{c+1}$, there must be some $s' \in \supp \bR$ with $\Ham(s') = c$ and for all $i \in [r]$ either $s'_i = 0$ or $s'_i = s_i$. Now $s' \neq t$ as $t \notin \supp \bR$, and it follows that $\supp(s') \neq \supp(t)$ as well.
However, $\supp(s') \cup \supp(t) \subseteq \supp(s)$. Therefore there exists an $i \in \supp(s) \setminus \supp(t)$ for which $(i,s'_i) = (i,s_i) \not\in I$. This contradicts the facts that $s \sim_I t$. We can thus conclude $\bR(t) = \bR_1(t)$.
\end{proof}

With this, we are now able to establish the following key claim:

\begin{claim}\label{clm:CaseAnalysisVCSP}
    For every $0^r$-dominant assignment $\psi \in D^{nr}$, any $\eps \in (0,1)$, and any $R$ for which $0^r$ is extreme, at least one of the following holds:
    \begin{enumerate}
        \item $(1 + \eps) \mathrm{sat}_{\bR_0,C}(\psi) \geq \mathrm{sat}_{\bR_1,C}(\psi)$.
        \item $\mathrm{sat}_{\bR_0,C}(\psi) = \Omega_{\bR}( \eps \cdot n^{r-c+ 1})$.
    \end{enumerate}
\end{claim}

The proof is nearly identical to \cref{clm:CaseAnalysisVCSP}, but we include it for completeness.

\begin{proof}
    Assume that (1) is false. This means that there must be some $t \in \vD^r$ for which $\bR_1(t) > \bR_0(t)$ yet $\mathrm{sat}_{t,C}(\psi) \ge \Omega_{\bR}(\eps \cdot \mathrm{sat}_{R_0}(\psi))$. Since $\bR_1$ and $\bR_0$ are decomposable, we may pick $t$ arbitrarily from its $\sim_I$ equivalence class. In particular, we may assume for all $i \in [r]$ that either $(i,t_i) \not\in I$ or $t_i = 0$. By \cref{claim:Ham-c-VCSP}, we have that $t$ is non-zero in at least $c+1$ coordinates. Thus, we see that
\begin{align*}
    \mathrm{sat}_{t,C} (\psi) = \prod_{i \in [r]: t_i =0} a_{i, 0} \cdot \prod_{i \in [r]: (i, t_i) \in \bar{I}} a_{i, t_i} \geq \Omega_{\bR}(\eps\cdot  \mathrm{sat}_{\bR_0,C}(\psi)).
\end{align*}
Now, note that since $\mathrm{Ham}(t) \ge c+1$ but the largest arity $\mathbf{AND}$ contained in $\bR$ is of size $c$, there must exist $0 \le s \le t$ with $\mathrm{Ham}(s) \leq c$ and $s \in R$. In fact, because $0^r$ is extreme for $R$, it must be the case that there is such an $s$ with $\mathrm{Ham}(s) = c$. By \cref{claim:Ham-c-VCSP}, we know that $s \in \supp \bR_0$, so
\begin{align*}
    \mathrm{sat}_{\bR_0,C}(\psi) \ge \mathrm{sat}_{s,C}(\psi) = \prod_{i \in [r]: s_i = 0} a_{i, 0} \cdot \prod_{i \in [r]: (i, s_i) \in \bar{I}} a_{i, s_i}.
\end{align*}

Let $K' = \supp(t) \setminus \supp(s)$. By dividing the previous two equations, we get that
\[
\prod_{i \in K'} a_{i, t_i} \ge \eps \prod_{i \in K'} a_{i,0} = \Omega_\bR(\eps n^{|K'|}).
\]
In particular, since for each $i \in K'$, $a_{i,t_i} \le n$, we must have that $a_{i,t_i} = \Omega_\bR(\eps n)$ for all $i \in K'$. 

Pick $i \in K'$ arbitrarily and note there must be $t' \in \cS_{0^r}$ with $0 \le t' \le t$ and $t'_i = t_i$. Then, note that
\[
\mathrm{sat}_{\bR_0, C}(\psi) \geq \mathrm{sat}_{t', C}(\psi) = \prod_{i \in [r]} a_{i,t'_i} \geq a_{i, t'_i} \cdot \prod_{i \in [r]: t'_i = 0} a_{i,0} = \Omega_{\bR}(\eps n) \cdot \left ( \frac{n}{|D|}\right )^{r-c} = \Omega_{\bR}(\eps n^{r-c+1}),  
\]
where we use the fact $a_{i,t'_i} \neq 0$ for all $i \in [r]$ as $t'_i \in \{0, t_i\}$, where $\sat_{t,C}(\psi) > 0$.
\end{proof}

We may now complete the proof of \cref{thm:RpartiteVCSP}.

\begin{proof}[Proof of \cref{thm:RpartiteVCSP}, \cref{item:rpartiteVCSP1}]
The case $\widehat{c} = c+1$ was handled by \cref{lem:c+1-ub}, so we assume that $\widehat{c} =c$. Sample $\ell$ constraints uniformly at random with replacement, where $\ell := \frac{\kappa(\bR) \log(n) \cdot n^c}{\eps^3}$ and $\kappa$ is a large constant that depends solely on $\bR$. In the resulting sample, every surviving constraint is given weight $\frac{n^r}{\ell}$. Thus, the sampling procedure is unbiased (i.e., the expected number of satisfied constraints for each assignment is preserved). 

We denote the resulting sparsifier by $\widehat{C}$. All that remains to be shown is that with high probability \emph{for every} assignment $\psi \in D^{nr}$, 
\[
\mathrm{sat}_{\bR, \widehat{C}}(\psi) \in (1 \pm \eps)\mathrm{sat}_{\bR,C}(\psi).
\]
We assume here that in every $V_i$, $\psi$ has at least $1$ coordinate equal to $d$ for each $d \in D$. If this does not hold true, we delete $d$ from $D_i$ and restrict $R$ accordingly. We then apply our analysis to this restricted sub-relation. In our final union bound, we will pay this factor of $\leq 2^{|D_1| + \cdots + |D_r|}$ (the number of possible restrictions). We have several cases. We also assume that $\psi$ is $0^r$-dominant, which occurs an extra factor of $|\vD^r|$ to the union bound.

\begin{enumerate}
    \item If the relation $R$ is a decomposable relation, then recall that by \cref{clm:countingBoundVCSP}, for every $\lambda \in \mathbb{Z}^+$, the number of distinct codewords of weight $\leq \lambda \cdot n^{r-c}$ is at most $n^{O_{\bR}(\lambda)}$. In particular, we know that by \cref{cor:concentration}, for any codeword $x$ of weight $[(\lambda / 2)\cdot n^{r-c}, \lambda \cdot n^{r-c})]$, the probability that $x$ has its weight preserved to a $(1 \pm \eps)$ factor under our sampling procedure is at least 
\[
1 - 2\exp\left(-\frac{\eps^2 (\lambda / 2)\cdot  n^{r-c} \cdot \ell}{3Wn^r}\right) \geq 1 - n^{-\lambda \cdot \kappa(\bR) / (6W)}.
\]
By setting $\kappa(\bR)$ to be sufficiently large, we can ensure that this probability is sufficiently large to survive a union bound. That is, we have, 
\begin{align}
\Pr&[\exists \psi \text{ $0^r$-dominant}: \mathrm{sat}_{\bR, \widehat{C}}(\psi) \notin (1 \pm \eps)\mathrm{sat}_{\bR,C}(\psi)]\nonumber\\
&\leq \sum_{\psi \in D_1^n \times \cdots \times D_r^n} \Pr[\mathrm{sat}_{\bR, \widehat{C}}(\psi) \notin (1 \pm \eps)\mathrm{sat}_{\bR, C}(\psi)]\nonumber\\
&\leq \sum_{\lambda \in \Z^+} \sum_{\psi: \mathrm{Ham}(\psi) \in [(\lambda / 2)\cdot  n^{r-c}, \lambda \cdot n^{r-c})]}\Pr[\mathrm{sat}_{\bR, \widehat{C}}(\psi) \notin (1 \pm \eps)\mathrm{sat}_{\bR,C}(\psi)]\nonumber\\
&\leq \sum_{\lambda \in \Z^+} n^{O_\bR(\lambda)} \cdot n^{-\lambda \cdot \kappa(\bR) / (6W)} \leq \frac{1}{2^{4|D|r} \cdot \mathrm{poly}(n)} \label{eq:decompUnionBoundVCSP}, 
\end{align}
for a sufficiently large choice of $\kappa(\bR)$. Thus, sampling these $w$ constraints will indeed yield a sparsifier. 
\item Otherwise, let us suppose that $\bR$ is not decomposable. Recall that $\psi$ is $0^r$-dominant. By \cref{clm:notExtremeGeneral}, if $0^r$ is not an extreme tuple for the relation $\supp \bR$, then we immediately know that $\mathrm{sat}_{\bR}(\psi)$ is $\Omega_r(n^{r - c + 1})$. In particular, upon sampling $\ell$ constraints (and re-weighing), \cref{cor:concentration} guarantees that
\[
\Pr\left[\mathrm{sat}_{\bR,\widehat{C}}(\psi) \in (1 \pm \eps ) \cdot \mathrm{sat}_{\bR,C}(\psi) \right] \geq 1 - 2^{-4n|D|r},
\]
where we use that $\kappa$ is a sufficiently large constant depending on $\bR$. Thus, taking a union bound over all possible $|\vD^r|^{n}$ such assignments, every such assignment has its weight preserved with probability $\geq 1 - 2^{-3n|D|r}$.

\item We now assume that $\bR$ is not decomposable and $0^r$ is extreme. Define $\bR_0$ and $\bR_1$ as in \cref{claim:Ham-c-VCSP}. By \cref{clm:CaseAnalysisVCSP}, we either have that \begin{align}
(1 + \eps/3) \mathrm{sat}_{\bR_0}(\psi) \geq\mathrm{sat}_{\bR_1}(\psi), \label{eq:sandwichVCSP}
\end{align}
or that
\[
\mathrm{sat}_{\bR}(\psi) \geq \mathrm{sat}_{\bR_0}(\psi) \geq \Omega_{\bR}(\eps/3 \cdot n^{r-c+1}).
\]
In the second case, we can immediately see (again via \cref{cor:concentration}), that 
\[
\Pr\left[\mathrm{sat}_{\bR, \widehat{C}}(\psi) \in (1 \pm \eps ) \cdot \mathrm{sat}_{\bR, C}(\psi) \right] \geq 1 - 2^{-4n|D|r},
\]
and taking a union bound over all $|D|^{nr}$ such possible strings yields a sparsifier with probability $1 - 2^{-3n|D|r}$.

Otherwise, we let $C_0(\psi), C(\psi), C_1(\psi) \in (\Z^{\ge 0})^{n^r}$ denote the codewords for $\psi$ with respect to $\bR_0, \bR,$ and $\bR_1$. Note that $C_0(\psi) \le C(\psi) \le C_1(\psi)$, in a coordinate-wise sense.
Furthermore, we know that $\bR_0$ and $\bR_1$ are both decomposable and that by \cref{claim:Ham-c-VCSP}, both $\bR_0$ and $\bR_1$ satisfy the counting bound of \cref{clm:countingBoundVCSP} for assignments that are $0^r$-dominant. In particular, we can use the exact same union bound of \cref{eq:decompUnionBoundVCSP} to show that with probability $\geq 1 - \frac{1}{2^{3|D|r} \cdot \mathrm{poly}(n)}$, both $C_0(\psi)$ and $C_1(\psi)$ are $(1 \pm \eps/3)$ sparsified when sampling $\ell$ constraints 
for all $0^r$-dominant assignments $\psi$. Now, because $\bR_0 \le \bR \le \bR_1$, we have that
\[
\mathrm{sat}_{\bR_0, \widehat{C}}(\psi) \leq \mathrm{sat}_{\bR, \widehat{C}}(\psi) \leq \mathrm{sat}_{\bR_1, \widehat{C}}(\psi).
\]
Simultaneously, because $\widehat{C}$ yields a $(1 \pm \eps/3)$ sparsifier of $C$ with respect to both $\bR_0$ and $\bR_1$, we see that 
\[
(1 - \eps / 3) \cdot \mathrm{sat}_{\bR_0, C}(\psi) \leq \mathrm{sat}_{\bR, \widehat{C}}(\psi) \leq (1 + \eps / 3) \cdot \mathrm{sat}_{\bR_1, C}(\psi).
\]
Now, we can plug in the inequality from \cref{eq:sandwichVCSP}, namely that $ \mathrm{sat}_{\bR_0,C}(\psi) \geq \frac{\mathrm{sat}_{\bR_1,C}(\psi)}{(1 + \eps/3)} \geq \frac{\mathrm{sat}_{\bR,C}(\psi)}{(1 + \eps/3)}$ and that $\mathrm{sat}_{\bR_1,C}(\psi) \leq (1 + \eps/3) \cdot \mathrm{sat}_{\bR_0,C}(\psi) \leq (1 + \eps/3) \cdot \mathrm{sat}_{\bR,C}(\psi)$ to conclude that
\[
(1 - \eps)\mathrm{sat}_{R, C}(\psi) \leq  \mathrm{sat}_{R, \widehat{C}}(\psi)\leq (1 + \eps)\mathrm{sat}_{R, C}(\psi),
\]
as we desire.

In particular, in this case we see that every assignment with dominant tuple $t$ has its weight preserved with probability $\geq \min(1 - \frac{1}{2^{4|D|r} \cdot \mathrm{poly}(n)}, 1 - 2^{-4n|D|r})$.
\end{enumerate}

Finally, we take a union bound over all $|D|^r$ choices of dominant tuple $t$, and all $2^{|D_1| + \cdots + |D_r|}$ choices of possible restrictions and conclude that $\widehat{C}$ is indeed a $(1 \pm \eps)$ sparsifier of $C$ with probability $\geq 9/10$, as we desire.
\end{proof}

\section{Sparsifiability of the Complete, Uniform Instance with Relations over Arbitrary Domains}\label{sec:completeArbitrary}

In this section, we completely classify the sparsifiability of the complete uniform instance for any CSP. For technical reasons that we shall soon see, it is of no additional difficulty to complete this classification for Valued CSPs. In particular, we build off the notation established in \cref{subsec:vcsp}.

Given a valued relation $\bR : D^r \to \Z^{\ge 0}$, in this section we let $C$ be the complete instance on $n$ variables corresponding to all $r$-tuples of $[n]$ without repetition (denoted by $P([n], r)$). Naturally, we are back to a single domain, as opposed to the $r$ different domains from the previous sections.
All clauses have weight one.  Explicitly, given $\psi \in D^n$, we let 
\[
\sat_{\bR, C}(\psi) := \sum_{(x_1, \hdots, x_r) \in C} \bR(\psi_{x_1}, \hdots, \psi_{x_r}).
\]

\subsection{Symmetric VCSP}

An important feature of the complete uniform instance is that if $(x_1, \hdots, x_r) \in C$, then any permutation of $(x_1, \hdots, x_r)$ is also in $C$. As such, the sparsifiability of $C$ with respect to $\bR$ is largely governed by the structure of the ``symmetrized'' version of $\bR$, which we now define.

\begin{definition}
We define a $D$-histogram to be a $D$-tuple $h \in \Z^D_{\ge 0}$. Given $t \in D^r$, we define its histogram $\hist(t) \in \Z^{D}_{\ge 0}$ by $\hist(t)_d := \#_d(t)$ for all $d \in D$. Given $h \in \Z^D_{\ge 0}$, we define its arity $\ar(h)$ to be the sum of its coordinates. For $r \ge 0$, we let $\Hist^D_r$ be the set of all $D$-histograms whose arity is $r$. One can observe that $|\Hist^D_r| = {{r+|D|-1} \choose {|D|-1}}$.
\end{definition}

\begin{definition}\label{def:sym}
We define an arity-$r$ \emph{symmetric valued relation} (SVR) to be a function $\bS : \Hist_r^D \to \Z_{\ge 0}$.  We let $\supp \bS$ denote the set of $h \in \Hist_r^D$ with $\bS(h) \neq 0$. Given a valued relation $\bR : D^r \to \Z_{\ge 0}$, we let $\Sym(\bR) : \Hist_r^D \to \Z_{\ge 0}$ be the \emph{symmetrization} of $\bR$. That is, for all $h \in \Hist_r^D$, we have that
\[
  \Sym(\bR)(h) := \prod_{d \in D} h_d!\sum_{\substack{t \in D^r\\\hist(t) = h}} \bR(t).
\]
\end{definition}

An instance of a symmetric VCSP on $n$ variables is a set of constraints $C \subseteq \binom{[n]}{r}$ with weights. Given $\bS : \Hist^D_r \to \Z_{\ge 0}$, the value of an assignment $\psi \in D^n$ is defined to be
\[
  \sat_{\bS,C}(\psi) := \sum_{T \in C} w_T \cdot \bS(\hist(\psi|_{T})).
\]
 The complete symmetric instance $C_{\sym}$ gives every $T \in \binom{[n]}{r}$ a weight of $1$. The notion of a $(1\pm \eps)$ sparsifier $C_{\sym}$ is analogous to that on VCSPs, i.e., a reweighting of the clauses which preserves the value of every assignment up to a $(1 \pm \eps)$ factor.

To justify that our definitions are accurate, we now explain that the complete symmetric  instance behaves identically to the uniform complete instance. This will eventually be useful for establishing \cref{thm:UniformGeneral}.

\begin{proposition} \label{prop:uniform-sym}
Let $\bR : D^r \to \Z_{\ge 0}$ be a valued predicate. Let $C$ be the complete uniform instance of $\bR$ on $n$ variables. Let $C_{\sym}$ be the complete symmetric instance of $\Sym(\bR)$ on $n$ variables. Then, for every $\psi \in D^n$, we have that $\sat_{\bR,C}(\psi) = \sat_{\Sym(\bR),C_{\sym}}(\psi)$.
\end{proposition}
\begin{proof}
Given a clause $x \in C$, let $\set(x) \in \binom{[n]}{r}$ denote its corresponding set of elements. For every $\psi \in D^n$, we have that
\begin{align*}
  \sat_{\bR, C}(\psi) = \sum_{x \in C} \bR(\psi_x) &= \sum_{T \in \binom{[n]}{r}} \sum_{\substack{x \in C\\\set(x) = T}} \bR(\psi_x)\\
&= \sum_{T \in \binom{[n]}{r}} \sum_{\substack{t \in D^r\\\hist(t) = \hist(\psi|_T)}} \sum_{\substack{x \in C\\\set(x) = T\\\psi_x=t}} \bR(t)\\
&= \sum_{T \in \binom{[n]}{r}} \sum_{\substack{t \in D^r\\\hist(t) = \hist(\psi|_T)}} \left(\prod_{d \in D} \#_d(t)!\right) \bR(t)\\
&= \sum_{T \in \binom{[n]}{r}} \left(\prod_{d \in D} \hist(\psi|_{T})_d!\right) \sum_{\substack{t \in D^r\\\hist(t) = \hist(\psi|_T)}}  \bR(t)\\
&= \sum_{T \in \binom{[n]}{r}} \Sym(\bR)(\hist(\psi|_T))= \sat_{\Sym(\bR),C_{\sym}}(\psi),
\end{align*}
as desired.
\end{proof}

\subsection{Coarse Classification: \texorpdfstring{$k$}{k}-Plentiful}

Recall in \cref{sec:rpartiteVCSP}, that the largest $\AND_c$ appearing in $\bR$ mostly governed the sparsifiability of the $r$-partite complete instance (with a final factor of $n$ being decided by more subtle properties). We now construct an analogous combinatorial property in the complete setting. 

To start, given $D$-histograms $v, w \in \Z_{\ge 0}^D$ and $E \subseteq D$, we let $v(E) := \sum_{d \in E} v(e)$.  We also say that $v \succ_{E} w$ if $v_d \ge w_d$ for all $d \in E$. In particular, $v(E) \ge w(E)$, so if $v(D) = w(D)$ (i.e., $\ar(v) = \ar(w)$), then $v(D \setminus E) \le w(D \setminus E)$. Often, we let $E = D \setminus \{d\}$ for some $d \in D$, in which case, we denote $\succ_{D \setminus \{d\}}$ by $\succ_{\bar{d}}$.

Given an SVR $\bS : \Hist_r^D \to \Z_{\ge 0}$, a key concept with define is that of being \emph{$k$-plentiful} which captures to what extent one coordinate of the histogram can be increased while only decreasing the other coordinates.

\begin{definition}\label{def:plentiful}
We say that $\bS : \Hist_r^D \to \Z_{\ge 0}$ is \emph{$k$-plentiful} if for all $d \in D$ and $h \in \supp \bS$, there exists $g \in \supp \bS$ with $g \prec_{\bar{d}} h$ such that $g_d \ge k$. In other words, for any histogram $h \in \supp \bS$ and any $d \in D$, there is $h' \in \supp \bS$ (possibly $h$ itself) with $h'_d \ge k$ and $h'_e \le h_e$ for all $e \in D \setminus \{d\}$.

We further say that $\bS$ is \emph{precisely} $k$-plentiful if it is $k$-plentiful but not $k+1$-plentiful. Finally, we say a valued relation $\bR : D^r \to \Z_{\ge 0}$ is (precisely) $k$-plentiful if and only if $\Sym(\bR)$ is (precisely) $k$-plentiful.
\end{definition}

To gain some intuition, consider the Boolean domain $D = \{0,1\}$. For a valued relation $\bR : \{0,1\}^r \to \Z_{ge 0}$, let $w_{\max}$ denote the maximum Hamming weight in $\supp \bR$ and let $w_{\min}$ denote the minimum Hamming weight in $\supp \bR$. We can see that $\bR$ is precisely $k$-plentiful for $k = \min(w_{\max}, r - w_{\min})$, as we can always take $h'$ in \Cref{def:plentiful} to correspond to the element of $\bR$ with maximum or minimum Hamming weight. This expression for $k$ is quite similar to the exponent appearing in \Cref{thm:BooleanUniform}, and as we shall soon see, this is no coincidence.

A concept related to plentifulness is that of \emph{tightness}. Given an SVR $\bS : \Hist_r^D \to \Z_{\ge 0}$, we say that $h \in \supp \bS$ is \emph{$d$-tight} if the only $g \in \supp \bS$ with $h \succ_{\bar{d}} g$ is $h$ itself. We say $h$ is \emph{$(d,k)$-tight} if $h$ is $d$-tight and $h_d = k$. We let $\sT_d(\bS)$ and $\sT_{d,k}(\bS)$ denote the set of $d$-tight and $(d,k)$-tight histograms with respect to $\bS$.

It is not hard to show that $\bS : \Hist_r^D \to \Z_{\ge 0}$ is $k$-plentiful if $\sT_{d,k'}(\bS)$ is empty for all $d\in D$ and $k' < k$. Further, $\bS$ is precisely $k$-plentiful if $\bS$ is $k$-plentiful and $\sT_{d,k}(\bS)$ is nonempty for some $d \in D$. 

Given these definitions, we can now do our first step of the classification, which is to show a ``coarse'' lower bound. This argument can be viewed as a generalization of \cref{thm:BooleanUniform}, \cref{BooleanUniformItem2}.

\begin{lemma}[Coarse lower bound]\label{lem:complete-lb-coarse}
Let $\bR : D^r \to \Z_{\ge 0}$ be a valued relation which is not $k+1$-plentiful. Then, any $(1\pm \eps)$ sparsifier of $C$ requires $\Omega_{\bR}(n^{r-k})$ clauses.
\end{lemma}

\begin{proof}
Let $\bS := \Sym(\bR)$. Since $\bR$ is not $k+1$-plentiful, there exists $d \in D$ and $h \in \sT_{d,k'}(\bS)$ for some $k' \le k$.
 Let $\Phi$ be the set of assignments $\psi \in D^{n}$ such that $\#_e \psi = \#_e h$ for all $e \in D \setminus \{d\}$. (Note we are assuming that $n \ge k \ge k'$.)

Consider $x \in C$ such that $\bR(\psi_x) > 0$. By definition of $\psi$, we must have that $\hist(\psi_x) \prec_{\bar{d}} h$. However, since $h$ is $(d,k')$-tight, this implies that $\hist(\psi_x) = h$. As such, any clause $x \in C$ which $\psi$ satisfies must completely contain $\psi^{-1}(D \setminus \{d\})$. Since $|\psi^{-1}(D \setminus \{d\})| = h(D \setminus \{d\}) = r -k',$ there are $\binom{n}{r-k'} = \Omega_{\bR}(n^{r-k'})$ choices for $\psi^{-1}(D \setminus \{d\})$. In any $(1 \pm \eps)$ sparsifier $\widehat{C}$, any $x \in \widehat{C}$ can overlap at most $\binom{r}{r-k'} = O_{\bR}(1)$ such sets. Thus, any $(1 \pm \eps)$ sparsifier of $\widehat{C}$ requires $\Omega_{\bR}(n^{r-k'}) = \Omega_{\bR}(n^{r-k})$ clauses.
\end{proof}

Likewise, we can show a coarse upper bound. This argument is similar to that of \cref{thm:BooleanUniform}, \cref{BooleanUniformItem1}.
\begin{lemma}[Coarse upper bound]\label{lem:complete-ub-coarse}
Let $\bR : D^r \to \Z_{\ge 0}$ be a $k$-plentiful valued relation. Then, there exists a $(1\pm \eps)$ sparsifier of $C$ with $O_{\bR}(n^{r-k+1} / \eps^2)$ clauses.
\end{lemma}

\begin{proof}
As in previous constructions, we sample $\ell$ constraints uniformly at random (with replacement) from $C$, where $\ell = \frac{\kappa(\bR) \cdot n^{r-k+1}}{\eps^2}$. Every surviving constraint is given weight $n^{r}/\ell$, so the sampler is unbiased. We denote the resulting sparsifier by $\widehat{C}$. We now show that with high probability \emph{for every} assignment $\psi \in D^n$, 
\[
\mathrm{sat}_{\bR, \widehat{C}}(\psi) \in (1 \pm \eps)\mathrm{sat}_{\bR,C}(\psi),
\]
If $\sat_{\bR,C}(\psi) = 0$, then we automatically have that $\sat_{\bR, \widehat{C}}(\psi) = 0$ as $\widehat{C}$ is a reweighting of $C$. Henceforth, we assume that $\sat_{\bR,C}(\psi) > 0$. As such, there is $h \in \supp \Sym(\bR)$ with $h \prec_D \psi$.

Pick $d \in D$ such that $\#_d(\psi)$ is maximal. In particular, $\#_d(\psi) \ge n/|D|$. Since $\bR$ is $k$-plentiful, there exist $g \in \supp \Sym(\bR)$ with $g \prec_{\bar{d}} h$ and $g_d \ge k$. As long as $n/|D| \ge k$, we thus have that $g \prec_D \psi$. Therefore, 
\[
  \sat_{\bR,C}(\psi) \ge \sat_{g,C}(\psi) \ge \prod_{e \in D} \binom{\#_e(\psi)}{g_e} \ge \binom{\#_d(\psi)}{g_d} \ge \frac{n^{k}}{\eta(\bR)},
\]
where $\eta(\bR)$ depends only on $\bR$ for $n$ sufficiently large.

As in \cref{sec:rpartiteVCSP}, we let $W$ denote the maximum output of $\bR$. 
We now apply \cref{cor:concentration} with $x \in [0,W]^{n(n-1) \cdots (n-r+1)}$ being the vector of $\bR(\psi_{x_1}, \hdots, \psi_{x_r}$ for all $(x_1, \hdots, x_r) \in C$,
$\ell = \frac{\kappa(\bR) \cdot n^{r-k+1}}{\eps^2}$, and $X_1, \hdots, X_\ell$ be the clause values in $\widehat{C}$ to see that
\begin{align*}
  \Pr_{\widehat{C}}\left [\sat_{\bR,\widehat{C}}(\psi)  \in (1 \pm \eps) \sat_{\bR,C}(\psi)\right ] &\geq 1 - 2 \exp\left(- \frac{\eps^2 \ell \cdot \wt(x)}{3Wm} \right )\\
&= 1 - 2 \exp\left(- \frac{\kappa(\bR)\cdot n^{r-k+1} \cdot n^{k}/\eta(\bR)}{3W n^r} \right )\\
&= 1 - 2\exp\left(-\frac{\kappa(\bR)}{3\eta(\bR)W}n\right). 
\end{align*}
In particular, for $\kappa(\bR) := 3\eta(\bR)W \log(3|\vD^r|)$, we have that
\[
  \Pr_{\widehat{C}}\left [\sat_{\bR,\widehat{C}}(\psi)  \in (1 \pm \eps) \sat_{\bR,C}(\psi)\right ] \ge 1 - \frac{2}{3|\vD^r|^n},
\]
so by a union bound over all $|\vD^r|^n$ assignments $\psi$, we have that a $1\pm \eps$ sparsifier $\widehat{C}$ of $C$ exists with $O_{\bR}(n^{k-r+1}/\eps^2)$ clauses.
\end{proof}

If we assume that $\bR$ is precisely $k$-plentiful, note that \cref{lem:complete-lb-coarse} and \cref{lem:complete-ub-coarse} are currently off by a factor of $n$. However, based on the current information extracted from $\bR$, either may be tight. To determine which is the correct answer for a particular $\bR$, we need to further probe the structure of $\bR$.

\subsection{Rigidity, Uncontrolled Domains, and Marginal Balance}\label{subsec:marginal-uniform}

In order to sharpen the upper bound of \cref{lem:complete-ub-coarse}, one needs a suitable counting bound like that of \cref{clm:countingBoundVCSP}. In \cref{sec:rpartiteVCSP}, such a counting bound was established for a special family of valued predicates which are \emph{decomposable}. In order to define what a decomposable relation is, we also needed a notion of \emph{irrelevant} coordinates. For the complete uniform model, we introduce an analogous notion of \emph{uncontrolled} domain elements. However, we must first define \emph{rigidity}.

\begin{definition}
Fix a symmetric valued relation $\bS : \Hist^D_r \to \Z_{\ge 0}$. Given $d \in E \subseteq D$, we say that $h \in \supp \bS$ is \emph{$(d,E)$-rigid} if (i) $h(e)= 0$ for all $e \in E \ \setminus \ \{d\}$, and (ii) for any $g \in \supp \bS$ with $g \prec_{\bar{E}} h$, we have that $g(E) = h(d)$. In particular, this means $g(e) = h(e)$ for all $e \in D \setminus E$ for such $g$.
\end{definition}

If $h$ is $(d,E)$-rigid, it means that we cannot move mass of the histogram into $E$ from $\bar{E}$, but we may distribute mass within $E$. 
Observe that rigidity is monotone in the following sense: if $d \in E' \subseteq E$, then $h$ being $(d,E)$-rigid implies that $h$ is $(d,E')$-rigid.

\begin{definition}
Given $h \in \supp \bS$ and $d \in D$, we let $U_{h,d} \subseteq D$ be the set of all $e \in D$ for which $h$ is $(d, \{d,e\})$-rigid. We say all $e \in U_{h,d}$ are \emph{uncontrolled} by $h$.
\end{definition}

Note that if $h$ is $d$-tight, then $h$ is $(d, \{d\})$-rigid, so $d \in U_{h,d}$. Intuitively, $U_{h,d}$ captures all domain elements which look ``identical'' to $d$ from the perspective of $h$. To capture this, given $d, e \in D$ and $h \in \Z^{D}_{\ge 0}$, let $h_{d \tot e}$ be the histogram in which $h_d$ and $h_e$ are swapped (with everything else unchanged). 

\begin{proposition}\label{prop:uncontrolled}
Assume that $\bS : \Hist^D_r \to \Z_{\ge 0}$ is $k$-plentiful and $h \in \sT_{d,k}(\bS)$. For all $e \in U_{h,d}$, we have that $h_{d \tot e} \in \sT_{e,k}(\bS)$.
\end{proposition}

\begin{proof}
If $d = e$, the result is trivial, so assume $d \neq e$. Since $h$ is $(d, \{d,e\})$-rigid, we have that $h(e) = 0$. Since $\bS$ is $k$-plentiful, there exists $g \in \supp \bS$ with $g \prec_{\bar e} h$ and $g_e \ge k$. Likewise, there exists $g' \in \supp \bS$ with $g' \prec_{\bar d} g$ and $g'_d \ge k$. By transitivity, we have that $g' \prec_{\overline{\{d,e\}}} h$.

Since $g'_d \ge k$, in order for $h$ to be $(d,\{d,e\})$-rigid, we must have that $g'_d = k$ and $g'_e = 0$. In particular, this implies that $g' \prec_{\bar d} h$. Thus, since $h$ is $(d,k)$-tight, we have that $g' = h$.

As such, for all $f \in D \setminus \{d,e\}$, we have that $h_f \ge g_f \ge g'_f \ge h_f$. Therefore, $g_d + g_e = g'_d + g'_e = h_d + h_e = k$. Since $g_e \ge k$, we have that $g_e = k$ and $g_e = 0$. Thus, $h_{d \tot e} = g \in \supp \bS$.

Finally, to show that $h_{d \tot e}$ is $(e,k)$-tight, note that if instead there exists $g'' \prec_e h_{d \tot e}$ with $g''_e \ge k+1$, then $g'' \prec_{\overline{\{d,e\}}} h$, showing that $h$ is not $(d, \{d,e\})$ rigid, a contradiction. Thus, $h_{d \tot e}$ is $(e,k)$-tight.
\end{proof}

We now define the notion of a \emph{marginal predicate.}

\begin{definition}
Assume that $\bS : \Hist_r^D \to \Z_{\ge 0}$ is precisely $k$-plentiful and $h \in \sT_{d,k}(\bS)$. Let $d \in E \subseteq D$ be such that $h$ is $(d,E)$-rigid.
Note this implies that $E \subseteq U_{h,d}$. For all $g \in \Hist^{E}_k$, let $g \sqcup_E h \in \Hist^D_r$ be the histogram such that for all $e \in D$,
\[
    (g \sqcup_E h)(e) = \begin{cases}
    g(e) & e \in E\\
    h(e) & \text{otherwise}.
    \end{cases}
\]
Let $\bS_{h,E} : \Hist^E_k \to \Z_{\ge 0}$ be defined by $\bS_{h,E}(g) := \bS(g \sqcup_E h)$ for all $g \in \Hist^E_k$. Note that $h$ being $(d,E)$-rigid is necessary for this definition to be well-defined. We say that $\bS_{h,E}$ is the \emph{marginal predicate} of $\bS$ induced by $h$ and $E$.
\end{definition}

\begin{remark}\label{remark:marginal}
As an example, let $D = \{0,1,2,3\}$ and let $\bR : D^4 \to \Z_{\ge 0}$ satisfy $\bR(0123) = \bR(0023) = \bR(1123) = 1$ with $\bR(t) = 0$ for all other $t \in D^4$. Note by \cref{def:sym} that $\Sym(\bR)(\hist(0023)) = \Sym(\bR)(\hist(1123)) = 2$ while $\Sym(\bR)(\hist(0123)) = 1$. 

Let $h = \hist(0023)$ and note that $h \in \sT_{d,k}(\Sym(\bR))$ for $d = 0$ and $k=2$ and that $h$ is $(0, E)$-rigid for $E= \{0,1\}$. For every $g \in \Hist^E_{2},$ we can then compute $g \sqcup_E h$. More precisely,
\begin{align*}
g = \hist(00) &\implies g \sqcup_E h = \hist(0023)\\
g = \hist(01) &\implies g \sqcup_E h = \hist(0123)\\
g = \hist(11) &\implies g \sqcup_E h = \hist(1123).
\end{align*}
Thus, $\Sym(\bR)_{h,E}(\hist(00)) = \Sym(\bR)_{h,E}(\hist(11)) = 2$ while $\Sym(\bR)_{h,E}(\hist(01)) = 1$.
\end{remark}

Recall in \cref{thm:BooleanUniform}, when $c=0$, we had separate behavior for $R = \{0,1\}^r$ and $R \neq \{0,1\}^r$. This distinction in the non-Boolean domain is captured by the \emph{marginal balance} of $\bS$. This definition can also be viewed as an analogue of \cref{def:generalized-AND} in the uniform model.
\begin{definition}\label{def:balance}
Assume that $\bS : \Hist_r^D \to \Z_{\ge 0}$ is precisely $k$-plentiful and $h \in \sT_{d,k}(\bS)$.  We say that $\bS_{h,E}$ is \emph{balanced} if for all $g \in \Hist^E_k$ we have that $\bS_{h,E}(g)$ is the same value; otherwise $\bS_{h,E}$ is \emph{imbalanced}. We say that $\bS$ itself is \emph{marginally balanced} if for every $d \in D$ and $h \in \sT_{d,k}(\bS)$ and $E \subseteq D$ such that $h$ is $(d,E)$-rigid, we have that $\bS_{h,E}$ is balanced. 
\end{definition}

\begin{remark}
Continuing \cref{remark:marginal}, note that $\Sym(\bR)_{h,E}$ for $h = \hist(0023)$ and $E = \{0,1\}$ is not constant. Therefore, it may appear that $\Sym(\bR)$ is not marginally balanced. However, perhaps unintuitively, $\Sym(\bR)$ is indeed marginally balanced. The reason is that $\bR$ is precisely $1$-plentiful as $2$ and $3$ appear exactly once in every $t \in \supp(\bR) = \{0123,0023,1123\}$. We can then compute that
\begin{align*}
\sT_{0,1}(\Sym(\bR)) &= \emptyset,\\
\sT_{1,1}(\Sym(\bR)) &= \emptyset,\\
\sT_{2,1}(\Sym(\bR)) &= \supp \Sym(\bR),\\
\sT_{3,1}(\Sym(\bR)) &= \supp \Sym(\bR).
\end{align*}
We can compute that each $h \in \Sym(\bR)$ is $(d,E)$-rigid only for $(d,E) = (2, \{2\})$ and $(d,E) = (3,\{3\})$. As such, these are the only pairs for which we need to check for marginal balance. Note that $\Hist^{\{2\}}_1$ consists of a single element $\hist(2)$. Thus, the marginal predicate $\Sym(\bR)_{h,\{2\}}$ is automatically constant for all $h \in \Sym(\bR)$. Likewise, the marginal predicate $\Sym(\bR)_{h,\{3\}}$ is constant for all $h \in \Sym(\bR)$. Therefore, $\Sym(\bR)$ \emph{is} marginally balanced.
\end{remark}

\begin{remark}
The marginal balance of a valued relation $\bR : D^r \to \Z_{\ge 0}$ is different. See \cref{subsec:7.6} for details.
\end{remark}

We can finally now state the main classification theorem.

\begin{theorem}\label{thm:UniformGeneral}
Let $\bR : D^r \to \Z_{\ge 0}$ be a valued relation which is precisely $k$-plentiful.
\begin{enumerate}
\item If $\Sym(\bR)$ is marginally balanced, for any $\eps \in (0,1)$, there is a choice of $\ell =\widetilde{O}_{\bR}(n^{r-k}/\eps^3)$ such that an iid sampling of $\ell$ constraints from $C_{\mathrm{sym}}$ is a $(1 \pm \eps)$ sparsifier of $C$ with high probability. \label{item:unif-ub}
\item If $\Sym(\bR)$ is marginally balanced, for any $\eps \in (0,1)$, every $(1 \pm \eps)$ sparsifier of $C$ must preserve $\Omega_{\bR}(n^{r-k})$ constraints. \label{item:unif-lb}
\item If $\Sym(\bR)$ is not marginally balanced, for any $\eps \in (0,1)$, there exists\footnote{We warn the reader that for the first time in this paper, the constructed sparsifier is \emph{not} an i.i.d. sampling of the constraints. This discrepancy is explored in \cref{subsec:7.6}.} a $(1 \pm \eps)$ sparsifier of $C$ which preserves $O_{\bR}(n^{r-k+1}/\eps^2)$ constraints. \label{item:non-unif-ub} 
\item If $\Sym(\bR)$ is not marginally balanced, for any $\eps \in (0,O_{\bR}(1))$, every $(1 \pm \eps)$ sparsifier of $C$ must preserve $\Omega_{\bR}(n^{r-k+1})$ constraints.\label{item:non-unif-lb}
\end{enumerate}
\end{theorem}

We note that \cref{thm:UniformGeneral} \cref{item:unif-lb} is implied by \cref{lem:complete-lb-coarse} and \cref{thm:UniformGeneral} \cref{item:non-unif-ub} is implied by \cref{lem:complete-ub-coarse}. For the remaining items, we prove \cref{thm:UniformGeneral} \cref{item:non-unif-lb} and then \cref{thm:UniformGeneral} \cref{item:unif-ub} below.

\subsection{Proof of \cref{thm:UniformGeneral} \cref{item:non-unif-lb} (Better lower bound for non-uniform case)}

Note that the case $k=0$ is impossible as every predicate of arity zero is balanced (since there can only be one weight). Thus, we assume that $k\ge 1$.

A key step toward proving the lower bound is the following lemma.
\begin{lemma}\label{lem:non-unif-lb}
Let $\bS : \Hist_k^E \to \Z_{\ge 0}$ be an imbalanced predicate. Let $C_{\sym}$ be the complete $k$-ary symmetric instance on $n$ variables. For some $\delta = \Omega_{\bS}(1)$, there exist assignments $\psi, \psi' \in E^n$ such that $\Ham(\psi, \psi') \le n/2$ but $\sat_{\bS,C_{\sym}}(\psi) > (1+\delta)\sat_{\bS,C_{\sym}}(\psi').$ 
\end{lemma}

We will prove \cref{lem:non-unif-lb} at the end of the section and now explain why it implies \cref{thm:UniformGeneral} \cref{item:non-unif-lb}. The argument will be similar to \cref{lem:hatc=c+1}.

\smallskip
Since $\Sym(\bR)$ is not marginally balanced, there exists $d \in E \subseteq D$ and $h \in \sT_{d,k}(\Sym (\bR))$ such that $h$ is $(d,E)$ rigid, but $\Sym(\bR)_{h,E}$ is imbalanced. Let $\Phi$ be the set of assignments $\psi \in D^n$ such that $\#_e\psi = \#_e h$ for all $e \in D \setminus E$. Consider $x \in C$ such that $\bR(\psi_x) > 0$. By definition of $\psi$, we must have that $\hist(\psi_x) \prec_{\bar{E}} h$. However, since $h$ is $(d,E)$-rigid, we must have that $\hist(\psi_x)_e = h_e$ for all $e \in D \setminus E$. Therefore, any $x \in C$ which is satisfied by $\psi$ must contain $\psi^{-1}(D \setminus E)$ which has size $r-k$. Let $C_{\sym}$ be the complete symmetric instance for $\Sym(\bR)$.

By \cref{prop:uniform-sym}, for all $\psi \in D^n$, we have that $\sat_{\bR, C}(\psi) = \sat_{\Sym(\bR), C_{\sym}}(\psi)$. Furthermore, by the proof of \cref{prop:uniform-sym}, we have that any $T \in C_{\sym}$ for which $\Sym(\bR)(\hist(\psi|_{T})) > 0$ corresponds to an $x \in C$ with $\set(x) = T$ for which $\bR(\psi_{x}) > 0$. Thus, for any $\psi \in \Phi$, we must have that $\psi^{-1}(D \setminus E) \subseteq T$.

Fix $X \subseteq [n]$ of size $r-k$ and let $Y = [n] \setminus X$. Let $\Phi_X$ be the set of $\psi \in \Phi$ for which $\psi^{-1}(D \setminus E) = X$. A key observation is that for any $\psi \in \Phi_X$, we have that
\begin{align*}
  \sat_{\bR,C}(\psi) = \sat_{\Sym(\bR), C_{\sym}}(\psi) &= \sum_{T : X \subseteq T \in \binom{[n]}{r}} \Sym(\bR)(\psi|_{T})\\
&= \sum_{T' : T' \in \binom{[n] \setminus X}{k}} \Sym(\bR)_{h,E}(\psi|_{T'})\\
&= \sat_{\Sym(\bR)_{h,E}, C_{\sym}}(\psi|_{Y}).
\end{align*}

Let $\widehat{C}$ be a potential $(1\pm \eps)$ sparsifier of $C$. Let $\hat{Y} \subseteq Y$ be the set of $y \in Y$ such that there is some edge $x \in \widehat{C}$ with $T \cup \{y\} \subseteq \set(x)$. Assume for sake of contradiction that $|\hat{Y}| < (n-r)/2$.

Now we can invoke \cref{lem:non-unif-lb}. In particular, we can find $\hat{\psi}, \hat{\psi}' \in D^Y $ which differ in at most $|Y|/2$ coordinates for which 
\[
\sat_{\Sym(\bR)_{h,E},C_{\sym}}(\hat{\psi}) > (1+\delta)\sat_{\Sym(\bR)_{h,E},C_{\sym}}(\hat{\psi}')
\]
for some $\delta = \Omega_{\bR}(1)$. By permuting the coordinates of $\hat{\psi}$ and $\hat{\psi}'$ simultaneously, we can assume that $\hat{\psi}_y = \hat{\psi}'_y$ for all $y \in \hat{Y}$. Arbitrarily extend $\hat{\psi}$ and $\hat{\psi'}$ to $\psi, \psi' \in D^n$, respectively, such that $\psi, \psi' \in \Phi_Y$. Then, by the previous logic, we have that $\sat_{\bR,\widehat{C}}(\psi) = \sat_{\bR,\widehat{C}}(\psi')$ but $\sat_{\bR,C}(\psi) > (1+\delta) \sat_{\bR,C}(\psi')$. Thus, for $\eps \in (0, \delta/3) = (0, O_{\bR}(1))$, $\widehat{C}$ cannot be a $1 \pm \eps$ sparsifier, a contradiction.

In other words, $|\hat{Y}| \ge (n-r)/2$. That is, at least $\Omega_{\bR}(n)$ edges of $\widehat{C}$ contain $T$ for every $T \in \binom{[n]}{r-k}$. Thus, $\widehat{C}$ contains at least $\Omega_{\bR}(n^{r-k+1})$ edges, as desired. 

\smallskip
We now return to the deferred proof of \cref{lem:non-unif-lb}.

\begin{proof}[Proof of \cref{lem:non-unif-lb}]
Consider the polynomial ring $\R[z_e : e \in E]$ and construct the following polynomial:
\[
  p = \sum_{g \in \Hist_k^E} \prod_{e \in E}\frac{z_e^{g_e}}{g_e!}\bS(g).
\]
Also let
\[
  q := \left(\sum_{e\in E} z_e\right)^k = \sum_{g \in \Hist_k^E} \prod_{e \in E}\frac{z_e^{g_e}}{g_e!} \cdot k!
\]
Since $\bS$ is imbalanced, we have that $p$ is not a scalar multiple of $q$. Let $\Delta_E := \{a \in \R^E_{\ge 0} : \sum_{e \in E} a_e = 1\}$. We claim there exists $a,b \in \Delta_E$ such that $p(a) \neq p(b)$.

Assume for sake of contradiction that $p$ is constant on $\Delta_E$. In other words, there exists $\lambda \in \mathbb R_{\ge 0}$ such that $p - \lambda q$ is identically zero in $\Delta_E$. 

Consider any line segment (of nonzero length) in $\Delta_E$. Since $p - \lambda q$ has bounded degree, the polynomial must equal zero on the entire line. By iterating this argument, we can show that $p - \lambda q$ is constant on the \emph{affine hull} of $\Delta_E$, i.e., on all points $a \in \R^E$ with $\sum_{e \in E} a_e = 1$.

Next, observe that $p - \lambda q$ is a homogeneous polynomial of degree $k$, thus if we scale all inputs by some $\lambda \in \R$, the output will scale by $\lambda^n$. Thus, since $p - \lambda q$ evaluates to zero for all $a \in \R^E$ with $\sum_{e \in E} a_e = 1$, $p - \lambda q$ always evaluates to zero. Thus, by the Schwarz-Zippel Lemma \cite{zippel1979probabilistic, schwartz1980fast}, $p = \lambda q$, a contradiction of that fact that $\bS$ is imbalanced.

Thus, there exists $a, b \in \Delta_E$ with $p(a) > p(b) > 0$. Let $\delta = \frac{p(a)-p(b)}{3 p(b)} > 0$. If we pick $a$ and $b$ such that $p(a)/p(b)$ is maximized while ensuring that $\sum_{e \in E} |a_e - b_e| \le \frac{1}{3}$, then $\delta = \Omega_{\bS}(1)$.

Next, let $\alpha, \beta \in E^n$ be such that $\#_e(\alpha) \in (na_e - 1, na_e + 1)$ and $\#_e(\beta) \in (nb_e - 1, nb_e + 1)$. Since $\sum_{e \in E} |a_e - b_e| \le \frac{1}{3}$, we have that $\Ham(\alpha, \beta) \le n/2$ for $n$ sufficiently large. We now estimate that
\[
  \sat_{\bS,C_{\sym}}(\alpha) = \sum_{g \in \Hist_k^E} \prod_{e \in E}\binom{\#_e(\alpha)}{g_e} \bS(g) = n^k \sum_{g \in \Hist_k^E} \prod_{e \in E}\frac{(na_e)^k}{g_e!} \bS(g) + O_{\bS}(n^{k-1}) = n^k p(a) + O_{\bS}(n^{k-1}).
\]
Likewise, $\sat_{\bS,C_{\sym}}(\beta) = n^k p(b) + O_{\bS}(n^{k-1}).$

Thus, \[
\frac{\sat_{\bS,C_{\sym}}(\alpha) }{\sat_{\bS,C_{\sym}}(\beta)} = 1 + 3\delta \pm O_{\bS}(1/n).
\]
Pick $n$ large enough such that $O_{\bS}(1/n) < \delta$, then $\alpha$ and $\beta$ satisfy the desired properties.
\end{proof}

\subsection{Proof of \cref{thm:UniformGeneral} \cref{item:unif-ub} (Better upper bound for the balanced case)}\label{subsec:7.5}

Assume $n$ sufficiently large. We now prove \cref{thm:UniformGeneral} \cref{item:unif-ub}, except our sparsifier will not independently sample each edge of $C$. Rather, we will independently sample $\ell = \frac{n^{r-k}\log(n) \kappa(\bR)}{\eps^3}$ sets $T \in \binom{[n]}{r}$ of $C_{\sym}$ and include all $r!$ edges $x \in C$ with $\set(x) = T$ in the sparsifier. The weight of each of these edges will be ${\binom{n}{r}}/\ell$ in order to create an unbiased sampling procedure.
Let $\widehat{C}_{\sym}$ be the sparsifier of $C_{\sym}$ and let $\widehat{C}$ be the sparsifier of $C$. 

For notational convenience, we let $\bS := \Sym(\bR)$. With a computation similar to that of \cref{prop:uniform-sym}, we have that for every $\psi \in D^n$ that
\begin{align}
\sat_{\bS,\widehat{C}_{\sym}}(\psi) = \sat_{\bR,\widehat{C}}(\psi)\label{eq:sym-eq}
\end{align}
Since $\sat_{\bS,C_{\sym}}(\psi) = \sat_{\bR,C}(\psi)$, it suffices prove that $\widehat{C}_{\sym}$ is a $(1\pm \eps)$ sparsifier of $C_{\sym}$ with respect to the predicate $\bS$. In other words, we seek to prove for all $\psi \in D^n$ that
\begin{align}
  \sat_{\bS,\widehat{C}_{\sym}}(\psi) \in [1-\eps, 1+\eps] \cdot \sat_{\bS,C_{\sym}}(\psi).\label{eq:sym-spar}
\end{align}

Fix an assignment $\psi \in D^n$. Let $S_{\psi} \subseteq \supp \bS$ be the set of $h \in \bS$ for which $\sat_{h,C_{\sym}}(\psi) > 0$. In particular, we may assume that $S_{\psi} \neq \emptyset$ or else $\sat_{\bS,C_{\sym}}(\psi) = 0$, which is maintained by our sparsifier. Assume further that $\psi$ is $d$-dominant. That is, $\#_d(\psi)$ is maximal and thus at least $\Omega_{\bR}(n)$.

As a first observation, if there exists $h \in S_{\psi}$ with $h_{d} \ge k+1$, then
\[
  \sat_{h,C_{\sym}}(\psi) \ge \sym(\bR)(h) \prod_{d \in D} \binom{\#_d\psi}{h_d} = \Omega_{\bR}(n^{k+1}).
\]
Thus, by \cref{cor:concentration}, the probability that \cref{eq:sym-spar} fails is at most $\exp(-\Omega_{\bR}(n^{r-k}\cdot n^{k+1} / n^r)) \le \frac{1}{|D|^{2n}}$. In particular, we can easily union bound over all such $\psi$. Otherwise, we say that $\psi$ is \emph{tight}.

We want to understand which domain elements $e \in D$ are ``under control'' in our union bound argument. For one if, a domain element appears very few times, say $\#_e(\psi) < r$, then we know $e$ can appear in at most $\binom{n}{\le r-1} = n^{r-1}$ ways. Likewise, if $(d, \{d,e\})$-rigidity cannot occur, then $e$ can be handled easily (see the end of the proof of \cref{lem:sandwich-bound}).

\begin{definition}
Assume $\psi \in D^n$ is $d$-dominant and tight. Let $U_{\psi} \subseteq D$ satisfy (1) $d \in U_{\psi}$ and (2) for all $e \in D \setminus \{d\}$, we have that $e \in U_{\psi}$ iff $\#_e(\psi) \ge r$ and there does not exist $h \in S_{\psi}$ with $h_d = k$ and $h_e \ge 1$.
\end{definition}

We now have a key claim which states that $(d, U_{\psi})$-rigid for \emph{all} $h \in S_{\psi}$ with $h_d = k$. In particular, we can use the marginal balance framework of \cref{subsec:marginal-uniform} to understand these ``uncontrolled'' domain elements.

\begin{claim}\label{claim:U-rigid}
For all $h \in S_{\psi}$ with $h_d = k$, we have that $h$ is $(d,k)$-tight and $(d,U_{\psi})$-rigid (with respect to $\bS$).
\end{claim}
\begin{proof}
Note that $h$ is $(d,k)$-tight since $\psi$ is tight. By property (2) of $U_{\psi}$, we have that $h(U_{\psi}) = h_d = k$.  Assume $h$ is not $(d, U_{\psi})$-rigid. Thus, there exists $h' \in \supp\bS$ such that $h' \prec_{\bar{U}_{\psi}} h$ and $h'(U_{\psi}) \ge k+1$. Since $\#_e(\psi) \ge r$ for all $e \in U_{\psi}$,
we have that $h' \in S_{\psi}$ as well. Furthermore, since $\bR$ is $k$-plentiful, there exists $h'' \in \supp \bS$ such that $h'' \prec_{\bar{d}} h'$ and $h''_d \ge k$. By similar logic as before, we have that $h'' \in S_{\psi}$. Thus, $h''_d = k$ (by assumption on $\psi$). However, 
\[h''(U_{\psi}) = r - h''(D \setminus U_{\psi}) \ge r - h'(D \setminus U_{\psi}) \ge h'(U_{\psi}) \ge k+1,\]
so $h''_e \ge 1$ for some $e \in U_{\psi} \setminus \{d\}$, a contradiction of the definition of $U_{\psi}$. Thus $h$ is $(d, U_{\psi})$-rigid.
\end{proof}

\subsubsection{Sandwiching}

Fix $d \in E \subseteq D$. Given $s,t \in \Hist^D_r$, we say that $s \sim_E t$ if $s_e = t_e$ for all $e \in D \setminus E$. Given $\bS$, we can define two sandwiching relations $\bS_{E,0}$ and $\bS_{E,1}$ as follows. For all $s \in \Hist^D_r$, we have that
\begin{align*}
\bS_{E,0}(s) &:= \min_{t \sim_{E} s} \bS(t),\\
\bS_{E,1}(s) &:= \max_{t \sim_{E} s} \bS(t).
\end{align*}
Note the similarity to the sandwiching relations from \cref{eq:relation-sandwich}.
To make sure this definition is sensible, we prove the following structural claim.

\begin{claim}\label{claim:tight-sandwich}
Let $\psi \in D^n$ be $d$-dominant and tight and assume that $U_{\psi} = E$. Then, for all $h \in S_{\psi}$ with $h_d = k$, we have that $\bS_{E,0}(h) = \bS_{E,1}(h) = \bS(h)$.
\end{claim}
\begin{proof}
First observe that any $h \in S_{\psi}$ with $h_d = k$ is $(d,k)$-tight, for if there is any $h' \prec_{\bar{d}} h$ with $h'_{d} \ge k+1$, then $h' \in S_{\psi}$ showing that $\psi$ is not tight. As previously shown, $h$ is $(d, E)$-rigid, so by the hypothesis of \cref{thm:UniformGeneral} \cref{item:unif-ub}, we have that $\bS_{h,E}$ is balanced. In particular, this means for all $t \sim_{E} h$, we have that $\bS(t) = \bS(h)$. Thus, $\bS_{E,0}(h) = \bS_{E,1}(h) = \bS(h)$.
\end{proof}
Next, we show the key sandwiching claim.
\begin{lemma}\label{lem:sandwich-bound}
Let $\bR$ be precisely $k$-plentiful, let $\psi \in D^n$ be $d$-dominant and tight and assume that $U_{\psi} = E$. Then for sufficiently large $n$, at least one of the following holds:
\begin{enumerate}
\item $(1+\eps / 3) \cdot \sat_{\bS_{E,0}, C_{\sym}}(\psi) \ge \sat_{\bS_{E,1}, C_{\sym}}(\psi)$
\item $\sat_{\bS_{E,0}, C_{\sym}}(\psi) \ge \Omega_{\bR}(\eps) \cdot n^{k+1}$.
\end{enumerate}
\end{lemma}
\begin{proof}
Assume (1) is false. Then, there is $g \in \supp \bS_{E,1}$ with $\bS_{E,1}(g) > \bS_{E,0}(g)$ and \[\sat_{g, C_{\sym}}(\psi) \ge \Omega_{\bR}(\eps) \cdot \sat_{\bS_{E,0}, C_{\sym}}(\psi).\]
By definition of $\bS_{E,1}$, there is $h \in \supp \bS$ with $h \sim_E g$ and $\bS_{E,1}(g) = \bS(h)$. Since $\bS$ is $k$-plentiful, there is $h' \prec_{\bar{d}} h$ with $h'_d \ge k$. Note in particular that $h' \in S_{\psi}$, so $h'_d = k$ since $\psi$ is tight. Thus, $h'$ is $(d,k)$-tight and $(d, E)$-rigid by \cref{claim:U-rigid}.  Now observe that
\[
\sat_{g, C_{\sym}}(\psi) \ge \Omega_{\bR}(\eps) \cdot  \sat_{h', C_{\sym}}(\psi).
\]
In other words,
\[
\prod_{e \in D} \binom{\#_e\psi}{g_e} \ge \Omega_{\bR}(\eps) \cdot  \prod_{e \in D} \binom{\#_e\psi}{h'_e}.
\]
Since both sides are nonzero, we can simplify the above inequality to
\begin{align}
  \prod_{e \in D \setminus \{d\}} (\#_e \psi)^{g_e - h'_e} \ge \Omega_{\bR}(\eps) \cdot n^{h'_d - g_d}.\label{eq:eps-prod}
\end{align}

Since $h'$ is $(d,k)$-tight and $(d, E)$-rigid by \cref{claim:U-rigid}, we can deduce that (1) $h' \prec_{\bar{d}} g$ as $h'(E \setminus \{d\}) = 0$ and (2) $\bS_{h',E}$ is balanced. The latter consequence implies that for all $g' \sim_{E} h'$ we have that $\bS(g') = \bS(h')$. Thus, we cannot have that $g \sim_E h'$. In other words, there exists $e \in D \setminus E$ for which $h'_e < g_e = h_e$.

Since $h' \prec_{\bar{d}} g$, we have that $h'_d - g_d = \sum_{e' \in D \setminus \{d\}} g_{e'} - h'_{e'}$ with the terms in the sum nonnegative. Thus, by \cref{eq:eps-prod}, we have that $\#_e \psi = \Omega_{\bR}(\eps n).$ Since $e \not\in U_{\psi}$, there is $h'' \in S_{\psi}$ with $h''_d = k$ and $h''_e \ge 1$. Thus, $h''$ is $(d,k)$-tight so $\bS_{h'',E}$ is balanced. Thus,
\[
  \sat_{\bS_{E,0}, C_{\sym}}(\psi) \ge   \sat_{h'', C_{\sym}}(\psi) \ge \Omega_{\bR}(1) \cdot (\#_e \psi)^{h_e} \cdot (\#_d \psi)^{h_d} \ge \Omega_{\bR}(\eps) n^{k+1},
\]
as desired.
\end{proof}

Finally, we show the two ends of the sandwich are sparsified correctly by $\widehat{C}_{\sym}$.

\begin{lemma}\label{lem:sandwich-spar}
Fix $d\in E \subseteq D$. With high probability over the chosen constraints of $\widehat{C}_{\sym}$, for all $\psi \in D^n$ which are $d$-dominant and tight with $U_{\psi} = E$, we have that
\begin{align*}
\sat_{\bS_{E,0},\widehat{C}_{\sym}}(\psi) \in [1-\eps/3, 1+\eps/3] \cdot \sat_{\bS_{E,0},C_{\sym}}(\psi)\\
\sat_{\bS_{E,1},\widehat{C}_{\sym}}(\psi) \in [1-\eps/3, 1+\eps/3] \cdot \sat_{\bS_{E,1},C_{\sym}}(\psi)
\end{align*}
\end{lemma}
\begin{proof}
Let $\psi' \in D^n$ be defined by
\[
  \psi'_x = \begin{cases}
    d &  \psi_x \in E\\
     \psi'_x & \text{otherwise}.
  \end{cases}
\]
By definition of $\bS_{E,0}$ and $\bS_{E,1}$, we have that $\psi'$ and $\psi$ have identical codewords (in the sense of \cref{def:codeword}). Thus, it suffices to consider assignments of the form $\psi'$ (i.e., supported only on $(D \setminus E) \cup \{d\}$). Since $E = U_{\psi'}$ for all $e \in D \setminus E$, we either have (1) $\#_e\psi' \le r$ or (2) there exists $h^e \in S_{\psi'}$ with $h^e_d = k$ and $h^e_e \ge 1$. Note that by \cref{claim:tight-sandwich}, $\bS_{E,0}(h^e) = \bS_{E,1}(h^e) = \bS(h^e)$. Let $E' \subseteq D \setminus E$ be the set of domain elements $e$ for which $h^e$ exists. Therefore,
\begin{align*}
\sat_{\bS_{E,1},C_{\sym}}(\psi') &\ge \sat_{\bS_{E,0},C_{\sym}}(\psi')\\
&\ge \sum_{e \in E'} \sat_{h^e,C_{\sym}}(\psi')\\
&= \sum_{e \in E'} \#_e\psi' \cdot \Omega_{\bR}(n^{k})\\
&\ge (n - \#_d \psi' - r|D|) \cdot \Omega_{R}(n^k).
\end{align*}
In particular, for all $\lambda \ge 1$, the number of $\psi'$ for which $\sat_{\bS_{E,0},C_{\sym}}(\psi')\le \lambda \cdot n^k$ is $n^{O_{\bR}(\lambda)}$.

Using \cref{cor:concentration}, we have that for any $\lambda \ge 1$, if a codeword $x$ has weight $[\lambda /2 , \lambda] \cdot \Omega_{\bR}(n^k)$, then by randomly sampling $\ell = \frac{n^{r-k}\log(n) \kappa(\bR)}{\eps^3}$ edges from $C_{\sym}$ (as we do in the construction of $\widehat{C}_{\sym}$), then
the probability that $x$ is sparsified to within a factor of $(1 \pm \eps)$ is at least 
\[
1 - 2e^{- \left ( \frac{\eps^2\ell \cdot \wt(x)}{3W \cdot m}\right )} \geq 1 - n^{-\Omega_{\bR}(\kappa(\bR) \lambda)},
\] 
where we have used that $m = \binom{n}{r}$ and that $W = O_{\bR}(1)$.

In particular, by our counting bound, the probability that all codewords of weight in the range $[\lambda /2 , \lambda] \cdot \Omega_{\bR}(n^k)$ are sparsified correctly is also $1 - n^{-\Omega_{\bR}(\kappa(\bR) \lambda)}$ for sufficiently large $\kappa(\bR)$. By summing over all $O(r\log(n))$ choices of $\lambda$, the probability this works for all $\lambda$ is at least $1 - n^{-\Omega_{bR}(1)} \ge 1 - \frac{1}{2^{4|D|} \cdot \poly(n)}$, as desired.
\end{proof}

\subsubsection{The Final Union Bound}

Finally, we can bring these pieces together and show an improved sparsification bound:

\begin{proof}[Proof of \cref{thm:UniformGeneral} \cref{item:unif-ub}]

Fix $\psi \in D^n$ with $n$ sufficiently large. If $\sat_{\bS,C_{\sym}}(\psi) \ge \Omega_{\bR}(\eps n^{k+1})$ (where the hidden constant is smaller than the one from \cref{lem:sandwich-bound}), then by \cref{cor:concentration}, we have that the probability that $\psi$ is sparsified correctly is at least $1 - \frac{1}{|D|^{4n}}$, which is sufficient for our union bound. Thus, we now assume that $\sat_{\bS,C_{\sym}}(\psi) = O_{\bR}(\eps n^{k+1})$. In particular, this means that $\psi$ is tight.

For all $d \in E \subseteq D$, condition on \cref{lem:sandwich-spar} holding for $\bS_{E,0}$ and $\bS_{E,1}$. This altogether happens with probability at least $1 - \frac{1}{n^{\Omega_{\bR}(1)}}$. Thus, for our particular $\psi$, we are guaranteed by \cref{lem:sandwich-bound} that
\begin{align*}
(1 + \eps / 3) \cdot \sat_{\bS_{E,0},\widehat{C}_{\sym}}(\psi) &\ge \sat_{\bS_{E,1},C_{\sym}}(\psi)\\
\sat_{\bS_{E,0},\widehat{C}_{\sym}}(\psi) &\in [1-\eps/3, 1+\eps/3] \cdot \sat_{\bS_{E,0},C_{\sym}}(\psi)\\
\sat_{\bS_{E,1},\widehat{C}_{\sym}}(\psi) &\in [1-\eps/3, 1+\eps/3] \cdot \sat_{\bS_{E,1},C_{\sym}}(\psi)\\
\sat_{\bS_{E,0},C_{\sym}}(\psi) &\le \sat_{\bS,C_{\sym}}(\psi) \le \sat_{\bS_{E,1},C_{\sym}}(\psi)\\
\sat_{\bS_{E,0},\widehat{C}_{\sym}}(\psi) &\le \sat_{\bS,\widehat{C}_{\sym}}(\psi) \le \sat_{\bS_{E,1},\widehat{C}_{\sym}}(\psi)
\end{align*}

Combining these inequalities, we can deduce that \[
\sat_{\bS,\widehat{C}_{\sym}}(\psi) \in [1-\eps, 1+\eps] \cdot \sat_{\bS,C_{\sym}}(\psi)\\
\]
as desired. 
Finally, recall that by \cref{prop:uniform-sym} and \cref{eq:sym-eq}, we have for every $\psi \in D^n$ that
\begin{align*}
\sat_{\bS,C_{\sym}}(\psi) &= \sat_{\bR,C}(\psi)\\
\sat_{\bS,\widehat{C}_{\sym}}(\psi) &= \sat_{\bR,\widehat{C}}(\psi),
\end{align*}
Thus, $\widehat{C}$ is a $(1\pm \eps)$ sparsifier of $C$ for the valued relation $\bR$, as desired.
\end{proof}

\subsection{Independent Sparsifier Analysis}\label{subsec:7.6}

Together, the previous sections provide tight bounds on the sparsifiability of the \emph{complete} CSP instance. Ultimately however, our goal is to understand the sparsifiability of a \emph{random} CSP instance. In most of the cases we saw before, this distinction is without consequence, as the sparsifiers we built for the complete instance were achieved by independently, randomly sampling constraints. However, in the marginally balanced case for $\Sym(\bR)$, recall that we did not build our sparsifiers by independent random sampling (and instead we grouped constraints before sampling). In this section, we will identify a sub-case when i.i.d sampling suffices for sparsification, and also show that the associated condition is necessary for any non-trivial sparsification of random instances.

To start, we require some extra definitions:

Given $E \subseteq D$ and $s,t \in D^r$, we say that $s \sim_E t$ if $s_i \in E$ iff $t_i \in E$ for all $i \in [r]$.

\begin{definition}
Assume that $\bR : D^r \to \Z_{\ge 0}$ is precisely $k$-plentiful. We say that $\bR$ is \emph{marginally balanced} if for all $d \in E \subseteq D$ and $h \in \sT_{d,k}(\Sym(\bR))$ for which $h$ is $(d,E)$-rigid, we have that for all $s \in D^r$ with $\hist(s) = h$ and all $t \sim_E s$, we have that $\bR(s) = \bR(t)$.
\end{definition}

Note that $\bR$ being marginally balanced implies $\Sym(\bR)$ is as well and so is a stronger condition. Below, we show that it is a \emph{strictly} stronger condition (i.e., that there are relations $\bR$ such that $\Sym(\bR)$ is marginally balanced, but $\bR$ is not).

\begin{remark}\label{rem:balance-examples}
A very important fact is that there exists $\bR$ which are not marginally balanced, but $\Sym(\bR)$ is marginally balanced. For VCSPs, we can consider $D = \{0,1\}$ and $r=2$ and the following predicate $\bR_1 : \{0,1\}^2 \to \Z_{\ge 0}$.
\begin{align*}
    \bR_1(0,0) &= 1\\\bR_1(0,1) &= 2\\
    \bR_1(1,0) &= 0\\\bR_1(1,1) &= 1.
\end{align*}
Clearly both $\bR_1$ and $\Sym(\bR_1)$ are $2$-plentiful. Further, $\Sym(\bR_1)$ is marginally balanced since it is the constant function on $\Hist^{\{0,1\}}_2$. However, $\bR_1$ is not marginally balanced. To see why, consider $h \in \Hist^D_r$ with $h_0 = 2$ and $h_1 = 0$. It is straightforward to verify that $h$ is $(0,\{0,1\})$-rigid. However, $(0,0) \sim_{\{0,1\}} (0,1)$ but $\bR_1(0,0) \neq \bR_1(0,1)$.

A more complex example shows that this discrepancy can occur for ordinary (non-Boolean) CSPs. Consider $D = \{0,1,2\}$ and $r = 4$. We define \[\bR_2 := \{0022,1122,0222,1222,0122,2201\}\]
with $\bR_2 : D^r \to \{0,1\}$ 
the corresponding VCSP. It is straightforward to verify that $\bR_2$ is precisely $2$-plentiful. With the only $(d,2)$-tight histograms being $h = (2,0,2)$ 
for $d=0$ and $h' = (0,2,2)$ for $d=1$. We have that $h$ is $(0, \{0,1\})$-rigid and $h'$ is $(1, \{0,1\})$-rigid. Both histograms are in the same $\{0,1\}$-equivalence class and it is straightforward to verify that
\[
    \Sym(\bR_2)(2,0,2) = \Sym(\bR_2)(1,1,2) = \Sym(\bR_2)(0,2,2) = 4.
\]
Thus, $\Sym(\bR_2)$ is marginally balanced. However, $\hist(0022) = h$ and $0022 \sim_{\{0,1\}} 1022$, but $\bR_2(0022) \neq \bR_2(1022)$. Thus, $\bR_2$ is not marginally balanced.

Examples like $\bR_1$ and $\bR_2$ have a gap between the performance of the iid sparsifier and ``bundled'' sparsifier considered in the previous section. We establish this gap in \cref{subsec:7.7}.
\end{remark}

\subsubsection{Theorem Statement}

The goal of this section is to prove the following theorem.

\begin{theorem}\label{thm:UniformNonSym}
Let $\bR : D^r \to \Z_{\ge 0}$ be a valued relation which is precisely $k$-plentiful and marginally balanced. Then, for any $\eps \in (0,1)$ the i.i.d. sparsifier of $C$ which preserves $\widetilde{O}_{\bR}(n^{r-k}/\eps^3)$ constraints is a $(1\pm \eps)$ sparsifier with high probability.
\end{theorem}

Here by an i.i.d. sparsifier, we mean that we sample $\ell = \frac{n^{r-k}\log(n) \kappa(\bR)}{\eps^3}$ constraints uniformly at random (and give weight $m / \ell$ to each surviving constraint, where $m$ is the number of constraints in the starting instance). 

The proof proceeds similarly to \cref{subsec:7.5}. In particular, fix an assignment $\psi \in D^n$ which is $d$-dominant. Define $S_{\psi} \subseteq \Hist^D_r$ and $U_{\psi} \subseteq D$ as in \cref{subsec:7.5} If $\psi$ is not tight, then $\sat_{\bR,C}(\psi) \ge \Omega_{\bR}(n^{k+1}),$ so the probability that $\psi$ is sparsified correctly is at least $1-\frac{1}{2^{4|D|n}}$. Note that \cref{claim:U-rigid} still applies: any $h \in S_{\psi}$ with $h_d = k$ is $(d,k)$-tight and $(d,U_{\psi})$-rigid.

\subsubsection{Sandwiching}

The main technical difference between this section and \cref{subsec:7.5} is that we need a VR sandwich instead of an SVR sandwich. We construct this sandwich as follows.

Fix $d \in E \subseteq D$. Given $\bR$, we can define two sandwiching relations $\bR_{E,0}$ and $\bR_{E,1}$ as follows. For all $s \in D^r$, we have that
\begin{align*}
\bR_{E,0}(s) &:= \min_{t \sim_{E} s} \bR(t),\\
\bR_{E,1}(s) &:= \max_{t \sim_{E} s} \bR(t).
\end{align*}
We start with ensuring that the sandwich does not change the value of any $(d,k)$-tight histograms.

\begin{claim}\label{claim:tight-sandwich-vr}
Let $\psi \in D^n$ be $d$-dominant and tight and assume that $U_{\psi} = E$. Then, for all $h \in S_{\psi}$ with $h_d = k$, we have that for all $s \in D^r$ with $\set(s) = h$ that $\bR_{E,0}(s) = \bR_{E,1}(s) = \bR(s)$.
\end{claim}

\begin{proof}
First observe that any $h \in S_{\psi}$ with $h_d = k$ is $(d,k)$-tight, for if there is any $h' \prec_{\bar{d}} h$ with $h'_{d} \ge k+1$, then $h' \in S_{\psi}$ showing that $\psi$ is not tight. By \cref{claim:U-rigid} $h$ is $(d, E)$-rigid. Therefore, since $\bR$ is marginally balanced, for all $s \in D^r$ with $\set(s) = h$, for all $t \sim_E s$, we have that $\bR(s) = \bR(t)$. Thus, $\bR_{E,0}(s) = \bR_{E,1}(s) = \bR(s)$, as desired.
\end{proof}

Next, we prove an analogue of \cref{lem:sandwich-bound}.

\begin{lemma}\label{lem:sandwich-bound-vr}
Let $\psi \in D^n$ be $d$-dominant and tight and assume that $U_{\psi} = E$. Then for sufficiently large $n$, we have at least one of
\begin{enumerate}
\item $(1+\eps / 3) \cdot \sat_{\bR_{E,0}, C}(\psi) \ge \sat_{\bR_{E,1}, C}(\psi)$
\item $\sat_{\bR_{E,0}, C}(\psi) \ge \Omega_{\bR}(\eps) \cdot n^{k+1}$.
\end{enumerate}
\end{lemma}

\begin{proof}
Assume (1) is false. Then, there is $s \in \supp \bR_{E,1}$ with $\bR_{E,1}(s) > \bR_{E,0}(s)$ and
\[\sat_{s, C}(\psi) \ge \Omega_{\bR}(\eps) \cdot \sat_{\bR_{E,0}, C}(\psi).\]

By definition of $\bR_{E,1}$, there is $t \in \supp \bR$ with $t \sim_E s$ and $\bR_{E,1}(s) = \bR(t)$. Since $\bR$ is $k$-plentiful, there is $t' \in \supp \bR$ with $\hist(t') \prec_{\bar{d}} \hist(h)$ with $\#_d t' \ge k$. Note in particular that $\hist(t') \in S_{\psi}$, so $\#_d t' = k$ since $\psi$ is tight. Thus, $\hist(t')$ is $(d,k)$-tight and $(d, E)$-rigid by \cref{claim:U-rigid}.  Now observe that
\[
\sat_{s, C}(\psi) \ge \Omega_{\bR}(\eps) \cdot  \sat_{t', C}(\psi).
\]
In other words,
\[
\prod_{e \in D} \binom{\#_e\psi}{\#_es} \ge \Omega_{\bR}(\eps) \cdot  \prod_{e \in D} \binom{\#_e\psi}{\#t'_e}.
\]
Since both sides are nonzero, we can simplify the above inequality to
\begin{align}
  \prod_{e \in D \setminus \{d\}} (\#_e \psi)^{\#_es - \#_et'} \ge \Omega_{\bR}(\eps) \cdot n^{\#_ds - \#_dt'}.\label{eq:eps-prod-vr}
\end{align}

Since $\hist(t')$ is $(d,k)$-tight and $(d, E)$-rigid by \cref{claim:U-rigid}, we can deduce that (1) $\hist(t') \prec_{\bar{d}} \hist(s)$ as $\hist(t')(E \setminus \{d\}) = 0$ and (2) since $\bR$ is marginally balanced for all $s' \sim_{E} t'$ we have that $\bR(s') = \bR(t')$. Thus, we cannot have that $s \sim_E t'$. In other words, there exists $e \in D \setminus E$ for which $\#_e t' < \#_e s = \#_e t$.

Since $\hist(t') \prec_{\bar{d}} \hist(s)$, we have that $\#_d t' - \#_d s = \sum_{e' \in D \setminus \{d\}} \#_{e'} s - \#_{e'} t'$ with the terms in the sum nonnegative. Thus, by \cref{eq:eps-prod-vr}, we have that $\#_e \psi = \Omega_{\bR}(\eps n).$ Since $e \not\in U_{\psi}$, there is $h \in S_{\psi}$ with $h_d = k$ and $h_e \ge 1$. In particular, there is $t'' \in \supp \bR$ with $t'' = \hist(h)$. Since $h$ is $(d,k)$-tight, we have that $\bR_{E,0}(t'') = \bR(t'') > 0$. Thus,
\[
  \sat_{\bR_{E,0}, C}(\psi) \ge   \sat_{t'', C}(\psi) \ge \Omega_{\bR}(1) \cdot (\#_e \psi)^{h_e} \cdot (\#_d \psi)^{h_d} \ge \Omega_{\bR}(\eps) n^{k+1},
\]
as desired.
\end{proof}

Finally, like in \cref{lem:sandwich-bound-vr}, we show the two ends of the sandwich are sparsified correctly by $\widehat{C}$.

\begin{lemma}\label{lem:sandwich-spar-vr}
Fix $d\in E \subseteq D$. With high probability over the chosen constraints of $\widehat{C}$, for all $\psi \in D^n$ which are $d$-dominant and tight with $U_{\psi} = E$, we have that
\begin{align*}
\sat_{\bR_{E,0},\widehat{C}}(\psi) \in [1-\eps/3, 1+\eps/3] \cdot \sat_{\bR_{E,0},C}(\psi)\\
\sat_{\bR_{E,1},\widehat{C}}(\psi) \in [1-\eps/3, 1+\eps/3] \cdot \sat_{\bR_{E,1},C}(\psi)
\end{align*}
\end{lemma}

\begin{proof}
Let $\psi' \in D^n$ be defined by
\[
  \psi'_x = \begin{cases}
    d &  \psi_x \in E\\
     \psi'_x & \text{otherwise}.
  \end{cases}
\]
By definition of $\bR_{E,0}$ and $\bR_{E,1}$, we have that $\psi'$ and $\psi$ have identical codewords. Thus, it suffices to consider assignments of the form $\psi'$ (i.e., supported only on $(D \setminus E) \cup \{d\}$). Since $E = U_{\psi'}$ for all $e \in D \setminus E$, we either have (1) $\#_e\psi' \le r$ or (2) there exists $h^e \in S_{\psi'}$ with $h^e_d = k$ and $h^e_e \ge 1$. Let $E' \subseteq D \setminus E$ be the set of domain elements $e$ for which $h^e$ exists. For $e \in E'$, let $t^e \in \supp \bR$ for which $\hist(t^e) = h^e$. By \cref{claim:tight-sandwich-vr}, we have that $\bR_{E,0}(t^e) = \bR_{E,1}(t^e) = \bR(t^e)$. Therefore,

\begin{align*}
\sat_{\bR_{E,1},C}(\psi') &\ge \sat_{\bR_{E,0},C}(\psi')\\
&\ge \sum_{e \in E'} \sat_{t^e,C}(\psi')\\
&= \sum_{e \in E'} \#_e\psi' \cdot \Omega_{\bR}(n^{k})\\
&\ge (n - \#_d \psi' - r|D|) \cdot \Omega_{R}(n^k).
\end{align*}
In particular, for all $\lambda \ge 1$, the number of $\psi'$ for which $ \sat_{\bR_{E,0},C}(\psi') \le \lambda \cdot n^k$ is $n^{O_{\bR}(\lambda)}$.

Using \cref{cor:concentration}, we have that for any $\lambda \ge 1$, if a codeword $x$ has weight $[\lambda /2 , \lambda] \cdot \Omega_{\bR}(n^k)$, then the probability that $x$ is sparsified correctly is at least $1 - n^{-\Omega_{\bR}(\kappa(\bR) \lambda)}$. In particular, by our counting bound, the probability that all codewords of weight in the range $[\lambda /2 , \lambda] \cdot \Omega_{\bR}(n^k)$ are sparsified correctly is also $1 - n^{-\Omega_{\bR}(\kappa(\bR) \lambda)}$ for sufficiently large $\kappa(\bR)$. Therefore, the probability this works for all $\lambda$ is at least $1 - n^{-\Omega_{bR}(1)} \ge 1 - \frac{1}{2^{4|D|} \cdot \poly(n)}$, as desired.
\end{proof}

\subsubsection{Final Union Bound}

Fix $\psi \in D^n$ with $n$ sufficiently large. If $\sat_{\bR,C}(\psi) \ge \Omega_{\bR}(\eps n^{k+1})$ (where the hidden constant is smaller than the one from \cref{lem:sandwich-bound-vr}), then by \cref{cor:concentration}, we have that the probability that $\psi$ is sparsified correctly is at least $1 - \frac{1}{|D|^{4n}}$, which is sufficient for our union bound. Thus, we now assume that $\sat_{\bR,C}(\psi) = O_{\bR}(\eps n^{k+1})$. In particular, this means that $\psi$ is tight.

For all $d \in E \subseteq D$, condition on \cref{lem:sandwich-spar-vr} holding for $\bR_{E,0}$ and $\bR_{E,1}$. This altogether happens with probability at least $1 - \frac{1}{n^{\Omega_{\bR}(1)}}$. Thus, for our particular $\psi$, we are guaranteed by \cref{lem:sandwich-bound-vr} that
\begin{align*}
(1 + \eps / 3) \cdot \sat_{\bR_{E,0},C}(\psi) &\ge \sat_{\bR_{E,1},C}(\psi)\\
\sat_{\bR_{E,0},\widehat{C}}(\psi) &\in [1-\eps/3, 1+\eps/3] \cdot \sat_{\bR_{E,0},C}(\psi)\\
\sat_{\bR_{E,1},\widehat{C}}(\psi) &\in [1-\eps/3, 1+\eps/3] \cdot \sat_{\bR_{E,1},C}(\psi)\\
\sat_{\bR_{E,0},C}(\psi) &\le \sat_{\bR,C}(\psi) \le \sat_{\bR_{E,1},C}(\psi)\\
\sat_{\bR_{E,0},\widehat{C}}(\psi) &\le \sat_{\bR,\widehat{C}}(\psi) \le \sat_{\bR_{E,1},\widehat{C}}(\psi)
\end{align*}

Combining these inequalities, we can deduce that \[
\sat_{\bR,\widehat{C}}(\psi) \in [1-\eps, 1+\eps] \cdot \sat_{\bR,C}(\psi)\\
\]
as desired.

\subsection{Average case lower bound}\label{subsec:7.7}

The purpose of this section is to close one last gap in the analysis of random CSP instances. Assume that $\bR$ is precisely $k$-plentiful and $\Sym(\bR)$ is marginally balanced, but $\bR$ is \emph{not} marginally balanced. What is the sparsifiability of a random instance with $\lambda \cdot n^{r-k}$ clauses, where $\lambda = O_{\bR}(n)$? We show in this case no nontrivial sparsification can be achieved.

\begin{theorem}\label{thm:tighterLBRnotuniform}
Let $\bR$ be precisely $k$-plentiful and not marginally balanced. Consider a random instance $I$ of $\VCSP(\bR)$ with $\lambda \cdot n^{r-k}$ clauses, where $\lambda = O_{\bR}(n)$. Then any $(1\pm \eps)$ sparsifier $\hat{I}$ of $I$ for $\eps \in (0, O_{\bR}(1))$ requires at least $\Omega_{\bR}(\lambda \cdot n^{r-k})$ edges--even if $\hat{I}$ is not a subset of $I$.
\end{theorem}

\begin{proof}
Since $\bR$ is not marginally balanced, there is $d \in E \subseteq D$ such that $h \in \sT_{d,k}(\Sym(\bR))$ is $(d,E)$-rigid as well as $s, t \in D^r$ with $s \sim_E t$, $\set(s) = h$, and $\bR(s) \neq \bR(t)$.

Fix $T \in \binom{[n]}{r-k}$ and a tuple $p \in D^T$ such that $\hist(p) = h|_{D \setminus E}$. Let $\Psi_{T,p}$ be the set of all $\psi \in D^n$ for which $\psi|_{T} = p$ and $\psi_i \in E$ for all $i \in [n] \setminus T$. $h \in \sT_{d,k}(\Sym(\bR))$ any clause $x \in C$ (the complete instance) with $\bR(\psi|_{x}) > 0$ must have $T \subseteq \set(x)$.

Let $I_{T}$ be the set of $x \in I$ with $T \subseteq \set(x)$. From what we just discussed, if $|I_{T}| \ge 1$ then $|\hat{I}_T| \ge 1$ as well in order to be a $(1\pm \eps)$-sparsifier (even if $\hat{I}$ is not a subset of $I$). In particular, if $\lambda = O_{\bR}(1)$, we can first argue that
\[
  |\hat{I}| = \Omega_{\bR}(1) \cdot \sum_{T \in \binom{[n]}{r-k}} |\hat{I}_T| \ge \Omega_{\bR}(1) \cdot \sum_{T \in \binom{[n]}{r-k}} \one[|I_T| \ge 1],
\]
where the first equality comes from the fact that each set $x \in \hat{I}$ is of size $r$, and so can only cover $\leq 2^r = O_{\bR}(1)$ sets $T \in \binom{[n]}{r-k}$. 

Finally, we also observe that 
\[
\sum_{T \in \binom{[n]}{r-k}} \one[|I_T| \ge 1] = \Omega_{\bR}(1) \cdot |I|.
\]
To see why, we consider the procedure of sampling the sets $S \subseteq \binom{[n]}{r}$ that our predicates act on in the random CSP. Every time such a sampling is done, this set $S$ ``covers'' all of its subsets of size $\binom{r}{r-k}$. Because every set is equally likely to be covered, the probability that an arbitrary set of size $r-k$ is covered in a single sample is $\frac{\binom{r}{r-k}}{\binom{n}{r-k}} \geq \frac{1}{\binom{n}{r-k}}$. In fact, for our analysis, we even consider the more pessimistic sampling procedure which samples a constraint $S \subseteq \binom{[n]}{r}$, and then only covers a single random subset $T \subseteq S$ of size $r-k$, thus making the covering probability exactly $\frac{1}{\binom{n}{r-k}}$ for each set $T$ in each sample. 

We let $X_i$ denote the number of samples after the $i-1$st distinct set of size $r-k$ has been covered, before covering the $i$th distinct set of size $r-k$. Note that, if exactly $i-1$ distinct sets have been covered at some point, the number of additional samples until the $i$th set is covered is distributed as an independent geometric random variable with success probability $1 -\frac{i-1}{\binom{n}{r-k}}$. In particular, for $i \leq \frac{1}{2} \cdot \binom{n}{r-k}$, we can see that $X_i$ is stochastically dominated by a $\mathrm{Geom}(1/2)$ random variable. Now, the number of samples in the random CSP required before we find $\zeta \leq \frac{1}{2} \cdot \binom{n}{r-k}$ distinct sets $T$ is stochastically dominated by $\sum_{j = 1}^{\zeta} \mathrm{Geom}(1/2)$. By \cite{janson2018tail}, Theorem 2.1, we then observe that with probability $1 - 2^{- \Omega(\zeta)}$, this $\zeta$th distinct set occurs before sampling $4 \cdot \zeta$ random constraints. To conclude, this means that with very high probability, provided $|I| \leq 2 \cdot \binom{n}{r-k}$, then $I$ contains at least $|I|/4$ distinct sets $T \in \binom{[n]}{r-k}$, and if $|I| \geq 2 \cdot \binom{n}{r-k}$, then we simply use the lower bound that $I$ contains at least $\frac{1}{2} \cdot \binom{n}{r-k} = \Omega_{\bR} \left ( n^{r-k}\right )$ distinct sets $T \in \binom{[n]}{r-k}$. Because $|I| = \lambda \cdot n^{r-k}$, and $\lambda = O_{\bR}(1)$, we see that $\sum_{T \in \binom{[n]}{r-k}} \one[|I_T| \ge 1] = \Omega_{\bR}(n^{r-k}) = \Omega_{\bR}(1) \cdot |I|$, as claimed above.

So, all together, we see that 
\[
|\hat{I}|  \ge \Omega_{\bR}(1) \cdot \sum_{T \in \binom{[n]}{r-k}} \one[|I_T| \ge 1] = \Omega_{\bR}(|I|)
\]

Thus, we may now assume that $\lambda = \Omega_{\bR}(1)$. 

We let $\set(I_T)$ (and $\set(\hat{I}_T)$) denote the set of $i \in [n]$ for which $i \in \set(x)$ for some $x \in I_T$. Since $\lambda \in [\Omega_{\bR}(1), O_{\bR}(1)]$ we have for at least $2/3$ fraction of $T \in \binom{[n]}{r-k}$, we have that $|\set(I_T)| = \Theta_{\bR}(\lambda)$, call such a $T$ \emph{regular}.

We say that $x, y \in I_T$ are \emph{disjoint} if $\set(x) \cap \set(y) = T$. If $\lambda = O_{\bR}(n)$ and $T$ is regular, then the probability that any $x \in I_T$ is disjoint from every other $y \in I_T$ is at least $1 - q(\bR)$ for a value $q(\bR) \in [0,1]$ which we soon choose. 

To see this, suppose that there are $L$ sets in $I_T$. For each such set $x \in I_T$, we know that $T \subseteq x$, and the remaining $k \leq r$ elements in $x$ are uniformly random. Now, for a given set $x$ we want to understand the probability that $x$ is disjoint from every other set $y \in I_T$. In order for this to occur, it must be the case that for all other $L-1$ sets $y \in I_T$, that (besides the common set $T$ they share) these sets only contain elements from $[n] - T - x = [n] - x$, which is a set of size $n - r$. For a random $y$, the probability that this occurs (conditioned on sharing $T$) is exactly 
\[
\Pr[y \in I_T \text{ disjoint from }x] = \frac{\binom{n-r}{k}}{\binom{n}{k}} = \frac{(n-r)(n-r-1) \dots (n-r - k + 1)}{(n)(n-1) \dots (n-r+1)}
\]
\[
\geq \frac{(n-r - k +1)^k}{n^k } \geq \frac{(n-2r)^r}{n^r} \geq 1 - O_{\bR}(1/n).
\]
Now, in order for $x$ to be disjoint from \emph{all} possible sets $y$, this must occur for each of the $L-1$ sets, and thus
\[
\Pr[x \in I_T \text{ disjoint from all other }y] \geq \left ( 1 - O_{\bR}(1/n) \right )^{L-1}.
\]
Provided $L = O_{\bR}(n)$, this means that 
\[
\Pr[x \in I_T \text{ disjoint from all other }y] \geq \Omega_{\bR}(1).
\]
Now, if $\lambda = O_{\bR}(n)$, then observe that the expected number of sets in $I_T$ is 
\[
\E[|I_T|] = \Pr[x \in I_T] \cdot \lambda \cdot n^{r-k} = \frac{\binom{r}{r-k}}{\binom{n}{r-k}} \cdot \lambda \cdot n^{r-k} = \Theta_{\bR} \left ( \lambda \right ).
\]
By a simple Chernoff bound then, we see that with high probability, $L = \E[|I_T|] =\Theta_{\bR} \left ( \lambda \right ) = O_{\bR}(n)$, and conditioned on this, we obtain with high probability that $\Pr[x \in I_T \text{ disjoint from all other }y] \geq \Omega_{\bR}(1)$.

Given $x \in C$ with $T \subseteq \set(x)$, define the $T$-type of $x$, denoted by $\st_{T}(x) : T \to [r]$ such that $\st_T(x)(a) = i$ iff $x_i = a$. Recall there are  $s, t \in D^r$ with $s \sim_E t$, $\set(s) = h$, and $\bR(s) \neq \bR(t)$. We say that $x \in C$ with $T \subseteq \set(x)$ is \emph{$p$-consistent} if for all $a \in T$, $s_{\st_{T}(x)(a)} = p_a$. 

Observe the probability that a random edge $x \in C_T$ is $p$-consistent is $\Omega_{\bR}(1)$. Thus, in expectation, $I_T$ has at least $\Omega_{\bR}(\lambda)$ choices of $x \in I_T$ which are $p$-consistent \emph{and} $x$ is disjoint from every other $y \in I_T$. Call this set of clauses $I_{I,p}$. Call any regular $T \in \binom{[n]}{r-k}$ satisfying $|I_{T,p}| = \Omega_{\bR}(\lambda)$ \emph{good}. Note that at least $1/2$ fraction of $T \in \binom{[n]}{r-k}$ are good.

For such a good $T$, we seek to show that $|\hat{I}_T| \ge \Omega_{\bR}(\lambda)$. Assume for sake of contradiction that $|\hat{I}_T| = O_{\bR}(\lambda)$. Then, $\set(\hat{I}_T) = O_{\bR}(\lambda)$ as well. Let $I'_{T,p} \subseteq I_{T,p}$ be the set of $x \in I_{T,p}$ for which $\set(\hat{I}_T) \cap \set(x) \subseteq T$. We thus have that $|I'_{T,p}| = \Omega_{\bR}(\lambda)$.

To complete the argument, fix an arbitrary partial assignment $\psi' : \set(\hat{I}_T) \to D$ consistent with $p$, and consider all possible extensions $\psi \in \Psi_{T,p}$. In particular, since $\hat{I}$ is a $(1\pm \eps)$ sparsifier, all these $\psi$'s should be given the same weight by $I$. Since $T$ is regular, this common weight should be $\Theta_{\bR}(\lambda)$.

For each $x \in I_{T,p}$ by the disjointness property, we may assign $\psi|_{\set(x) \setminus T}$ independently. In particular, we can independently choose whether $\psi|_{x} = s$ or $\psi|_{x} = t$. Since $|I_{T,p}| = \Omega_{\bR}(\lambda)$, we can thus change the value of $\sat_{\bR,I}(\psi)$ by an additive $\Omega_{\bR}(\lambda)$. Let us denote these two assignments by $\psi$ and $\psi'$. This means that 
\[
\frac{\sat_{\bR,I}(\psi')}{\sat_{\bR,I}(\psi)} = \frac{\sat_{\bR,I}(\psi) + \Omega_{\bR}(\lambda)}{\sat_{\bR,I}(\psi)} = 1 + \frac{\Omega_{\bR}(\lambda)}{\sat_{\bR,I}(\psi)}= 1 + \frac{\Omega_{\bR}(\lambda)}{O_{\bR}(\lambda)} = 1 + \Omega_{\bR}(1).
\]

Hence, because $\hat{I}$ is a $(1 \pm \eps)$ sparsifier of $I$, we have that 
\[
(1 - \eps)\sat_{\bR,I}(\psi') \leq \sat_{\bR,\hat{I}}(\psi') = \sat_{\bR,\hat{I}}(\psi) \leq (1 + \eps) \cdot \sat_{\bR,I}(\psi),
\]
which means that 
\[
\frac{\sat_{\bR,I}(\psi')}{\sat_{\bR,I}(\psi)} \leq \frac{1 + \eps}{1 - \eps} \leq 1 + 3 \eps.
\]
So, for a constant choice of $\eps = O_{\bR}(1)$, we get a contradiction with the fact that 
\[
\frac{\sat_{\bR,I}(\psi')}{\sat_{\bR,I}(\psi)} =1 + \Omega_{\bR}(1),
\]
and hence $\hat{I}$ cannot be a $(1 \pm \eps)$ sparsifier of $I$. Thus, since at least $\frac{1}{2}\binom{n}{r-k}$ choices of $T$ are good, we have that $|\hat{I}| \ge \Omega_{\bR}(\lambda n^{r-k})$, as desired.
\end{proof}

\section{Translating Sparsifiability Bounds to Random Instances}\label{sec:random}

In this section, we relate sparsifiability bounds for \emph{complete} instances to (high probability) sparsifiability bounds for \emph{random} instances. Our formal models of random CSPs are defined below:

\begin{definition}[Random Uniform Model]
    For a relation $R \subseteq D^r$, a random CSP instance with $m$ constraints over $n$ variables is said to be sampled uniformly at random if each of the $m$ constraints is sampled by applying the predicate to the variables of an ordered set $S$ chosen at random (with replacement) from $P([n], r)$.
\end{definition}

\begin{definition}[Random $r$-Partite Model]
    For a relation $R \subseteq D^r$, a random CSP instance with $m$ constraints over $nr$ variables (broken into $r$ groups $V_1, \dots V_r$) is said to be a random $r$-partite instance if each of the $m$ constraints is sampled by applying the predicate to the ordered set of variables $S$ chosen at random (with replacement) from $V_1 \times V_2 \times \dots \times V_r$.
\end{definition}

With these formal definitions, our primary theorems from this section are the following:

\begin{theorem}\label{thm:uniformRandomFinal}
    Let $R \subseteq D^r$ be a relation, and let $C$ be a random uniform CSP with relation $R$ with $m$ constraints over $n$ variables. Then, with high probability:
    \begin{enumerate}
        \item If $R$ is \textbf{precisely} $k$\textbf{-plentiful}, and the \textbf{symmetrization} of $R$ is \textbf{not marginally balanced}, then (1) $C$ is sparsifiable to $\widetilde{O}_{\bR}(\min(m, n^{r-k+1} / \eps^2))$ constraints, and (2) for a sufficiently small, constant $\eps$, any sparsifier of $C$ requires $\Omega_{\bR}(\min(m, n^{r-k+1}))$ constraints. \label{item:uniformRandomFinal1}
        \item If $R$ is \textbf{precisely} $k$\textbf{-plentiful}, and $R$ is \textbf{marginally balanced}, then (1) $C$ is sparsifiable to $\widetilde{O}_{\bR}(\min(m, n^{r-k} / \eps^3))$ constraints, and (2) for a sufficiently small, constant $\eps$, any sparsifier of $C$ requires $\Omega_{\bR}(\min(m, n^{r-k}))$ constraints. \label{item:uniformRandomFinal2}
        \item If $R$ is \textbf{precisely} $k$\textbf{-plentiful}, the \textbf{symmetrization} of $R$ is \textbf{ marginally balanced}, but $R$ itself is \textbf{not marginally balanced}, then (1)  if $m \geq \kappa(\bR) \cdot n^{r-k+1} / \eps^2$ for a sufficiently large constant $\kappa(\bR)$, $C$ is  $(1 \pm \eps)$-sparsifiable to $\widetilde{O}_{\bR}(n^{r-k} / \eps^3)$ constraints, but (2) if $m \leq n^{r-k+1} / \kappa'(\bR)$ for a sufficiently large $\kappa'(\bR)$, there is a small, constant $\eps$, such that any sparsifier of $C$ requires $\Omega_{\bR}(m)$ constraints. \label{item:uniformRandomFinal3}
    \end{enumerate}
\end{theorem}

\begin{theorem}\label{thm:rpartiteRandomInstances}
    Let $R \subseteq D^r$ be a relation, and let $C$ be a random, $r$-partite CSP with relation $R$ with $m$ constraints over $n$ variables. Then, for $c$ being computed as in \cref{thm:RpartiteArbDomain}, with high probability:
    \begin{enumerate}
        \item For any $\eps > 0$, $C$ admits a $(1 \pm \eps)$ sparsifier of size $\min(m, \widetilde{O}(n^c / \eps^2))$. \label{item:rpartiteRandom1}
        \item There exists constant $\eps > 0$ for which $C$ does not admit a $(1 \pm \eps)$ sparsifier of size $\Omega(\min(m, n^c))$. \label{item:rpartiteRandom2}
    \end{enumerate}
\end{theorem}

\subsection{Bounds for Random Uniform Instances}

We prove the items from \cref{thm:uniformRandomFinal} separately, starting with \cref{item:uniformRandomFinal1}.

\begin{proof}[Proof of \cref{thm:uniformRandomFinal}, \cref{item:uniformRandomFinal1}.]
As per \cref{lem:complete-ub-coarse}, for any precisely $k$-plentiful relation $R$, there is a choice of $\ell = O_{\bR}(n^{r-k+1}/\eps^2)$, such that a random sample of $m \geq \ell$ constraints (with weight $\binom{n}{r} \cdot r! / m$ each), is a $(1 \pm \eps/10)$ sparsifier of the complete CSP with high probability (at least $9/10$). So, if $C$ is a random uniform CSP with $m$ constraints, and $m \geq \ell$, then with high probability the CSP instance $C $ re-weighted by $\binom{n}{r} \cdot r! / m$ is a $(1 \pm \eps/10)$ sparsifier of the complete instance. At the same time, we know that we can construct a $(1 \pm \eps/10)$ sparsifier of the complete uniform CSP instance with $\widetilde{O}(n^{r - k +1} / \eps^2)$ constraints, which we denote by $\widehat{C}$. In particular then, we obtain that $C$ is a $(1 \pm \eps/10)$ sparsifier of the complete instance, for which $\widehat{C}$ is also a $(1 \pm \eps/10)$ sparsifier. By composition, we then obtain that $C$ admits a $(1 \pm \eps)$ sparsifier with $\widetilde{O}(n^{r - k +1} / \eps^2)$ constraints. 

Next, if $m  = O_{\bR}(n^{r-k+1})$, then for a sufficiently large choice of $\kappa'(\bR)$, \cref{thm:tighterLBRnotuniform} immediately tells us that there exists a constant $\eps$ such that any $(1 \pm \eps)$ sparsifier of $C$ must preserve $\Omega(m)$ re-weighted constraints (note that we can invoke \cref{thm:tighterLBRnotuniform} because if $\mathrm{Sym}(\bR)$ is not marginally balanced, then $\bR$ is not marginally balanced). Otherwise, if $m =\Omega_{\bR}(n^{r-k+1}) $, by the above, there is a choice of $\ell = O_{\bR}(n^{r-k+1}/\eps^2)$, such that a random sample of $m \geq \ell$ constraints (with weight $\binom{n}{r} \cdot r! / m$ each), is a $(1 \pm \eps/10)$ sparsifier of the complete CSP with high probability (at least $9/10$). In particular, any $(1 \pm \eps/10)$ sparsifier of $C$ would be (when weighted by $\binom{n}{r} \cdot r! / m$), a $(1 \pm \eps/10)^2$ sparsifier, and thus a $(1 \pm \eps)$ sparsifier of the complete CSP. But, by \cref{thm:UniformGeneral}, \cref{item:non-unif-lb}, there exists a choice of $\eps = \Omega_{\bR}(1)$ such that any $(1 \pm \eps)$ sparsifier of the complete instance requires $\Omega_{\bR}(n^{r-k+1})$ constraints. 
\end{proof}

\begin{proof}[Proof of \cref{thm:uniformRandomFinal}, \cref{item:uniformRandomFinal2}.]

As before, we start with the upper bound. By \cref{thm:UniformNonSym}, we know that there is a choice of $\ell =\widetilde{O}_{\bR}(n^{r-k}/\eps^3)$, such that the i.i.d. sparsifier of the complete instance which preserves $\widetilde{O}_{\bR}(n^{r-k}/\eps^3)$ constraints is a $(1\pm \eps/10)$ sparsifier with high probability. In particular, this means that if $m \geq \ell$, we have that $\frac{\binom{n}{r} r!}{m}\cdot C$ is a $(1 \pm \eps/10)$ sparsifier of the complete instance (with high probability). Likewise, a random sample of $\ell$ constraints from $C$ (denote this by $\hat{C}$) is still a uniformly random sample of $\ell$ constraints from the complete instance, and so is also a $(1 \pm \eps/10)$ sparsifier of the complete instance. By composition, we then see that $\hat{C}$ is a $(1 \pm \eps)$ sparsifier of $C$, and has $\ell = \widetilde{O}_{\bR}(n^{r-k}/\eps^3)$ constraints. 

Next, we proceed to the lower bound. Recall that because $\bR$ is not $k+1$ plentiful, there exists $d \in D$ and $h \in \sT_{d,k}(\bS)$. Let $\Phi$ be the set of assignments $\psi \in D^{n}$ such that $\#_e \psi = \#_e h$ for all $e \in D \setminus \{d\}$. (Note we are assuming that $n \ge k$.)
Let $\bS := \Sym(\bR)$. Since $\bR$ is not $k+1$-plentiful, there exists $d \in D$ and $h \in \sT_{d,k}(\bS)$. Let $\Phi$ be the set of assignments $\psi \in D^{n}$ such that $\#_e \psi = \#_e h$ for all $e \in D \setminus \{d\}$. (Note we are assuming that $n \ge k$.)

Now, consider a constraint $x$ such that $\bR(\psi_x) > 0$. By definition of $\psi$, we must have that $\hist(\psi_x) \prec_{\bar{d}} h$. However, since $h$ is $(d,k)$-tight, this implies that $\hist(\psi_x) = h$. As such, any clause $x$ which $\psi$ satisfies must completely contain $\psi^{-1}(D \setminus \{d\})$. 

Since $|\psi^{-1}(D \setminus \{d\})| = h(D \setminus \{d\}) = r -k,$ there are $\binom{n}{r-k} = \Omega_{\bR}(n^{r-k})$ choices for $\psi^{-1}(D \setminus \{d\})$. Any $x $ can overlap at most $\binom{r}{r-k} = O_{\bR}(1)$ such sets. 

Now, for our satisfying tuple $h \in D^r$, we consider the exact positions of the $r-k$ variables in $h$ which are not equal to $d$. I.e., there is some subset $W \subseteq [r]$ such that $h^{-1}(d) = [r] - W$. Now, as we sample random constraints $x$, we keep track of which variables occupy these positions $W$ in the constraint: naturally, there are $\binom{n}{|W|} \cdot |W|!$ possible ways to choose which variables fill these positions, and for each constraint $x$ chosen at random, we \emph{randomly} choose which variables are in these positions according to $W$. Now, we create sets $X_T: T \in P([n], |W|)$ corresponding to which variables are in positions $W$. For each constraint $x$ we add, we place it in the set $X_T$ such that $T = x|_W$.

Now, our claim is that after adding $m$ constraints, with high probability, at least $\Omega \left ( \min(m, n^{r-k}) \right )$ of the sets $X_T$ have at least one constraint. This follows because for any $T \in P([n], |W|)$, $\Pr[x|_W = T'] = \frac{1}{\binom{n}{r-k} \cdot (r-k)!}$. So, across our $m$ random constraints, 
\[
\Pr[\exists x: x|_W =T] = 1 - \left(1 - \frac{1}{\binom{n}{r-k} \cdot (r-k)!}\right)^m = \Omega \left (\frac{\min(m, n^{r-k})}{n^{r-k}} \right ).
\]
In particular, this means that 
\[
\E[\sum_{T} \mathbf{1}[|X_T| > 0] = \Omega(\min(m, n^{r-k})).
\]
In fact, we can even get stronger concentration. We know that there are $\binom{n}{r-k} \cdot (r-k)!$ different $X_T$. Let us consider the procedure of adding each $x_i$ one at a time. After $x_{1}, \dots x_{i-1}$ are added, if there are still $\leq \binom{n}{r-k} \cdot (r-k)!/2$ different $X_T$'s which have one set, then in the next iteration, the probability that $x_i$ occupies a previously empty $X_T$ is at least $1/2$. We consider the variables $Z_1, \dots Z_m$ which denote the indicator of whether the $i$th constraint $x_i$ occupies a previously empty set $X_T$. The above implies that, conditioned on $\sum_{T} \mathbf{1}[|X_T| > 0] \leq \binom{n}{r-k} \cdot (r-k)!/2$, then $\sum_{i \in [m]} Z_i$ is stochastically lower bounded by a sum of $m$ independent $\mathrm{Bern}(1/2)$. However, by a simple Chernoff bound, with probability $1 - 2^{-\Omega(m)}$, $\sum_{i \in [m]} Z_i \geq \frac{m}{4}$. Thus, with probability $1 - 2^{-\Omega(m)}$, $\sum_{T} \mathbf{1}[|X_T| > 0] $ is either at least $\binom{n}{r-k} \cdot (r-k)!/2$, or at least $m/4$, and hence is $\Omega(\min(m, n^{r-k}) )$.

Finally, for each set $X_T$ which has an $x \in X_T$, the assignment $\psi$ such that $\psi_T = h$ is satisfied by $x$. So, we must have that there are $\Omega(\min(m, n^{r-k}))$ different assignments $\psi \in \Phi$ which satisfy at least one constraint. To conclude, we recall that $x $ can only be satisfied by at most $\binom{r}{r-k} \cdot r!= O_{\bR}(1)$ many $\psi \in \Phi$, as $x$ must contain all of $\psi^{-1}(D \setminus \{ d\})$, and there are $\binom{r}{r-k} \cdot r!$ many such ordered subsets in $x$ of size $r-k$. So, any sparsifier must preserve at least
\[
\frac{|\{X_T: |X_T| \geq 1|}{O_{\bR}(1)} = \Omega(\min(m, n^{r-k}))
\]
many constraints, with high probability.
\end{proof}

\begin{proof}[Proof of \cref{thm:uniformRandomFinal}, \cref{item:uniformRandomFinal3}.]

If $m \geq \kappa(\bR) \cdot n^{r-k+1} / \eps^2$ for a sufficiently large constant $\kappa(\bR)$, then, as in \cref{lem:complete-ub-coarse}, we can observe that $C$ re-weighted by $\frac{\binom{n}{r}\cdot r!}{m}$ is a $(1 \pm \eps/10)$ sparsifier of the complete, uniform CSP instance, as these sparsifiers are created by uniform, random sampling with replacement. In particular, we can then invoke \cref{thm:UniformGeneral}, \cref{item:unif-ub} which states that if $\Sym(\bR)$ is marginally balanced, then for any $\eps \in (0,1)$, there exists a $(1 \pm \eps)$ sparsifier (denote this $\hat{C}$) of the complete instance which preserves $\widetilde{O}_{\bR}(n^{r-k}/\eps^3)$ constraints. Now, by composing sparsifiers, we obtain that $\hat{C}$ is a $(1 \pm \eps/10)^2$ and thus a $(1 \pm \eps)$ sparsifier of our random CSP $C$ (re-weighted by $\frac{\binom{n}{r}\cdot r!}{m}$). Multiplying by the inverse of the weighting then yields the upper bound. 

For the lower bound, we can directly invoke \cref{thm:tighterLBRnotuniform}.
\end{proof}

\subsection{Bounds for Random $r$-Partite Instances}

To start, we prove the lower bound of \cref{thm:rpartiteRandomInstances}.

\begin{proof}[Proof of \cref{thm:rpartiteRandomInstances}, \cref{item:rpartiteRandom2}]
First, note that if $c = 0$, then the lower bound is trivial. So, we focus only on the case when $c \geq 1$.

As before, because the value of $c$ represents the maximum arity \textbf{AND} that can be restricted to in $R$, this means that there exists a set of $D_1, \dots D_r \subseteq D$, such that $c$ of the $D_i$'s are of size $2$ and the remaining $D_i$'s are of size $1$. Now, let $T \subseteq [r]$ contain exactly the indices $i \in [r]$ for which $|D_i| = 2$.

Now, let $S_1, \dots S_m$ denote the ordered sets that are sampled in the random CSP $C$. I.e., each $S_{\ell} \in V_1 \times V_2 \times \dots \times V_r$, and each constraint is the result of applying the relation $R$ to the variables $x_{S_{\ell}}$. In order to derive our lower bound, we will be concerned with how many different $c$-tuples of variables will appear in indices corresponding to $T$. Formally, let $W \in [n]^T$ denote a choice of $|T|$ indices. We let $X_W = \mathbf{1}[\exists S_{\ell}: S_{\ell}|_T = W]$ denote the indicator vector of whether there exists an $S_{\ell}$ for which $S_{\ell}$ agrees with $W$ on the coordinates in the set $T$. 

Because each $S_{\ell}$ is chosen at random from the $r$-partite model (and for now, assuming with replacement), this means that for any $W$, $\Pr[S_{\ell}|_T = W] = \left ( \frac{1}{n} \right )^{|T|}= 1 / n^c$. Across all $m$ choices of $S_{\ell}$, we then get that 
\[
\Pr[\exists \ell: S_{\ell}|_T =W] = 1 - (1 - 1 / n^c)^m = \Omega \left (\frac{\min(m, n^c)}{n^c} \right ).
\]
In particular, this means that 
\[
\E[\sum_{W} X_W] = \Omega(\min(m, n^c)).
\]

We know that there are $n^{|T|}$ different $X_T$. Let us consider the procedure of adding each $S_{\ell}$ one at a time. After $S_{1}, \dots S_{i-1}$ are added, if there are still $\leq n^{|T|}/2$ different $X_W$'s which have one set, then in the next iteration, the probability that $S_i$ occupies a previously empty $X_W$ is at least $1/2$. We consider the variables $Z_1, \dots Z_m$ which denote the indicator of whether the $i$th constraint $S_i$ occupies a previously empty set $X_W$. The above implies that, conditioned on $\sum_{W} \mathbf{1}[|X_W| > 0] \leq n^{|T|}/2$, then $\sum_{i \in [m]} Z_i$ is stochastically lower bounded by a sum of $m$ independent $\mathrm{Bern}(1/2)$. However, by a simple Chernoff bound, with probability $1 - 2^{-\Omega(m)}$, $\sum_{i \in [m]} Z_i \geq \frac{m}{4}$. Thus, with probability $1 - 2^{-\Omega(m)}$, $\sum_{W} \mathbf{1}[|X_W| \geq 1] $ is either at least $n^{|T|}/2$, or at least $m/4$, and hence is $\Omega(\min(m, n^{r-k}) )$. We let $Q$ denote this set of $W$'s. I.e., $Q \subseteq [n]^{T}$, and $|Q| = \Omega(\min(m, n^c))$ with probability at least $9/10$.

To conclude then, we show that any sparsifier of $C$ must retain at least $|Q|$ constraints. As in \cref{sec:rpartiteGeneral}, we let the single satisfying assignment in $R \cap D_1 \times D_2 \times \dots \times D_r$ be denoted by $a_1 a_2 \dots a_r$. When $|D_i| = 2$, we let $b_i$ be the single element in $D_i - \{a_i \}$. To obtain our set of assignments $X \subseteq D^{nr}$, for every $i \in T$, we choose a special \emph{distinguished vertex} $x^{(i)}_{j^*_i} \in V_i$. The assignment $x \in D^{nr}$ is then given by setting $x^{(i)}_{j} = a_i$ for every $i \in [r] -T, j \in [n]$. For the remaining $i \in T$, we set $x_{j^*_i}^{(i)} = a_i$, and for all other $j \in V_i - \{j^*\}$, we set $x_j^{(i)} = b_i$. In this way, the assignment $x$ is completely determined by our choices of $j^*_i: i \in T$. In fact, we can immediately see that the number of possible assignments we generate is exactly $n^c$ (since $|T| = c$, and for each set in $T$, there are $n$ options for $j^*_i$). 

Next, as in \cref{sec:rpartiteGeneral}, we show that for any two assignments $x, x'$ generated in the above manner, there is no constraint $R_S$ which is satisfied by \emph{both} $x$ and $x'$. To see this, recall that any constraint $R_S$ operates on a set of variables $S \in V_1 \times V_2 \times \dots \times V_r$ (we denote this set by $(j_1, j_2, \dots j_r)$). For any assignments $x, x'$, by construction, we enforce that for each $i \in [r]$, and for each $j \in [n]$, $x^{(i)}_j \in D_i$ (and likewise for $x'$). In particular, this means that the only way for $R_S(x)$ to be satisfied is if $x^{(1)}_{j_1} = a_1, x^{(2)}_{j_2} = a_2, \dots x^{(r)}_{j_r} = a_r$. However, by construction of our assignment $x$, for any $i \in T$, there is only one choice of index $j_i \in [n]$ for which $x^{(i)}_{j_i} = a_i$, namely when $j_i = j^*_i$ (the distinguished index). For every $i \in T$, this forces $j_i^* = j_i$. However, this yields a contradiction, as this would imply that $x, x'$ both have the same set of distinguished indices, and would therefore be the same assignment.

The lower bound then follows because for every $W \in Q$, the assignment $x_W$ whose distinguished vertices are exactly $W$ will satisfy at least one constraint in $C$. Thus, there are $|Q|$ distinct satisfying assignments in $X$, and no single constraint can be satisfied by two assignments. So in order to get positive weight for each of the $|Q|$ distinct assignments, we must retain $|Q| = \Omega(\min(m, n^c))$ distinct constraints. 
\end{proof}

Next, we show the upper bound.

\begin{proof}[Proof of \cref{thm:rpartiteRandomInstances}, \cref{item:rpartiteRandom1}]

First, note that we can assume that $m \geq w$, where $w = \frac{\kappa(r, |D|) \log(n) \cdot n^c}{\eps^3}$ as set in the proof of \cref{thm:RpartiteArbDomain}, \cref{item:BooleanRpartiteArbDom1}. If $m$ is smaller than this value of $w$, then we simply return the CSP $C$ itself, as it is already meeting our desired sparsity bound. 

Otherwise, observe that if $m > w$, then sampling $m$ random constraints is strictly better than sampling $w$ random constraints in terms of the concentration of \cref{cor:concentration}. That is to say, if we sample $m$ random constraints yielding CSP $C$, and assign each sampled constraint weight $n^r / m$, then the resulting CSP (which we denote by $\frac{n^r}{m} \cdot C$) is a $(1 \pm \eps)$ sparsifier of the complete $r$-partite instance with probability at least $9/10$ (as per \cref{thm:RpartiteArbDomain}). 

At the same time, observe that if we first sample $m$ random constraints, and then further sub-sample these $m$ random constraints down to $w$ constraints, this yields the exact same distribution over $w$ constraints as randomly sampling $w$ constraints initially. So, if we let the complete $r$-partite instance be denoted by $\widehat{C}$ and let $C''$ denote the result of sub-sampling $C$ to $w$ constraints, this means that:
\begin{enumerate}
    \item With probability $\geq 9/10$, $\frac{n^r}{m} \cdot C$ is a $(1 \pm \eps)$ sparsifier of $\widehat{C}$.
    \item With probability $\geq 9/10$, $\frac{n^r}{w} \cdot C''$ is a $(1 \pm \eps)$ sparsifier of $\widehat{C}$.
\end{enumerate}

In particular, this means with probability $\geq 4/5$, $\frac{m}{w} \cdot C'' $ is a $(1 \pm \eps)^2$ sparsifier of $C$. By starting the argument instead with $\eps / 3$, we then obtain that with probability $\geq 4/5$, $C$ admits a $(1 \pm \eps)$ sparsifier with $\widetilde{O}(n^c / \eps^3)$ sub-sampled constraints. This yields the claim. 
\end{proof}

Note that using the same reasoning above, but instead with \cref{thm:RpartiteVCSP} yields an analogous characterization for $r$-partite \emph{valued} CSPs, yielding \cref{thm:introrpartiteVCSP}.

\bibliographystyle{alpha}
\bibliography{ref}

\end{document}